\documentclass[11pt,preprint]{aastex}

\usepackage{natbib}
\bibpunct{(}{)}{;}{a}{}{,}

\begin{document}

\title{Evolution of emission line activity in intermediate mass young stars}

\author{P. Manoj}
\affil{Academia Sinica, Institute of Astronomy \& Astrophysics, 
            P. O. Box 23-141, Taipei 10617, Taiwan}
\email{manoj@asiaa.sinica.edu.tw}

\author{H. C. Bhatt}
\affil{Indian Institute of Astrophysics, II Block, Koramangala,
             Bangalore 560034, India} 
\email{hcbhatt@iiap.res.in}

\author{G. Maheswar}

\affil{Aryabhatta Research Institute of Observational Sciences (ARIES)
Manora Peak, Nainital 263 129, India}
\email{maheswar@aries.ernet.in}

\and

\author{S. Muneer}
\affil{Indian Institute of Astrophysics, II Block, Koramangala,
             Bangalore 560034, India} 
\email{muni@itiltd.co.in}

\begin{abstract}
We present optical spectra of 45 intermediate mass Herbig Ae/Be
stars. Together with the multi-epoch spectroscopic and photometric
data compiled for a large sample of these stars and ages estimated for
individual stars by using pre-main sequence evolutionary tracks, we
have studied the evolution of emission line activity in them. We find
that, on average, the H$\alpha$ emission line strength decreases with
increasing stellar age in HAeBe stars, indicating that the accretion
activity gradually declines during the PMS phase. This would hint at a
relatively long-lived (a few Myr) process being responsible for the
cessation of accretion in Herbig Ae/Be stars. We also find that the
accretion activity in these stars drops substantially by $\sim$ 3
Myr. This is comparable to the timescale in which most intermediate
mass stars are thought to lose their inner disks, suggesting that
inner disks in intermediate mass stars are dissipated rapidly after
the accretion activity has fallen below a certain level.  We, further
find a relatively tight correlation between strength of the emission
line and near-infrared excess due to inner disks in HAeBe stars,
indicating that the disks around Herbig Ae/Be stars cannot be entirely
passive.  We suggest that this correlation can be understood within
the frame work of the puffed-up inner rim disk models if the radiation
from the accretion shock is also responsible for the disk heating.
\end{abstract}


\section{Introduction}

There has been considerable interest in the evolutionary studies of
circumstellar disks around young stars in the past few years. The
primary motivation for such studies has been the realisation that
young circumstellar disks are the potential sites of planet
formation. There is now overwhelming observational evidence for the
presence of circumstellar disks around a majority of young pre-main sequence
(PMS) stars \citep[e.g.][]{becksar96, mansar97, wein02, eisner03}. The
observed disk masses, sizes and chemical composition of young
circumstellar disks are analogous to that of the {\it protosolar
nebula} \citep[e.g.][]{beck99, hill05}. How these disks evolve into
planetary systems is one of the fundamental questions in astronomy.

In the early PMS phase, young stars actively accrete matter from the
surrounding optically thick, gas-rich disk \citep [e.g.][]{hart98} .
As the star evolves, the disk accretion rate goes down and eventually
the accretion is terminated. By the time the central stars reach the
main sequence, young circumstellar disks surrounding them lose most of
the material due to planet formation and other disk dispersal
processes. Characterising this evolution of circumstellar disks,
therefore, is critical to improve our understanding of the planet
formation process and disk dispersal mechanisms and their associated
timescales.

Temporal evolution studies of circumstellar disks use different
observational diagnostics of the disks such as emission lines,
continuum excess in the UV, near-infarred (NIR) excesses and
millimiter and sub-mm excesses which trace different properties of
the disks with varying efficacy. For example, emission lines and
continuum excess in the UV trace level of accretion activity in young
stars, NIR excess traces the hot inner disk and millimeter and sub-mm
excesses trace cold dust in the outer disk.

Most disk evolution studies in the literature have been based on the
NIR excess which traces the evolution of inner accretion disks
\citep[e.g.][]{strom89, skrut90, sek93, haisch01, haisch01b, hill03, hill05}. 
From the JHKL imaging studies of embedded and revealed clusters of
different ages, \citet{haisch01} have found that essentially all the
stars in a cluster lose their disks in $\sim$ 6 Myr. \citet{hill05}, 
 for a much larger sample of young clusters and associations obtained
the median lifetime of the optically thick accretion disks to be 2-3
Myr, but with a large scatter, especially at young ages ($<$ 1 Myr).
All these studies are consistent with the fact that the inner disks in
most young stars do not last for more than 5 Myr.

How long does emission line activity persist in young stars? Emission
lines are more closely related to accretion activity than the JHKL
excess \citep[e.g.][]{muz04}. Excesses at the NIR wavelengths can be
produced even by passive disks if there is hot ( T $\sim$ 1000 K) dust
present close to the star. Therefore the evolution of accretion is
better diagnosed with emission lines. In this contribution we study
the evolution of emission line activity in intermediate mass young
stars. With the data compiled for a large sample of Herbig Ae/Be(
hereafter HAeBe) stars, which are PMS objects of intermediate mass
($1\leq M/M_{\odot} \leq 8$), we study the evolution of emission line
activity in these stars.

One of the prominent observational features of HAeBe stellar group and
that which distinguishes these stars from the normal main sequence
stars is the presence of emission lines in the
spectra. Apart from the H$\alpha$ line which is almost always in
emission, the other emission lines that are often observed in the
optical spectra of HAeBe stars are HeI ($\lambda\:5876
\AA$ \& $\lambda\:6678 \AA$), OI ($\lambda \:7774 \AA$ \& $\lambda\:
8446 \AA$) and CaII triplet ($\lambda \lambda \:8498, 8542, 8662 \AA$)
\citep{herbig60, cohenkuhi79, finkmun84, hp92, bohm95} and forbidden
emission lines such as [OI] ($\lambda\:6300$ and $\lambda\:6364$)
\citep{herbig60, ham94, bc94, bh97, corray98, vieira03, acke05}.

In the low mass PMS T Tauri stars, the origin of
emission lines is currently understood in the framework of
magnetospheric accretion model where emission lines are believed to be
formed in the magnetospheric accretion columns \citep{us85, kon91,
hhc94, mch98, muzerolle01}. However, in HAeBe  stars the
situation is less clear. Nevertheless, recent high spatial resolution
interferometric observations at millimeter (mm), sub-mm and NIR
wavelengths have revealed the presence of gas and dust circumstellar
disks with inner holes, and possibly with puffed up inner rims, around
a large number of HAeBe stars \citep{ngm00, mansar97, mansar00,
petu03, ohashi05, mill01, eisner03, eisner04, monn05}. Observed disk
properties are in general similar to those found for the disks
surrounding T Tauri stars. Spectropolarimetric observations of linear
polarization across emission lines in HAeBe stars have clearly
demonstrated that the emission lines originate in a compact region
close to the star \citep{vink02, vink05}. Magnetic fields also have
been detected in a few HAeBe stars \citep{hub04,
wade05}. Magnetospheric accretion model has now been succesfully
applied to HAeBe stars \citep{muz04}. This evidence
supports the fact that emission lines in HAeBe stars are related
to accretion.

We obtained optical spectra of 45 HAeBe stars in order to investigate
the evolutionary trends in the emission line activity of these
stars. We also compiled available photometric and spectroscopic data
for a large number of these stars. Based on the multi-epoch
photometric and spectroscopic data thus obtained for a sample of 91
stars, we studied the evolution of emission line activity in HAeBe
stars. In Section 2 of the paper we describe our sample. Spectroscopic
observations obtained are presented in Section 3. Main results of our
study are presented in Section 4 and discussed in detail in Section
5. Finally, in Section 6 we summarize our conclusions.

\section{Sample}

Our sample of stars is taken from the catalogues of \citet{the94} and
\citet{vieira03} and consists of {\it `bonafide'} HAeBe 
stars whose PMS  status has been established and
for which photometric data and reasonable distance estimates are
available in the literature. We have included only B, A \& F type stars in
our sample. Further, stars which are known to be classical Be stars
and for which there is no clear evidence for the presence of circumstellar dust
\citep[e.g.][]{van98} have been excluded. Our sample consists of 91
HAeBe stars.

We compiled BVR photometric data from the literature for the stars in
our sample except for 5 stars, for which R band data is not
available. For 76 stars in our sample (84\%) we have multi-epoch
photometric data compiled from \citet{hs99}, \citet{dewinter01},
\citet{oped01} and \citet{herbigbell88}. For each star, 
we use a combination of magnitudes and colors measured
quasi-simultaneously at a roughly mean magnitude in our analysis.

Recently, there have been attempts to accurately determine the
spectral types of HAeBe stars \citep[e.g.][]{hernand04,mora01}.
We have compiled these refined spectral type estimates for the stars
in our sample from \citet{hernand04}, \citet{mora01} and
\citet{vieira03}. We also compiled distances for all the stars in our 
sample from the literature.

In Table \ref{data} we list the BVR magnitudes, spectral types and
distances obtained from the literature for the 91 stars in our
sample. All the magnitudes and colors listed in Table \ref{data} are
in the Johnson system. Wherever the R magnitudes and/or V-R colors
were listed in the Cousins system in the literature, we converted
those into the Johnson system following the relation given in
\citet{bessell83}.

\section{Observations}

Optical CCD spectra centered at the H$\alpha$ line ($\lambda$ $\sim$
6562 $\AA$ ) were obtained for 45 stars in our sample with the
Optomechanics Research (OMR) spectrograph on the 2.34 meter Vainu
Bappu Telescope (VBT) at the Vainu Bappu Observatory, Kavalur, India
and with the Hanle Faint Object Spectrograph (HFOSC) on the 2 meter
Himalayan Chandra Telescope at the Indian Astronomical Observatory,
Hanle, India. The projected slit widths employed for the observations
ranged from 1-2 $\arcsec$ resulting in resolutions (R =
$\lambda/\Delta \lambda$) ranging from $1000\:-\: 3000$. 
The log of spectroscopic observations is presented in Table \ref{log}.

Spectral reduction was carried out in the standard manner using the
IRAF{\footnote{ IRAF is distributed by National Optical Astronomy
Observatories, USA.}} package. All the spectra were first bias
subtracted and then flat-field corrected. Thereafter, the sky
subtracted spectra were extracted using the task APALL. Sky
subtraction was achieved by interpolating the sky spectra on both
sides of the target spectrum. The spectra were then wavelength
calibrated using the FeNe comparison spectra obtained during the
observing run. Spectra obtained on some nights with VBT/OMR showed
strong telluric absorption lines of H$_2$O and O$_2$. Telluric lines
from these spectra were removed using the standard stars observed on
those nights and using the task TELLURIC in IRAF. Standard star
spectrum was first scaled and shifted to match the intensity and
wavelength of the telluric lines in the object spectra and then object
spectra were divided by the standard spectrum. Removal of the telluric
lines has been satisfactory for most stars although in a few spectra
some residuals remain. Final reduced spectra of 45 stars are presented
in Fig.~\ref{spectra}.

\section{Results}

\subsection{Ages of Herbig Ae/Be stars}

We derived the ages of the stars in our sample by placing them in the
Hertzsprung- Russel (HR) diagram and comparing with theoretical PMS
evolutionary tracks. To compute the luminosities of the
stars and to locate them in the HR diagram, a good estimation of the
extinction towards these stars is very important. It has been shown by
several authors that the value of the ratio of total-to-selective
extinction R$_V$ towards HAeBe stars is higher than that for
the diffuse interstellar medium
\citep{strom72, herbst82, gortibhatt93, hernand04}. Recently,
\citet{hernand04} have demonstrated that a value of R$_V$ = 5 fits the
observations better than the average interstellar value of 3.1 for a
large sample of HAeBe stars. This high value of R$_V$ strongly
suggests that the average grain size in the circumstellar environment
around HAeBe stars is larger than that in the diffuse interstellar
medium. In the following we calculate R$_V$ for the stars in our
sample from their B - V and V - R colors.

For the 86 stars in our sample for which we have both B - V and V - R
colors, the color excesses E(B - V) and E(V - R) were calculated using
the intrinsic colors for the spectral types given in
\citet{kenhart95}. For 4 stars the color excesses turned out to be negative 
and we have called it to zero.  For the remaining 82 stars we computed
the ratio of total-to-selective extinction R$_V$ from the ratio of E(B - V) and
E(V - R) color excesses using the following relation from \citet{ccm89}. 
\[\frac{E(V - R)}{E(B - V)}\:=\:\frac{0.1314 R_V\:+\:0.3660}{1.0495\:-\:0.0018 R_V}\]

For most of the photometric data found in literature the errors in the
measurements are not quoted. For measurements where errors are quoted,
typical photometric errors are $\sim$ 1\% or 0.01 mag. Taking the
uncertainity in the spectral type estimation to be 2 sub-classes, rms
errors in E(B - V) and E(V - R) are 0.06 and 0.05 respectively. For 77
stars in our sample the values of color excesses E(B-V) and E(V-R) are
$\ge$ 1$\sigma$ error. We computed R$_V$ towards these stars using the
relation above. For the remaining 14 objects we assign and R$_V$ value
of 5.0. We then computed extinction A$_V$ for all the stars in our
sample using the relation
\[A_V\:=\:\frac{E(B-V)} {\left(-0.0018\:+\:\frac{1.0495}{R_V}\right)}\] \citep{ccm89}.

From the mean V magnitudes and the distances listed in Table
\ref{data} and together with the extinction computed as described above, we
calculated the absolute V magnitudes of the stars. We then computed
the absolute bolometric magnitudes using the values of bolometric
corrections from \cite{kenhart95}. The bolometric luminosity of the
star is then computed from the equation
\[\frac{L_{\star}}{L_{\odot}}\:=\: 10^{\left(\frac{4.74\:-\:M_{bol}}{2.5}\right)}\]
where 4.74 is the absolute bolometric magnitude of the sun. The
effective temperatures were assigned from the spectral types listed in
Table \ref{data} and using the calibration of \cite{kenhart95}.

We thus computed the luminosities and assigned effective temperatures
for all the 91 stars in our sample.  Fig. \ref{hr} shows the locations
of HAeBe stars in the HR diagram.  Also plotted are the evolutionary
tracks of \citet{pallastahler93}.  We derive masses and ages for the
HAeBe stars from their positions in the HR diagram and by comparing 
with the evolutionary tracks. For stars which fall above the birth
line we quote lower limits on mass and upperlimits on age and for
stars that fall below the main sequence we quote limits on ages and
masses corresponding to the ZAMS value for that spectral type. Masses
and ages thus derived are listed in Table \ref{age}.

\subsection{Evolution of emission line activity}

The most prominent emission feature in HAeBe stars is the
H$\alpha$ line. To study  the temporal evolution of emission line
activity in HAeBe stars, we use the equivalent width of
H$\alpha$ emission line (W(H$\alpha$)) as a measure of emission
line activity. We measured the equivalent widths of the H$\alpha$
line from the observed spectra.  We also compiled H$\alpha$
equivalent width W(H$\alpha$) measurements available from the
literature for the stars in our sample. Multi-epoch measurements of
W(H$\alpha$) obtained from our observations and compiled from the
literature are tabulated in Table \ref{ew}.

In Fig. \ref{ewage} we plot the equivalent widths of H$\alpha$
emission line W(H$\alpha$) for the HAeBe stars listed in
Table \ref{ew} against their ages that have been derived and listed in
Table \ref{age}, to look for possible evolutionary trend.  Wherever
more than one measurement of W(H$\alpha$) are available we have
plotted the mean value of those measurements. The error bars plotted
for W(H$\alpha$) represent the dispersion in different measurements
and are a measure of variability of H$\alpha$ emission
strength. 

It can be seen from Fig. \ref{ewage} that there is an overall decrease
in the equivalent width of H$\alpha$ emission line seen in HAeBe stars
with their PMS age. Out of the 43 stars which show
$|$~W(H$\alpha$)~$|$~$\ge$~20~$\AA$, 84\% have ages $<$ 3~Myr while
77\% of the stars older than 3~Myr show
$|$~W(H$\alpha$)~$|$~$\le$~20~$\AA$. The dashed line shown in the
figure is an upper envelope to the distribution of W(H$\alpha$) with
the ages of HAeBe stars and is of the functional form
$W(age)=W(0)e^{-\:age/\tau}$ with $W(0)=-100$ and $\tau$ = 3 Myr. More
than 83\% of the data points fall below the line indicating that the
strength of H$\alpha$ emission in HAeBe stars decline relatively
rapidly on a timescale of about 3~Myr.

Although there is an overall decrease in the H$\alpha$ emission
strength with age, there is a large spread in the equivalent widths at
young ages ( $<$ 3 Myr). A significant fraction ($\sim$ 41\%) of the
stars younger than 3 Myr show low H$\alpha$ emission (
$|$~W(H$\alpha$)~$|$~$\le$~20~$\AA$) and in $\sim$ 22\% of these stars
emission line activity appears to have weakened considerably (
$|$~W(H$\alpha$)~$|$~$\le$~10~$\AA$) in about 3 Myr.

It is interesting to note from Fig \ref{ewage} that though most stars
seem to lose emission line activity by 3 - 5 Myr, there are a few
relatively old stars with ages $\ge$ 10 Myr which show appreciable
H$\alpha$ equivalent widths ($|$W(H$\alpha$)$|$ $\ge$ 5 $\AA$).
H$\alpha$ emission appear to persist for long ( $\ge$ 10 Myr) atleast
in a few young stars.

In Fig \ref{ewage} we have identified stars of different spectral
types with solid brown circles representing  B stars, black circles
 A stars and blue F stars. The distribution of equivalent width with
age for stars of different spectral types show indications for a
relatively faster evolution of emission line activity in B type stars
as compared to A and F type stars. However, this could also be due to a
selection effect since for a given emission line strength (assuming
emission lines originate independent of the star) B type stars would
have smaller equivalent widths than the late type stars because of the
much stronger continuum levels in B stars.

We note here that determining the PMS age of a star by
placing it on the HR diagram and comparing with the theoretical
evolutionary tracks are prone to two different kinds of errors -
random error introduced while converting the observables to stellar
luminosity and systematic error due to the variation between the
predictions of different theoretical evolutionary tracks
\citep[e.g.][]{hill05}. By using a large sample for our analysis and
obtaining multi-epoch photometry and best spectral type and distance
estimates from the literature we have tried to minimize the random
errors in the derived ages. However, systematic error mentioned above
is hard to deal with and all evolutionary studies in literature suffer
from this uncertainity. Nevertheless, even if we were to use different
sets of PMS evolutionary tracks, the overall trend of the decline of
emission line activity would still remain but the time in which the
emission line activity falls significantly would be different by
20-50\%.

\subsection{Emission line - inner disk connection}

In this section we present the results of our study on the connection
between emission line activity and the inner disks of HAeBe
stars. To characterise the inner disk we compiled the
near-infrared (NIR) J, H, and $K_s$ magnitudes for our sample stars
from the 2MASS All-Sky Point Source Catalogue.  We have considered
only those sources for which the 2MASS catalogue gives the optical
association with a Tycho-2 star or an USNO star and the positional
offset between the 2MASS source and the optical counterpart is $\leq$
1 arcsec. Since the effective resolution of the 2MASS system is
approximately 5 arcsec, only those 2MASS sources were included where
the distance between the source and its nearest neighbor in the 2MASS
PSC is $\ge$ 5 arcsec to avoid any source confiusion. Further, we
picked up only those sources where the signal-to-noise ratio in J, H,
and $K_s$ measurements is $\ge$ 10 and uncertainity in the magnitudes
are $\leq$ 0.1 (2MASS photometric quality flag = A). Only those
sources which are unaffected by known artifact or source confusion
(Contamination and confusion flag = 0) are chosen. Also we have
excluded sources which have extended source contamination flag set to
non-zero values in the 2MASS catalogue.  NIR J, H, and $K_s$
magnitudes obtained from the 2MASS catalogue for 62 HAeBe stars
in our sample is listed in Table \ref{2mass}.

In our analysis we used H - K color excess to study the characteristics
of inner disks.We computed intrinsic color excesses $\Delta$(H-K) due
to thermal emission from the inner disk by subtracting the photospheric
color and color due to interstellar and large scale circumstellar
reddening from the observed colors. We used the photospheric colors corresponding
to the spectral types from those given by \citet{kenhart95}. Since the
NIR and optical colors in \citet{kenhart95} are essentially in the
\citet{bb88} system,all the observed colors are converted into that system.
The  intrinsic color excess $\Delta$(H-K) due to inner disks is defined  as

\[ \Delta(H-K)\:=\:(H-K)_{observed}\:-\:(H-K)_{photosphere}\:-\:\left(0.0871\:-\:\frac{0.08}{R_V}\right)\:\times\:A_V\]

where in calculating the reddening term we have used the relation from
\citet{ccm89} with the effective wavelengths for filters V, J, H \& K 
taken from \citet{bb88}. The last term on the rhs is the color
contribution due to interstellar reddening.  Intrinsic color excess
$\Delta$(H-K) thus obtained and estimated error in excess are listed
in Table \ref{2mass}.

In Figure \ref{ewhk} we plot equivalent widths of H$\alpha$ emission
line W(H$\alpha$) against the color excess due to inner disk
$\Delta$(H-K) for HAeBe stars. Clearly, W(H$\alpha$) is well
correlated with $\Delta$(H-K). In general, stars with low emission
line activity have smaller (H-K) excess and stars with strong emission
line activity exhibit large $\Delta$(H-K). A Spearman's rank test
gives a $\rho$ of 0.6 with probabilities of being drawn from a random
distribution of 1.2 $\times$ 10$^{-6}$ implying a robust
correlation between the H$\alpha$ equivalent width and color excess
due to inner disk $\Delta$(H-K). It is interesting to note that for
most HAeBe stars $\Delta$(H-K) values are in a narrow range of 0.4 -
0.9 mag, quite similar to that seen in T Tauri stars \citep{mch97}.

\section{Discussion}

We studied the evolution of emission line activity in HAeBe stars by
following the evolution of the equivalent width of H$\alpha$ emisison
in them with the stellar age. Our results show that on average the
emission line strength in these stars decreases with the increasing
age during the PMS lifetimes of these stars. Additionally, we find
that the strength of H$\alpha$ emission falls substantially in
relatively short timescales.  The emission line strength decreases by
a factor of more than 2 in as short a timescale as $\sim$3 Myr.
However, we also find that in a few objects, emission line activity
persists even at $\sim$ 10 Myr.

As discussed in Section  1, NIR studies of disk frequency in young clusters
have shown that most young stars lose their inner disks in about
$\sim$ 5 Myr \citep{haisch01, hill05}. In intermediate mass stars the
inner disk dissipation is even faster with timescale as short as
$\sim$ 3 Myr \citep{hernand05, haisch01b}. However, this is a trend
for the fraction of stars still showing NIR excess. The nature of
the excess itself is generally believed to be binary: either it is
there, or it isn't. Several authors have noted that there is a
paucity of `transition' objects with observational properties
intermediate between disk-bearing young stars and disk-less ones
\citep[e.g][]{skrut90, hart05, aw05}. These results have led to the
conclusion that the disks are dispersed very rapidly and that the
dispersal process is stochastic in nature. Statistically, the disk
dispersal time is estimated to be of the order of $\sim$ 10$^5$ yr
\citep{skrut90, sp95, ww96}. In short, these disk evolutionary studies
point to a `two-timescale' behaviour where the disks surrounding young
stars are dispersed rapidly in $\sim$ 10$^5$ yr after a disk lifetime
of a few Myr.

Most of the earlier disk evolution studies in the NIR
\citep[e.g.][]{haisch01, hill05, hernand05} have concentrated on the trends
in disk fractions in young clusters and associations with known ages. Our
analysis is different from this in that we study the evolutionary
trend among stars which have inner disks and show accretion
signatures. We use the individual ages of the stars to follow the
evolution of the strength of H$\alpha$ emission, which is a measure of
accretion activity. Our results indicate a
gradual decrease in the strength of accretion activity with stellar
age in HAeBe stars. This would hint at a relatively long-lived process
being responsible for the cessation of accretion (and persumably for
disk dissipation) in HAeBe stars.

Although there is an overall reduction in the strength of accretion
with stellar age, our results also show that there is a large scatter
at young ages with a significant fraction of stars in our sample with
ages $\le$ 3 Myr showing very low levels of accretion activity. In
general, we find that accretion activity in most HAeBe stars has
weakened considerably by $\sim$ 3 Myr. It is interesting to note that
this timescale is quite comparable to the inner disk lifetimes derived
for the intermediate mass stars from the NIR studies of disk
frequencies \citep{hernand05}. Most intermediate mass stars lose their
inner disks on a timescale similar to the one in which the accretion
activity in HAeBe stars drops significantly.

It is possible to qualitatively understand our results within the
framework of the `UV-switch' model proposed recently to explain the
`two-timescale' behaviour of disk dissipation \citep{clarke01,
alex06I, alex06II}. This model couples photoevaporation of the outer
disks to the viscous evolution of the  disk. In the early phase of
evolution, when the accretion rate through the disk is much larger
than the mass loss rate due to photoevaporative wind, the wind has a
negligible effect. At some point during the disk evolution the
accretion rate falls to a level comparable to the mass-loss rate from
the wind. Photoevaporation then becomes important, depriving the disk
of resupply inside the gravitational radius R$_g$
\citep[see][]{alex06I}. At this point, the inner disk drains on its 
own, short, viscous time-scale, giving a dispersal time much shorter
than the disk life time. Thus, this model predicts that as long as
accretion rate through the disks remain sufficiently high above the
photoevaporative mass-loss rate, the inner disks survive around young
stars. Our results show that on average accretion activity in HAeBe
stars drops off steadily with increasing stellar age during the PMS
evolution. This suggests that accretion is the relatively long-lived
(a few Myr) mechanism which controls the disk lifetimes in young stars
and that the disk dissipation is probably not stochastic in nature.
We further show that the accretion activity in HAeBe stars drops
siginificantly in $\sim$ 3 Myr. Most intermediate mass stars
also appear to lose their inner disks on similar timescale
\citep{hernand05}, supporting the idea that the rapid dispersal of the 
inner disk occurs when the accretion rate has dropped below the
critical level.

Next, we discuss the correlation found in Figure \ref{ewhk}. In recent
years the spectral energy distributions (SEDs) of HAeBe stars have
been successfully modelled as arising from passive irradiated disks
with inner holes \citep{ddn01, npnwgm01}. In this framework, the mid-
and far-IR emission arise in the flared outer regions of the disk. The
inner rim of the disk, which is directly exposed to the stellar
radiation, is heated very efficiently and is puffed up. Radiation from
the inner surface of the rim, if located at the dust sublimation
radius, can reproduce the shape and strength of the near-IR part of
the SEDs (the NIR bump) of HAeBe stars. The relatively narrow range of
$\Delta$(H-K) is supportive of such a scenario.

However, the correlation that we find between the excess color due to
the inner disks $\Delta$(H-K) and an accretion indicator W(H$\alpha$)
suggests that the disk is not entirely passive.  The equivalent width
of H$\alpha$ is known to scale with normalized accretion luminosity (L$_{acc}$/L$_{\star}$) and
$\Delta$(H-K) is the ratio  of excess flux in K band to that in H band, ie.
\[\frac{1+ (F_{K_{excess}}/  F_{K_{phot}})  }{ 1+ (F_{H_{excess}} / F_{H_{phot}} )  } = 10^{\left(\frac{\Delta(H-K)}{2.5}\right)}\]
  
The correlation found in Figure 2 then implies an increase in
normalized accretion luminosity with an increase in the excess flux in
the K band relative to that in the H band. The redness of NIR colors
has been known to increase as the relative contribution from the NIR
to the SED goes up in HAeBe and T Tauri stars
\citep[e.g.][]{corray98, cabrit90}. Such a correlation between
normalized accretion luminosity and excess (H-K) color is not
necessarily expected if the excess emission in the NIR wavelengths is
entirely due to disk irradiation. Accretion through the disks in
HAeBe stars must also be responsible for the excess emission in
the NIR.

However, it has been argued in literature that accretion alone cannot
explain the observed SED in the NIR wavelengths. If the disk heating
and thereby the excess is entirely due to the viscous dissipation in
the disk, the required accretion rates to reproduce the observed NIR
excess is relatively high ( $\ge 10^{-6} M_{\odot}yr^{-1}$) \citep{hill92,
ladadams92}. Also, these accreting disks have to have an inner hole
inorder to explain the lack of excess emission shortward of 2$\micron$
\citep{hill92}. \citet{hart93} have pointed out that at such high
accretion rates the innermost parts of disks will not remain optically
thin as is required to explain the shape and strength of the NIR
SED. Moreover, recent NIR interferometric observations of HAeBe stars
have confirmed the presence of inner holes in HAeBe disks, suggesting
that the accretion rates in HAeBe disks must be $\le 10^{-7}
M_{\odot}yr^{-1}$ so as to keep the gas within the dust sublimation
radius optically thin.

Our results suggests that even at the allowed low accretion rates,
strength of NIR excess and level of disk accretion are related. We recall
here that strong correlations between NIR colors and accretion
indicators (e.g. W(H$\alpha$)) / outflow signatures (e.g. W([OI]) in
HAeBe stars have been reported in the literature
\citep[e.g.][]{corray98, gseh94} (albeit for smaller samples). 
Such correlations are difficult to explain within the framework of
passive disks. Recently, passive disks with puffed up inner rim models
have been invoked to explain the infrared and longer wavelength
emission from T Tauri stars as well \citep[e.g.][]{ muz03,cieza05},
although the correlaton between excess infrared colors and accretion
luminosity is also found in these stars. This correlation is explained
by adding accretion shock radiation also to the central heating source
which effectively pushes the inner rim farther away from the
star. Also, higher accretion rates increases the physical height of
the dust rim as the vertical density structure is affected by the
accretion rate \citep{muz04}. Both these effects related to accretion
increase the effective area of the surface of the inner rim from where
the NIR excess emission arises.  It is conceivable that such a
mechanism is also acting in HAeBe stars even at relatively low
accretion rates which keep the inner hole optically thin. This can
explain the correlation that we find in Figure 2. However, detailed
modelling and more quantitative analysis are required to confirm such
a scenario.

\section{Conclusions}

In this paper we have studied the temporal evolution of emission line
activity in intermediate mass PMS stars by compiling  multi-epoch
spectroscopic and photometric observations for a large sample
of HAeBe stars. We find that, on average, the H$\alpha$ emission line
strength decreases with increasing stellar age in HAeBe stars,
suggesting that the accretion activity gradually declines during the
PMS phase. This would hint at a relatively long-lived ( a few Myr)
process being responsible for the cessation of accretion in
intermediate mass stars thereby implying that the evolution of
accretion and disk dissipation is probably not entirely stochastic in
nature. We also show that  in most HAeBe stars the H$\alpha$
emission has weakened considerably in $\sim$ 3 Myr, indicating that
the accretion activity in them has dropped significantly. Most
intermediate stars also appear to lose their inner disks on similar
timescale, suggesting that inner disks dissipate rapidly after the
accretion activity has fallen below certain level.

Futher, we find a relatively tight correlation between emission line
strength and NIR excess due to inner disks in HAeBe stars. Such a
correlation strongly suggests a physical connection between accretion
activity and the strength of emission from the pufffed-up inner rim of
the disks. We suggest that this correlation can be explained if the
radiation from the accretion shock is also responsible for heating the
inner rim of the disk. At higher accretion rates, more will be the
contribution from the accretion shock radiation which will increase
the effective surface area of the inner rim, thereby increasing the NIR
excess emission.

\acknowledgments
We thank the staff of VBO, Kavalur, IAO, Hanle and CREST, Hosakote
that made these obervations possible. The facilities at VBO, IAO and
CREST are operated by the Indian Institute of Astrophysics, Bangalore.
This publication makes use of data products from the Two Micron All
Sky Survey, which is a joint project of the University of
Massachusetts and the Infrared Processing and Analysis
Center/California Institute of Technology, funded by the National
Aeronautics and Space Administration and the National Science
Foundation.
             
\bibliography{/home/manoj/natbib/references} 

\begin{thebibliography}{106}
\expandafter\ifx\csname natexlab\endcsname\relax\def\natexlab#1{#1}\fi

\bibitem[{{Acke} \& {van den Ancker}(2004)}]{acke04}
{Acke}, B. \& {van den Ancker}, M.~E. 2004, \aap, 426, 151

\bibitem[{{Acke} {et~al.}(2005){Acke}, {van den Ancker}, \&
  {Dullemond}}]{acke05}
{Acke}, B., {van den Ancker}, M.~E., \& {Dullemond}, C.~P. 2005, \aap, 436, 209

\bibitem[{{Alencar} {et~al.}(2003){Alencar}, {Melo}, {Dullemond}, {Andersen},
  {Batalha}, {Vaz}, \& {Mathieu}}]{alencar03}
{Alencar}, S.~H.~P., {Melo}, C.~H.~F., {Dullemond}, C.~P., {et~al.} 2003, \aap,
  409, 1037

\bibitem[{{Alexander} {et~al.}(2006{\natexlab{a}}){Alexander}, {Clarke}, \&
  {Pringle}}]{alex06I}
{Alexander}, R.~D., {Clarke}, C.~J., \& {Pringle}, J.~E. 2006{\natexlab{a}},
  \mnras, 369, 216

\bibitem[{{Alexander} {et~al.}(2006{\natexlab{b}}){Alexander}, {Clarke}, \&
  {Pringle}}]{alex06II}
{Alexander}, R.~D., {Clarke}, C.~J., \& {Pringle}, J.~E. 2006{\natexlab{b}},
  \mnras, 369, 229

\bibitem[{{Andrews} \& {Williams}(2005)}]{aw05}
{Andrews}, S.~M. \& {Williams}, J.~P. 2005, \apj, 631, 1134

\bibitem[{{Baines} {et~al.}(2004){Baines}, {Oudmaijer}, {Mora}, {Eiroa},
  {Porter}, {Mer{\'{\i}}n}, {Montesinos}, {de Winter}, {Cameron}, {Davies},
  {Deeg}, {Ferlet}, {Grady}, {Harris}, {Hoare}, {Horne}, {Lumsden}, {Miranda},
  {Penny}, \& {Quirrenbach}}]{baines04}
{Baines}, D., {Oudmaijer}, R.~D., {Mora}, A., {et~al.} 2004, \mnras, 353, 697

\bibitem[{{Beckwith}(1999)}]{beck99}
{Beckwith}, S.~V.~W. 1999, in NATO ASIC Proc. 540: The Origin of Stars and
  Planetary Systems, 579

\bibitem[{{Beckwith} \& {Sargent}(1996)}]{becksar96}
{Beckwith}, S.~V.~W. \& {Sargent}, A.~I. 1996, \nat, 383, 139

\bibitem[{{Berrilli} {et~al.}(1992){Berrilli}, {Corciulo}, {Ingrosso},
  {Lorenzetti}, {Nisini}, \& {Strafella}}]{berrilli92}
{Berrilli}, F., {Corciulo}, G., {Ingrosso}, G., {et~al.} 1992, \apj, 398, 254

\bibitem[{{Beskrovnaya} {et~al.}(1999){Beskrovnaya}, {Pogodin},
  {Miroshnichenko}, {Th{\'e}}, {Savanov}, {Shakhovskoy}, {Rostopchina},
  {Kozlova}, \& {Kuratov}}]{beskro99}
{Beskrovnaya}, N.~G., {Pogodin}, M.~A., {Miroshnichenko}, A.~S., {et~al.} 1999,
  \aap, 343, 163

\bibitem[{{Bessell}(1983)}]{bessell83}
{Bessell}, M.~S. 1983, \pasp, 95, 480

\bibitem[{{Bessell} \& {Brett}(1988)}]{bb88}
{Bessell}, M.~S. \& {Brett}, J.~M. 1988, \pasp, 100, 1134

\bibitem[{{B\"{o}hm} \& {Catala}(1994)}]{bc94}
{B\"{o}hm}, T. \& {Catala}, C. 1994, \aap, 290, 167

\bibitem[{{B\"{o}hm} \& {Catala}(1995)}]{bohm95}
{B\"{o}hm}, T. \& {Catala}, C. 1995, \aap, 301, 155

\bibitem[{{B\"{o}hm} \& {Hirth}(1997)}]{bh97}
{B\"{o}hm}, T. \& {Hirth}, G.~A. 1997, \aap, 324, 177

\bibitem[{{Cabrit} {et~al.}(1990){Cabrit}, {Edwards}, {Strom}, \&
  {Strom}}]{cabrit90}
{Cabrit}, S., {Edwards}, S., {Strom}, S.~E., \& {Strom}, K.~M. 1990, \apj, 354,
  687

\bibitem[{{Calvet} {et~al.}(2004){Calvet}, {Muzerolle}, {Brice{\~ n}o},
  {Hern{\' a}ndez}, {Hartmann}, {Saucedo}, \& {Gordon}}]{calvet04}
{Calvet}, N., {Muzerolle}, J., {Brice{\~ n}o}, C., {et~al.} 2004, \aj, 128,
  1294

\bibitem[{{Cardelli} {et~al.}(1989){Cardelli}, {Clayton}, \& {Mathis}}]{ccm89}
{Cardelli}, J.~A., {Clayton}, G.~C., \& {Mathis}, J.~S. 1989, \apj, 345, 245

\bibitem[{{Cieza} {et~al.}(2005){Cieza}, {Kessler-Silacci}, {Jaffe}, {Harvey},
  \& {Evans}}]{cieza05}
{Cieza}, L.~A., {Kessler-Silacci}, J.~E., {Jaffe}, D.~T., {Harvey}, P.~M., \&
  {Evans}, II, N.~J. 2005, \apj, 635, 422

\bibitem[{{Clarke}(2001)}]{clarke01}
{Clarke}, C.~J. 2001, in IAU Symposium, 346--353

\bibitem[{{Cohen} \& {Kuhi}(1979)}]{cohenkuhi79}
{Cohen}, M. \& {Kuhi}, L.~V. 1979, \apjs, 41, 743

\bibitem[{{Corcoran} \& {Ray}(1998)}]{corray98}
{Corcoran}, M. \& {Ray}, T.~P. 1998, \aap, 331, 147

\bibitem[{{Coulson} \& {Walther}(1995)}]{coul95}
{Coulson}, I.~M. \& {Walther}, D.~M. 1995, \mnras, 274, 977

\bibitem[{{de Winter} {et~al.}(2001){de Winter}, {van den Ancker}, {Maira},
  {Th{\'e}}, {Djie}, {Redondo}, {Eiroa}, \& {Molster}}]{dewinter01}
{de Winter}, D., {van den Ancker}, M.~E., {Maira}, A., {et~al.} 2001, \aap,
  380, 609

\bibitem[{{Dullemond} {et~al.}(2001){Dullemond}, {Dominik}, \& {Natta}}]{ddn01}
{Dullemond}, C.~P., {Dominik}, C., \& {Natta}, A. 2001, \apj, 560, 957

\bibitem[{{Dunkin} \& {Crawford}(1998)}]{dunkin98}
{Dunkin}, S.~K. \& {Crawford}, I.~A. 1998, \mnras, 298, 275

\bibitem[{{Eisner} {et~al.}(2003){Eisner}, {Lane}, {Akeson}, {Hillenbrand}, \&
  {Sargent}}]{eisner03}
{Eisner}, J.~A., {Lane}, B.~F., {Akeson}, R.~L., {Hillenbrand}, L.~A., \&
  {Sargent}, A.~I. 2003, \apj, 588, 360

\bibitem[{{Eisner} {et~al.}(2004){Eisner}, {Lane}, {Hillenbrand}, {Akeson}, \&
  {Sargent}}]{eisner04}
{Eisner}, J.~A., {Lane}, B.~F., {Hillenbrand}, L.~A., {Akeson}, R.~L., \&
  {Sargent}, A.~I. 2004, \apj, 613, 1049

\bibitem[{{ESA}(1997)}]{esa97}
{ESA}. 1997, {The Hipparcos and Tycho Catalogues (ESA 1997)}, 1239

\bibitem[{{Finkenzeller} \& {Mundt}(1984)}]{finkmun84}
{Finkenzeller}, U. \& {Mundt}, R. 1984, \aaps, 55, 109

\bibitem[{{Fujii} {et~al.}(2002){Fujii}, {Nakada}, \& {Parthasarathy}}]{fuji02}
{Fujii}, T., {Nakada}, Y., \& {Parthasarathy}, M. 2002, \aap, 385, 884

\bibitem[{{Ghandour} {et~al.}(1994){Ghandour}, {Strom}, {Edwards}, \&
  {Hillenbrand}}]{gseh94}
{Ghandour}, L., {Strom}, S., {Edwards}, S., \& {Hillenbrand}, L. 1994, in ASP
  Conf. Ser. 62: The Nature and Evolutionary Status of Herbig Ae/Be Stars, 223

\bibitem[{{Gorti} \& {Bhatt}(1993)}]{gortibhatt93}
{Gorti}, U. \& {Bhatt}, H.~C. 1993, \aap, 270, 426

\bibitem[{{Haisch} {et~al.}(2001{\natexlab{a}}){Haisch}, {Lada}, \&
  {Lada}}]{haisch01b}
{Haisch}, K.~E., {Lada}, E.~A., \& {Lada}, C.~J. 2001{\natexlab{a}}, \aj, 121,
  2065

\bibitem[{{Haisch} {et~al.}(2001{\natexlab{b}}){Haisch}, {Lada}, \&
  {Lada}}]{haisch01}
{Haisch}, K.~E., {Lada}, E.~A., \& {Lada}, C.~J. 2001{\natexlab{b}}, \apjl,
  553, L153

\bibitem[{{Hamaguchi} {et~al.}(2005){Hamaguchi}, {Yamauchi}, \&
  {Koyama}}]{hamaguchi05}
{Hamaguchi}, K., {Yamauchi}, S., \& {Koyama}, K. 2005, \apj, 618, 360

\bibitem[{{Hamann}(1994)}]{ham94}
{Hamann}, F. 1994, \apjs, 93, 485

\bibitem[{{Hamann} \& {Persson}(1992)}]{hp92}
{Hamann}, F. \& {Persson}, S.~E. 1992, \apjs, 82, 285

\bibitem[{{Hartmann}(1998)}]{hart98}
{Hartmann}, L. 1998, {Accretion processes in star formation} (Cambridge, UK ;
  New York : Cambridge University Press, 1998.~(Cambridge astrophysics series ;
  32))

\bibitem[{{Hartmann} {et~al.}(1994){Hartmann}, {Hewett}, \& {Calvet}}]{hhc94}
{Hartmann}, L., {Hewett}, R., \& {Calvet}, N. 1994, \apj, 426, 669

\bibitem[{{Hartmann} {et~al.}(1993){Hartmann}, {Kenyon}, \& {Calvet}}]{hart93}
{Hartmann}, L., {Kenyon}, S.~J., \& {Calvet}, N. 1993, \apj, 407, 219

\bibitem[{{Hartmann} {et~al.}(2005){Hartmann}, {Megeath}, {Allen}, {Luhman},
  {Calvet}, {D'Alessio}, {Franco-Hernandez}, \& {Fazio}}]{hart05}
{Hartmann}, L., {Megeath}, S.~T., {Allen}, L., {et~al.} 2005, \apj, 629, 881

\bibitem[{{Herbig}(1960)}]{herbig60}
{Herbig}, G.~H. 1960, \apjs, 4, 337

\bibitem[{{Herbig} \& {Bell}(1988)}]{herbigbell88}
{Herbig}, G.~H. \& {Bell}, K.~R. 1988, {Catalog of emission line stars of the
  orion population : 3 : 1988} (Lick Observatory Bulletin, Santa Cruz: Lick
  Observatory, |c1988)

\bibitem[{{Herbst} \& {Racine}(1976)}]{hr76}
{Herbst}, W. \& {Racine}, R. 1976, \aj, 81, 840

\bibitem[{{Herbst} \& {Shevchenko}(1999)}]{hs99}
{Herbst}, W. \& {Shevchenko}, V.~S. 1999, \aj, 118, 1043

\bibitem[{{Herbst} {et~al.}(1982){Herbst}, {Warner}, {Miller}, \&
  {Herzog}}]{herbst82}
{Herbst}, W., {Warner}, J.~W., {Miller}, D.~P., \& {Herzog}, A. 1982, \aj, 87,
  98

\bibitem[{{Hern{\' a}ndez} {et~al.}(2004){Hern{\' a}ndez}, {Calvet}, {Brice{\~
  n}o}, {Hartmann}, \& {Berlind}}]{hernand04}
{Hern{\' a}ndez}, J., {Calvet}, N., {Brice{\~ n}o}, C., {Hartmann}, L., \&
  {Berlind}, P. 2004, \aj, 127, 1682

\bibitem[{{Hern{\'a}ndez} {et~al.}(2005){Hern{\'a}ndez}, {Calvet}, {Hartmann},
  {Brice{\~n}o}, {Sicilia-Aguilar}, \& {Berlind}}]{hernand05}
{Hern{\'a}ndez}, J., {Calvet}, N., {Hartmann}, L., {et~al.} 2005, \aj, 129, 856

\bibitem[{{Hillenbrand}(2002)}]{hill03}
{Hillenbrand}, L.~A. 2002, astro-ph/0210520

\bibitem[{{Hillenbrand}(2005)}]{hill05}
{Hillenbrand}, L.~A. 2005, ArXiv Astrophysics e-prints

\bibitem[{{Hillenbrand} {et~al.}(1992){Hillenbrand}, {Strom}, {Vrba}, \&
  {Keene}}]{hill92}
{Hillenbrand}, L.~A., {Strom}, S.~E., {Vrba}, F.~J., \& {Keene}, J. 1992, \apj,
  397, 613

\bibitem[{{H{\o}g} {et~al.}(2000){H{\o}g}, {Fabricius}, {Makarov}, {Urban},
  {Corbin}, {Wycoff}, {Bastian}, {Schwekendiek}, \& {Wicenec}}]{tych00}
{H{\o}g}, E., {Fabricius}, C., {Makarov}, V.~V., {et~al.} 2000, \aap, 355, L27

\bibitem[{{Hubrig} {et~al.}(2004){Hubrig}, {Sch{\"o}ller}, \& {Yudin}}]{hub04}
{Hubrig}, S., {Sch{\"o}ller}, M., \& {Yudin}, R.~V. 2004, \aap, 428, L1

\bibitem[{{Kawamura} {et~al.}(1998){Kawamura}, {Onishi}, {Yonekura}, {Dobashi},
  {Mizuno}, {Ogawa}, \& {Fukui}}]{kawa98}
{Kawamura}, A., {Onishi}, T., {Yonekura}, Y., {et~al.} 1998, \apjs, 117, 387

\bibitem[{{Kenyon} \& {Hartmann}(1995)}]{kenhart95}
{Kenyon}, S.~J. \& {Hartmann}, L. 1995, \apjs, 101, 117

\bibitem[{{Koenigl}(1991)}]{kon91}
{Koenigl}, A. 1991, \apjl, 370, L39

\bibitem[{{Kutner} {et~al.}(1980){Kutner}, {Machnik}, {Tucker}, \&
  {Dickman}}]{kut80}
{Kutner}, M.~L., {Machnik}, D.~E., {Tucker}, K.~D., \& {Dickman}, R.~L. 1980,
  \apj, 237, 734

\bibitem[{{Lada} \& {Adams}(1992)}]{ladadams92}
{Lada}, C.~J. \& {Adams}, F.~C. 1992, \apj, 393, 278

\bibitem[{{Lahulla}(1985)}]{lahulla85}
{Lahulla}, J.~F. 1985, \aaps, 61, 537

\bibitem[{{Leinert} {et~al.}(1997){Leinert}, {Richichi}, \& {Haas}}]{leinert97}
{Leinert}, C., {Richichi}, A., \& {Haas}, M. 1997, \aap, 318, 472

\bibitem[{{Leinert} {et~al.}(2004){Leinert}, {van Boekel}, {Waters},
  {Chesneau}, {Malbet}, {K{\"o}hler}, {Jaffe}, {Ratzka}, {Dutrey}, {Preibisch},
  {Graser}, {Bakker}, {Chagnon}, {Cotton}, {Dominik}, {Dullemond},
  {Glazenborg-Kluttig}, {Glindemann}, {Henning}, {Hofmann}, {de Jong},
  {Lenzen}, {Ligori}, {Lopez}, {Meisner}, {Morel}, {Paresce}, {Pel},
  {Percheron}, {Perrin}, {Przygodda}, {Richichi}, {Sch{\"o}ller}, {Schuller},
  {Stecklum}, {van den Ancker}, {von der L{\"u}he}, \& {Weigelt}}]{leinert04}
{Leinert}, C., {van Boekel}, R., {Waters}, L.~B.~F.~M., {et~al.} 2004, \aap,
  423, 537

\bibitem[{{Malfait} {et~al.}(1998){Malfait}, {Bogaert}, \& {Waelkens}}]{malf98}
{Malfait}, K., {Bogaert}, E., \& {Waelkens}, C. 1998, \aap, 331, 211

\bibitem[{{Mannings} \& {Sargent}(1997)}]{mansar97}
{Mannings}, V. \& {Sargent}, A.~I. 1997, \apj, 490, 792

\bibitem[{{Mannings} \& {Sargent}(2000)}]{mansar00}
{Mannings}, V. \& {Sargent}, A.~I. 2000, \apj, 529, 391

\bibitem[{{Meyer} {et~al.}(1997){Meyer}, {Calvet}, \& {Hillenbrand}}]{mch97}
{Meyer}, M.~R., {Calvet}, N., \& {Hillenbrand}, L.~A. 1997, \aj, 114, 288

\bibitem[{{Millan-Gabet} {et~al.}(2001){Millan-Gabet}, {Schloerb}, \&
  {Traub}}]{mill01}
{Millan-Gabet}, R., {Schloerb}, F.~P., \& {Traub}, W.~A. 2001, \apj, 546, 358

\bibitem[{{Miroshnichenko} {et~al.}(2001{\natexlab{a}}){Miroshnichenko},
  {Bjorkman}, {Chentsov}, {Klochkova}, {Gray}, {Garc{\'{\i}}a-Lario}, \& {Perea
  Calder{\'o}n}}]{mirosh01}
{Miroshnichenko}, A.~S., {Bjorkman}, K.~S., {Chentsov}, E.~L., {et~al.}
  2001{\natexlab{a}}, \aap, 377, 854

\bibitem[{{Miroshnichenko} {et~al.}(2004){Miroshnichenko}, {Gray}, {Klochkova},
  {Bjorkman}, \& {Kuratov}}]{mirosh04}
{Miroshnichenko}, A.~S., {Gray}, R.~O., {Klochkova}, V.~G., {Bjorkman}, K.~S.,
  \& {Kuratov}, K.~S. 2004, \aap, 427, 937

\bibitem[{{Miroshnichenko} {et~al.}(2001{\natexlab{b}}){Miroshnichenko},
  {Levato}, {Bjorkman}, \& {Grosso}}]{mirosh01b}
{Miroshnichenko}, A.~S., {Levato}, H., {Bjorkman}, K.~S., \& {Grosso}, M.
  2001{\natexlab{b}}, \aap, 371, 600

\bibitem[{{Monnier} {et~al.}(2005){Monnier}, {Millan-Gabet}, {Billmeier},
  {Akeson}, {Wallace}, {Berger}, {Calvet}, {D'Alessio}, {Danchi}, {Hartmann},
  {Hillenbrand}, {Kuchner}, {Rajagopal}, {Traub}, {Tuthill}, {Boden}, {Booth},
  {Colavita}, {Gathright}, {Hrynevych}, {Le Mignant}, {Ligon}, {Neyman},
  {Swain}, {Thompson}, {Vasisht}, {Wizinowich}, {Beichman}, {Beletic},
  {Creech-Eakman}, {Koresko}, {Sargent}, {Shao}, \& {van Belle}}]{monn05}
{Monnier}, J.~D., {Millan-Gabet}, R., {Billmeier}, R., {et~al.} 2005, \apj,
  624, 832

\bibitem[{{Mora} {et~al.}(2001){Mora}, {Mer{\'{\i}}n}, {Solano}, {Montesinos},
  {de Winter}, {Eiroa}, {Ferlet}, {Grady}, {Davies}, {Miranda}, {Oudmaijer},
  {Palacios}, {Quirrenbach}, {Harris}, {Rauer}, {Cameron}, {Deeg}, {Garz{\'
  o}n}, {Penny}, {Schneider}, {Tsapras}, \& {Wesselius}}]{mora01}
{Mora}, A., {Mer{\'{\i}}n}, B., {Solano}, E., {et~al.} 2001, \aap, 378, 116

\bibitem[{{Muzerolle} {et~al.}(1998){Muzerolle}, {Calvet}, \&
  {Hartmann}}]{mch98}
{Muzerolle}, J., {Calvet}, N., \& {Hartmann}, L. 1998, \apj, 492, 743

\bibitem[{{Muzerolle} {et~al.}(2001){Muzerolle}, {Calvet}, \&
  {Hartmann}}]{muzerolle01}
{Muzerolle}, J., {Calvet}, N., \& {Hartmann}, L. 2001, \apj, 550, 944

\bibitem[{{Muzerolle} {et~al.}(2003){Muzerolle}, {Calvet}, {Hartmann}, \&
  {D'Alessio}}]{muz03}
{Muzerolle}, J., {Calvet}, N., {Hartmann}, L., \& {D'Alessio}, P. 2003, \apjl,
  597, L149

\bibitem[{{Muzerolle} {et~al.}(2004){Muzerolle}, {D'Alessio}, {Calvet}, \&
  {Hartmann}}]{muz04}
{Muzerolle}, J., {D'Alessio}, P., {Calvet}, N., \& {Hartmann}, L. 2004, \apj,
  617, 406

\bibitem[{{Natta} {et~al.}(2000){Natta}, {Grinin}, \& {Mannings}}]{ngm00}
{Natta}, A., {Grinin}, V., \& {Mannings}, V. 2000, Protostars and Planets IV,
  559

\bibitem[{{Natta} {et~al.}(2001){Natta}, {Prusti}, {Neri}, {Wooden}, {Grinin},
  \& {Mannings}}]{npnwgm01}
{Natta}, A., {Prusti}, T., {Neri}, R., {et~al.} 2001, \aap, 371, 186

\bibitem[{{Ohashi} \& {Lin}(2005)}]{ohashi05}
{Ohashi}, N. \& {Lin}, S. 2005, in ASP Conf. Ser. 344: The Cool Universe:
  Observing Cosmic Dawn, 168--+

\bibitem[{{Oudmaijer} {et~al.}(2001){Oudmaijer}, {Palacios}, {Eiroa}, {Davies},
  {de Winter}, {Ferlet}, {Garz{\' o}n}, {Grady}, {Cameron}, {Deeg}, {Harris},
  {Horne}, {Mer{\'{\i}}n}, {Miranda}, {Montesinos}, {Mora}, {Penny},
  {Quirrenbach}, {Rauer}, {Schneider}, {Solano}, {Tsapras}, \&
  {Wesselius}}]{oped01}
{Oudmaijer}, R.~D., {Palacios}, J., {Eiroa}, C., {et~al.} 2001, \aap, 379, 564

\bibitem[{{Palla} \& {Stahler}(1993)}]{pallastahler93}
{Palla}, F. \& {Stahler}, S.~W. 1993, \apj, 418, 414

\bibitem[{{Pi{\' e}tu} {et~al.}(2003){Pi{\' e}tu}, {Dutrey}, \&
  {Kahane}}]{petu03}
{Pi{\' e}tu}, V., {Dutrey}, A., \& {Kahane}, C. 2003, \aap, 398, 565

\bibitem[{{Preibisch}(2003)}]{preib03}
{Preibisch}, T. 2003, \aap, 401, 543

\bibitem[{{Reipurth} {et~al.}(1996){Reipurth}, {Pedrosa}, \& {Lago}}]{bpl96}
{Reipurth}, B., {Pedrosa}, A., \& {Lago}, M.~T.~V.~T. 1996, \aaps, 120, 229

\bibitem[{{Shevchenko} {et~al.}(1999){Shevchenko}, {Ezhkova}, {Ibrahimov}, {van
  den Ancker}, \& {Tjin A Djie}}]{shev99}
{Shevchenko}, V.~S., {Ezhkova}, O.~V., {Ibrahimov}, M.~A., {van den Ancker},
  M.~E., \& {Tjin A Djie}, H.~R.~E. 1999, \mnras, 310, 210

\bibitem[{{Shevchenko} \& {Yakubov}(1989)}]{shev89}
{Shevchenko}, V.~S. \& {Yakubov}, S.~D. 1989, Soviet Astronomy, 33, 370

\bibitem[{{Simon} \& {Prato}(1995)}]{sp95}
{Simon}, M. \& {Prato}, L. 1995, \apj, 450, 824

\bibitem[{{Skrutskie} {et~al.}(1990){Skrutskie}, {Dutkevitch}, {Strom},
  {Edwards}, {Strom}, \& {Shure}}]{skrut90}
{Skrutskie}, M.~F., {Dutkevitch}, D., {Strom}, S.~E., {et~al.} 1990, \aj, 99,
  1187

\bibitem[{{Strom} {et~al.}(1989){Strom}, {Strom}, {Edwards}, {Cabrit}, \&
  {Skrutskie}}]{strom89}
{Strom}, K.~M., {Strom}, S.~E., {Edwards}, S., {Cabrit}, S., \& {Skrutskie},
  M.~F. 1989, \aj, 97, 1451

\bibitem[{{Strom} {et~al.}(1993){Strom}, {Edwards}, \& {Skrutskie}}]{sek93}
{Strom}, S.~E., {Edwards}, S., \& {Skrutskie}, M.~F. 1993, in Protostars and
  Planets III, 837--866

\bibitem[{{Strom} {et~al.}(1972){Strom}, {Strom}, {Yost}, {Carrasco}, \&
  {Grasdalen}}]{strom72}
{Strom}, S.~E., {Strom}, K.~M., {Yost}, J., {Carrasco}, L., \& {Grasdalen}, G.
  1972, \apj, 173, 353

\bibitem[{{Terranegra} {et~al.}(1994){Terranegra}, {Chavarria-K.}, {Diaz}, \&
  {Gonzalez-Patino}}]{terra94}
{Terranegra}, L., {Chavarria-K.}, C., {Diaz}, S., \& {Gonzalez-Patino}, D.
  1994, \aaps, 104, 557

\bibitem[{{Testi} {et~al.}(1998){Testi}, {Palla}, \& {Natta}}]{testi98}
{Testi}, L., {Palla}, F., \& {Natta}, A. 1998, \aaps, 133, 81

\bibitem[{{Th\'{e}} {et~al.}(1994){Th\'{e}}, {de Winter}, \& {Perez}}]{the94}
{Th\'{e}}, P.~S., {de Winter}, D., \& {Perez}, M.~R. 1994, \aaps, 104, 315

\bibitem[{{Uchida} \& {Shibata}(1985)}]{us85}
{Uchida}, Y. \& {Shibata}, K. 1985, \pasj, 37, 515

\bibitem[{{van Boekel} {et~al.}(2005){van Boekel}, {Min}, {Waters}, {de Koter},
  {Dominik}, {van den Ancker}, \& {Bouwman}}]{boekel05}
{van Boekel}, R., {Min}, M., {Waters}, L.~B.~F.~M., {et~al.} 2005, \aap, 437,
  189

\bibitem[{{van den Ancker} {et~al.}(2004){van den Ancker}, {Blondel}, {Tjin A
  Djie}, {Grankin}, {Ezhkova}, {Shevchenko}, {Guenther}, \& {Acke}}]{van04}
{van den Ancker}, M.~E., {Blondel}, P.~F.~C., {Tjin A Djie}, H.~R.~E., {et~al.}
  2004, \mnras, 349, 1516

\bibitem[{{van den Ancker} {et~al.}(1998){van den Ancker}, {de Winter}, \&
  {Tjin A Djie}}]{van98}
{van den Ancker}, M.~E., {de Winter}, D., \& {Tjin A Djie}, H.~R.~E. 1998,
  \aap, 330, 145

\bibitem[{{Vieira} {et~al.}(2003){Vieira}, {Corradi}, {Alencar}, {Mendes},
  {Torres}, {Quast}, {Guimar{\~ a}es}, \& {da Silva}}]{vieira03}
{Vieira}, S.~L.~A., {Corradi}, W.~J.~B., {Alencar}, S.~H.~P., {et~al.} 2003,
  \aj, 126, 2971

\bibitem[{{Vink} {et~al.}(2002){Vink}, {Drew}, {Harries}, \&
  {Oudmaijer}}]{vink02}
{Vink}, J.~S., {Drew}, J.~E., {Harries}, T.~J., \& {Oudmaijer}, R.~D. 2002,
  \mnras, 337, 356

\bibitem[{{Vink} {et~al.}(2005){Vink}, {Drew}, {Harries}, {Oudmaijer}, \&
  {Unruh}}]{vink05}
{Vink}, J.~S., {Drew}, J.~E., {Harries}, T.~J., {Oudmaijer}, R.~D., \& {Unruh},
  Y. 2005, \mnras, 359, 1049

\bibitem[{{Wade} {et~al.}(2005){Wade}, {Drouin}, {Bagnulo}, {Landstreet},
  {Mason}, {Silvester}, {Alecian}, {B{\"o}hm}, {Bouret}, {Catala}, \&
  {Donati}}]{wade05}
{Wade}, G.~A., {Drouin}, D., {Bagnulo}, S., {et~al.} 2005, \aap, 442, L31

\bibitem[{{Weinberger} {et~al.}(2002){Weinberger}, {Becklin}, {Schneider},
  {Chiang}, {Lowrance}, {Silverstone}, {Zuckerman}, {Hines}, \&
  {Smith}}]{wein02}
{Weinberger}, A.~J., {Becklin}, E.~E., {Schneider}, G., {et~al.} 2002, \apj,
  566, 409

\bibitem[{{Wolk} \& {Walter}(1996)}]{ww96}
{Wolk}, S.~J. \& {Walter}, F.~M. 1996, \aj, 111, 2066

\bibitem[{{Yonekura} {et~al.}(1997){Yonekura}, {Dobashi}, {Mizuno}, {Ogawa}, \&
  {Fukui}}]{yonekura97}
{Yonekura}, Y., {Dobashi}, K., {Mizuno}, A., {Ogawa}, H., \& {Fukui}, Y. 1997,
  \apjs, 110, 21

\end{thebibliography}
\bibliographystyle{/home/manoj/natbib/aa}

%%%%%%%%%%%%%%%%%%%%%%%
%%%%%%%%Figures%%%%%%%%
%%%%%%%%%%%%%%%%%%%%%%%

\begin{figure}
\centering
\resizebox{0.45\textwidth}{!}{\includegraphics{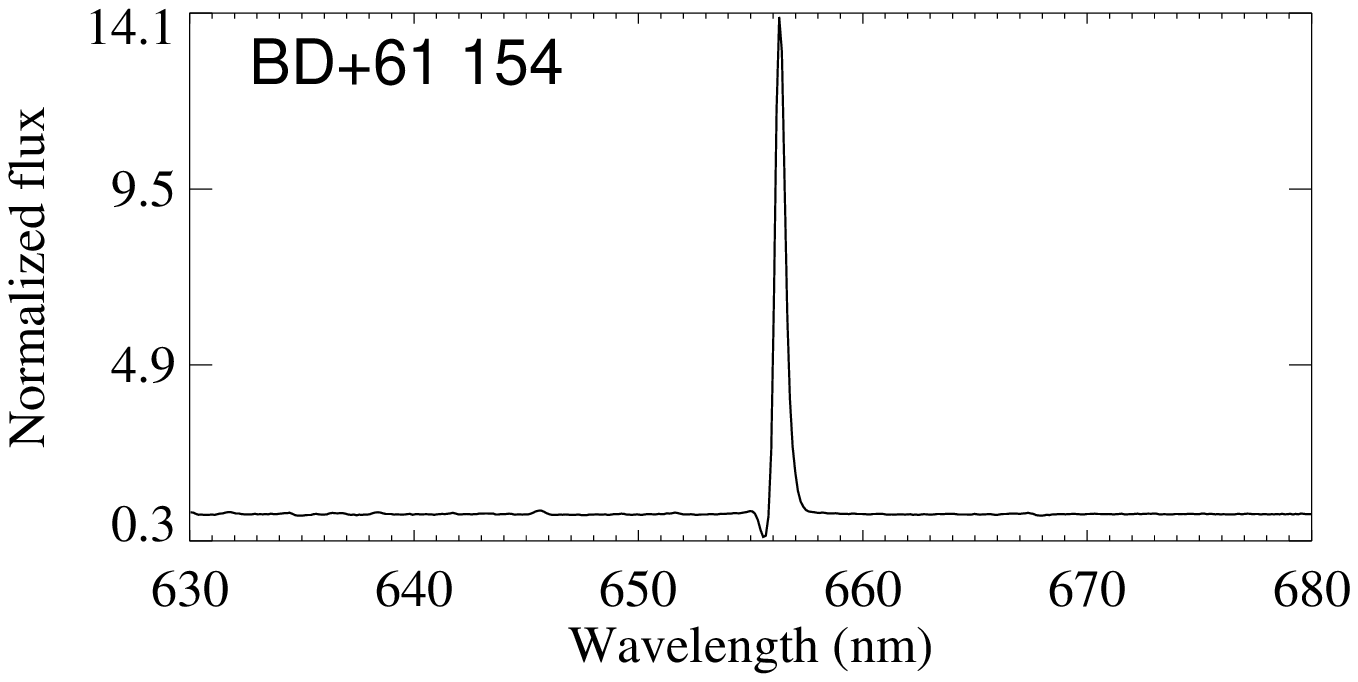}}
\resizebox{0.45\textwidth}{!}{\includegraphics{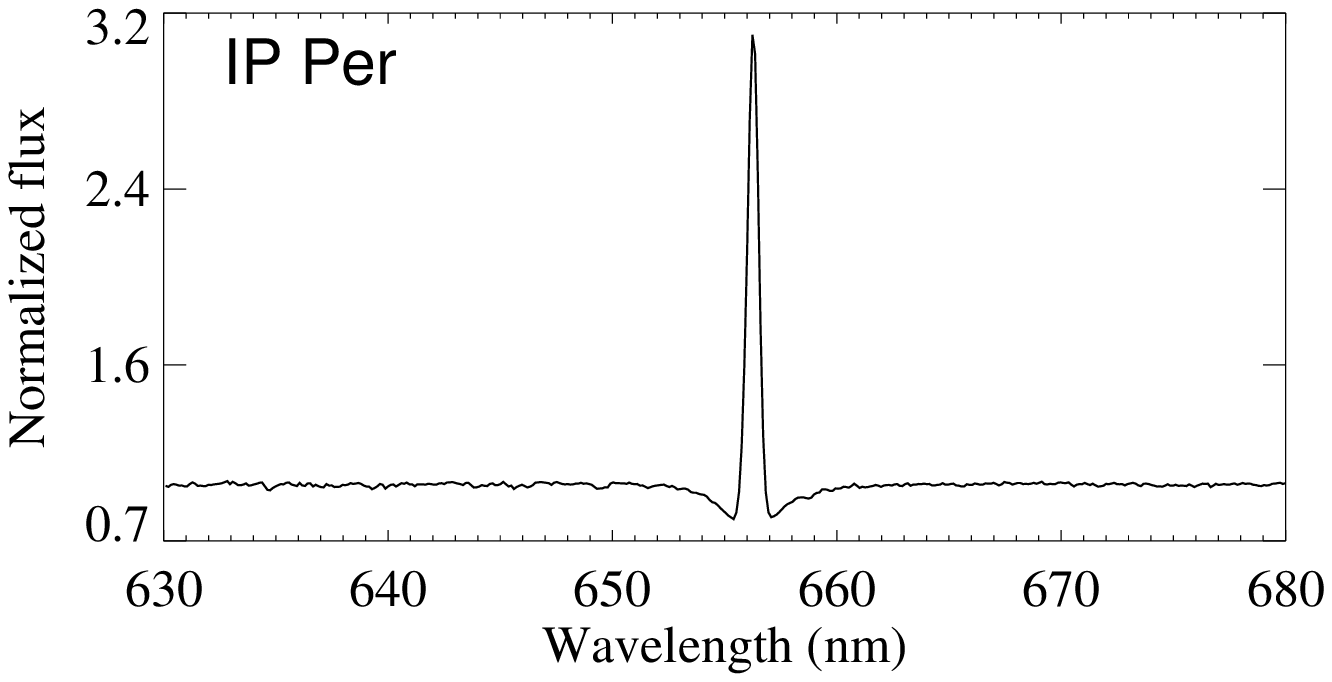}}
\resizebox{0.45\textwidth}{!}{\includegraphics{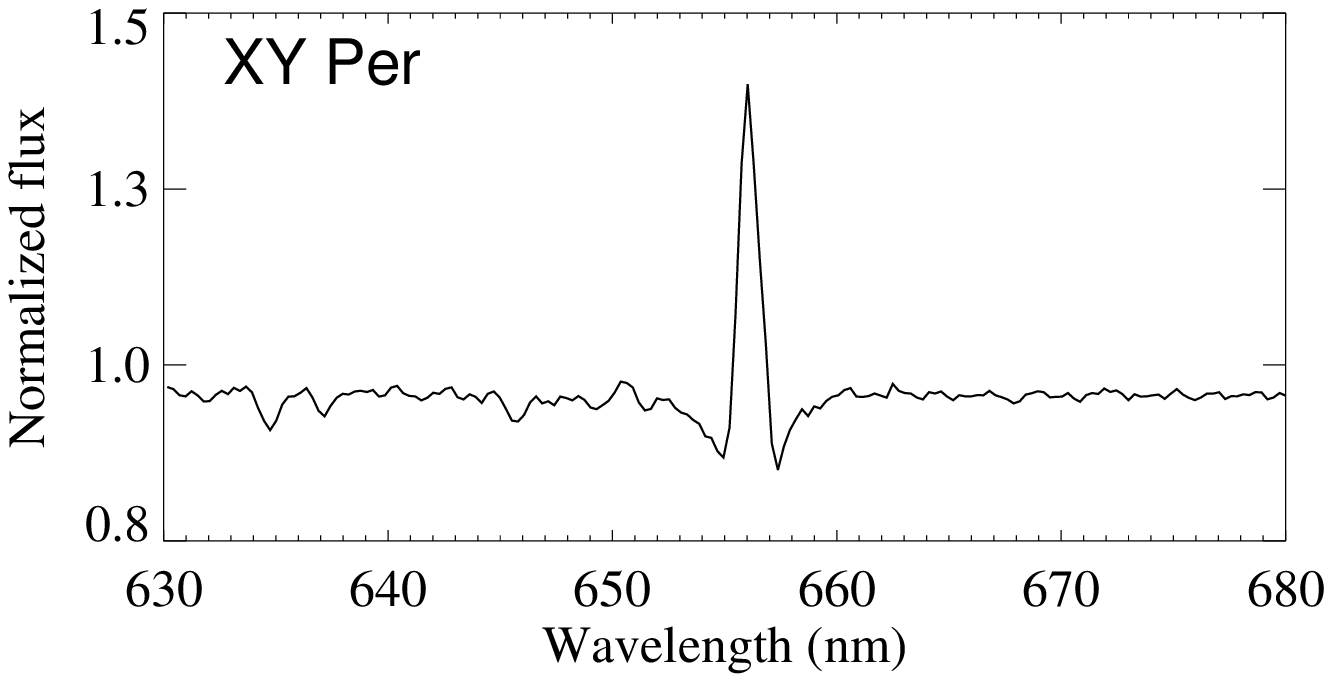}}
\resizebox{0.45\textwidth}{!}{\includegraphics{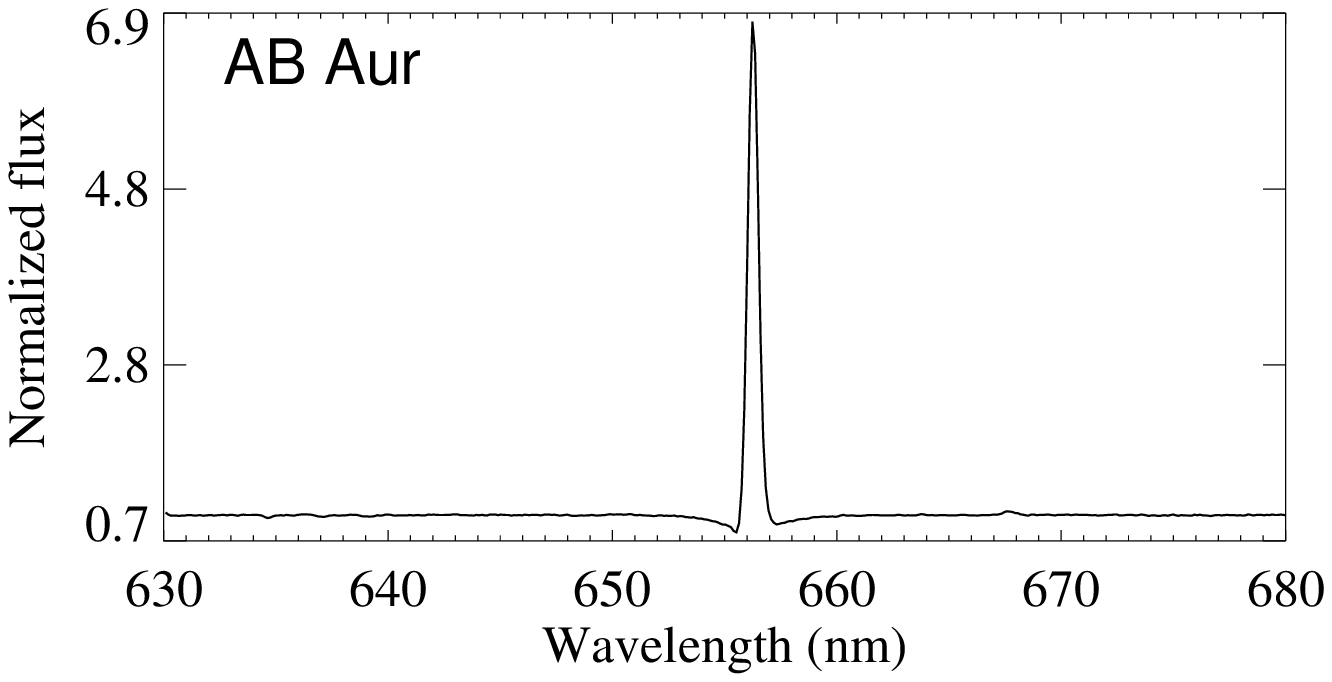}}
\resizebox{0.45\textwidth}{!}{\includegraphics{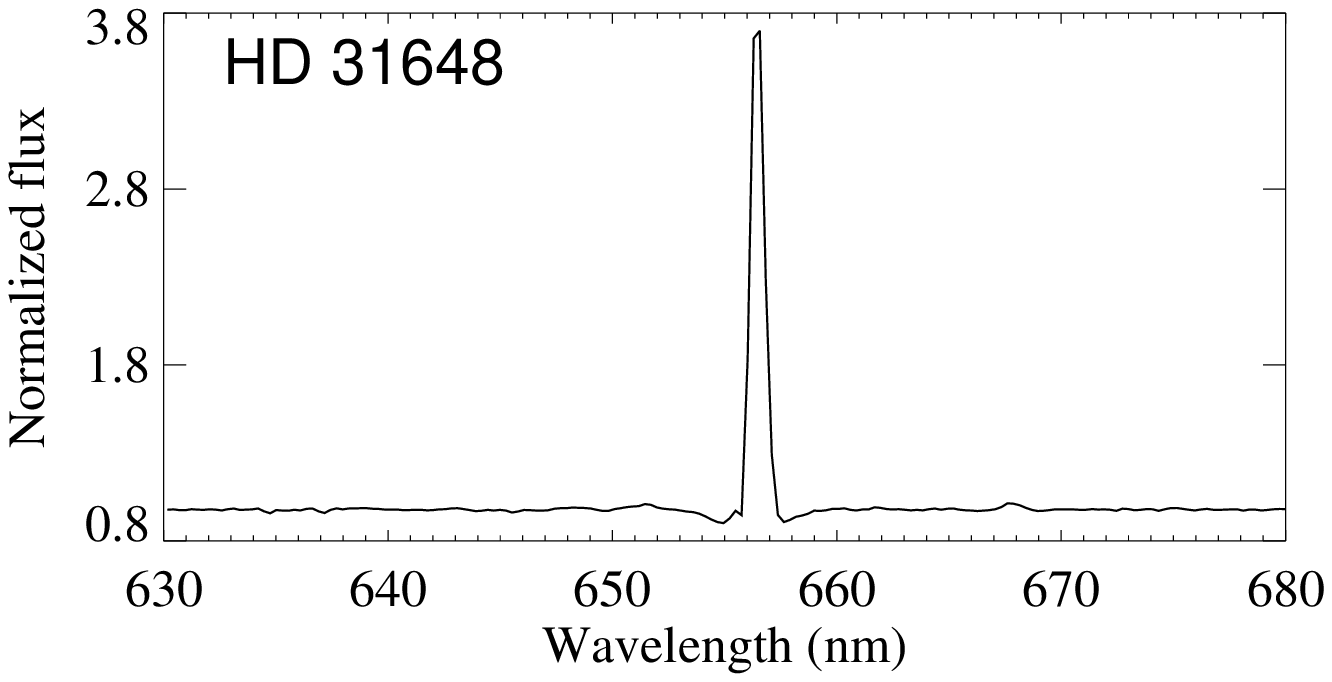}}
\resizebox{0.45\textwidth}{!}{\includegraphics{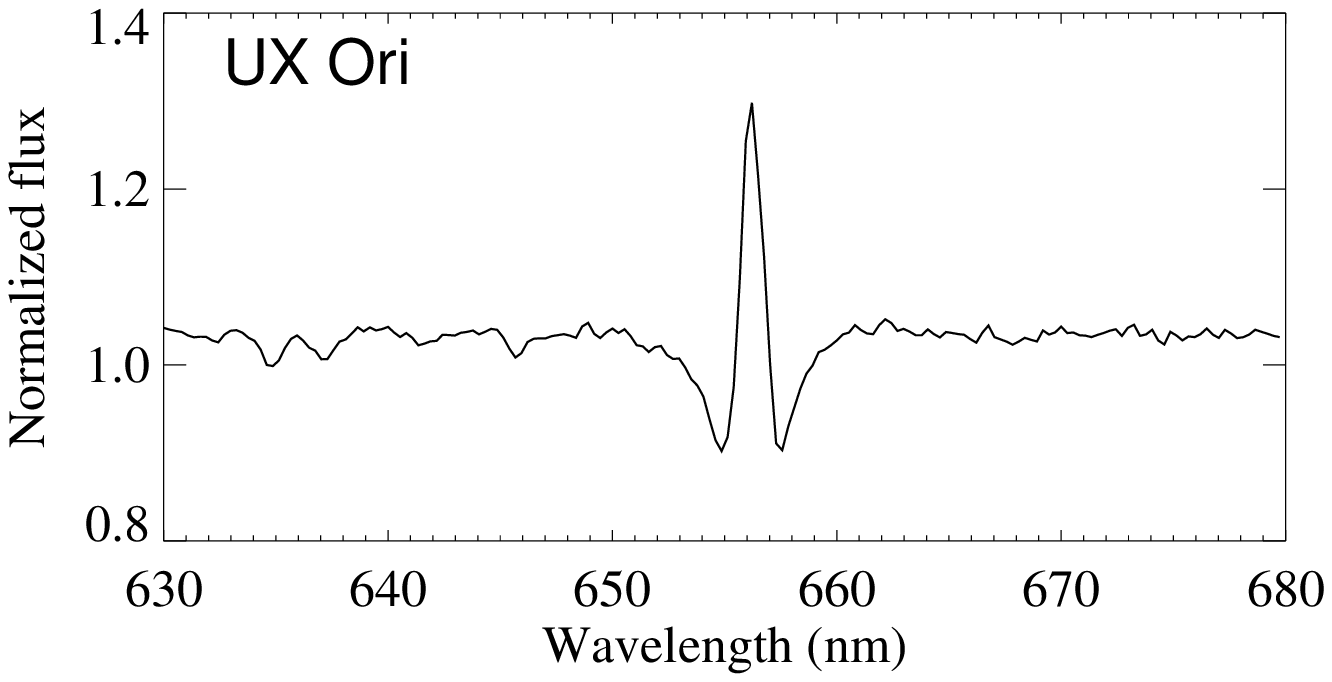}}
\resizebox{0.45\textwidth}{!}{\includegraphics{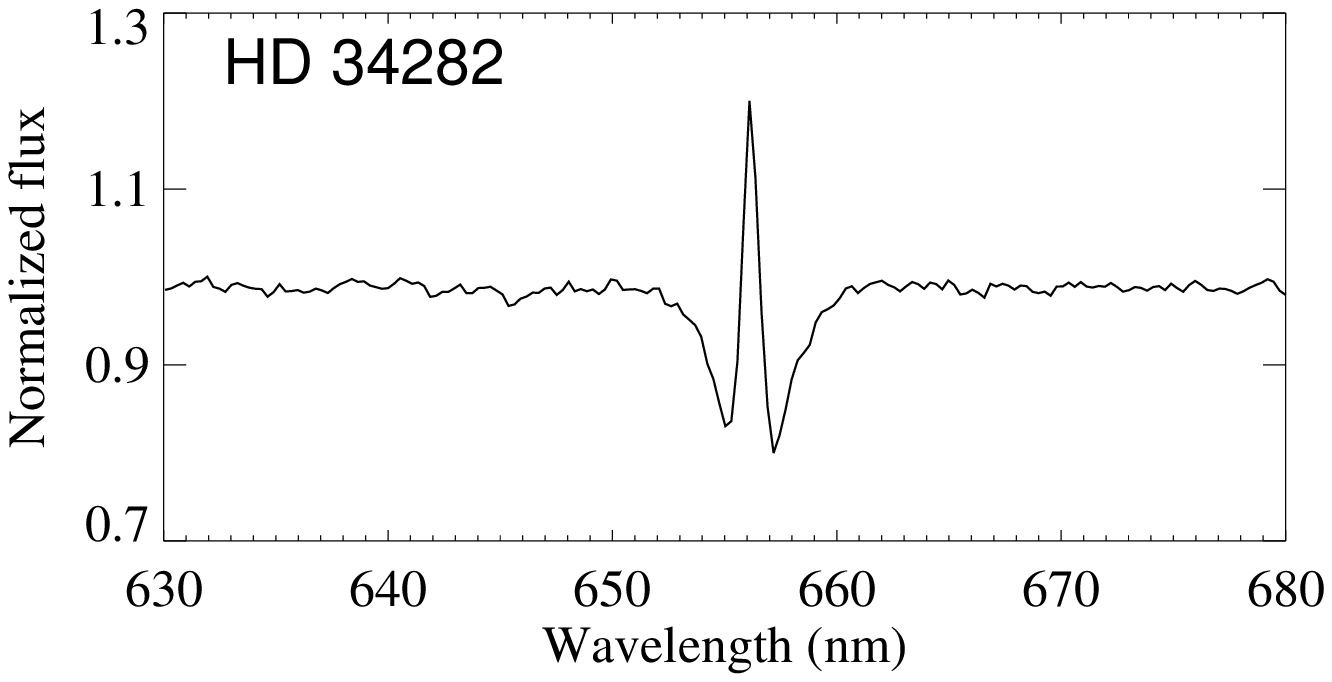}}
\resizebox{0.45\textwidth}{!}{\includegraphics{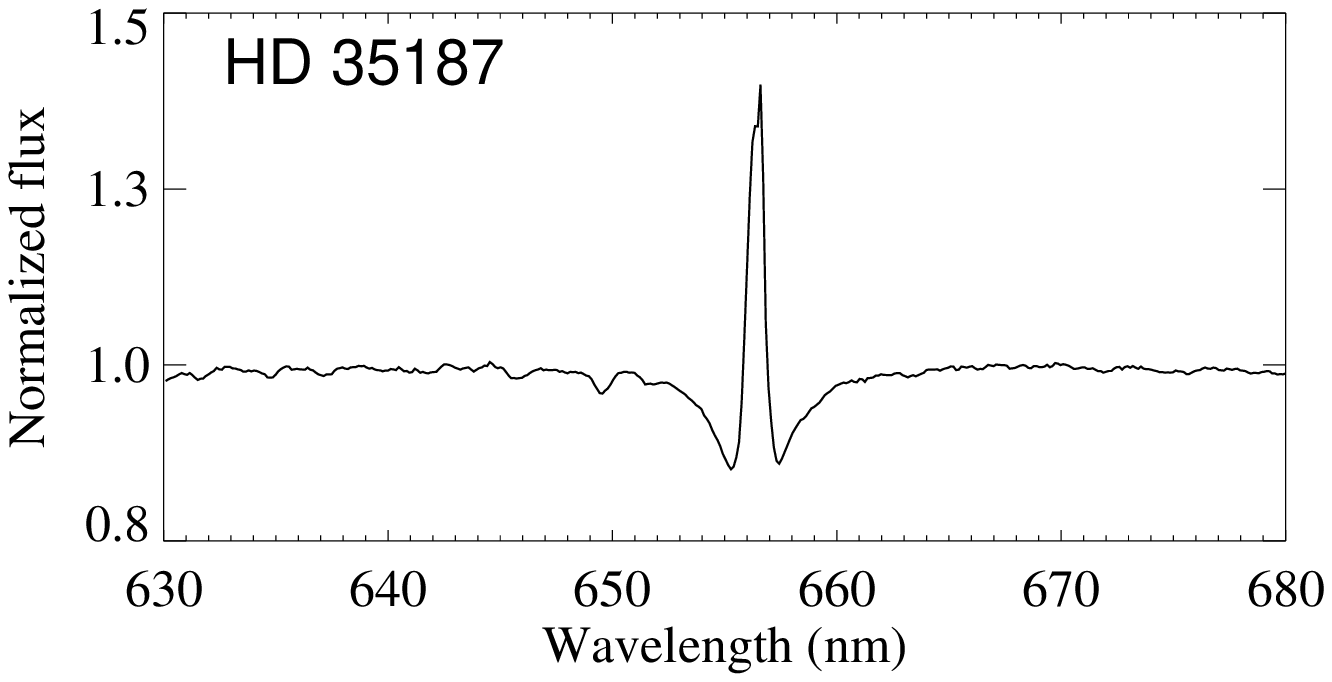}}
\resizebox{0.45\textwidth}{!}{\includegraphics{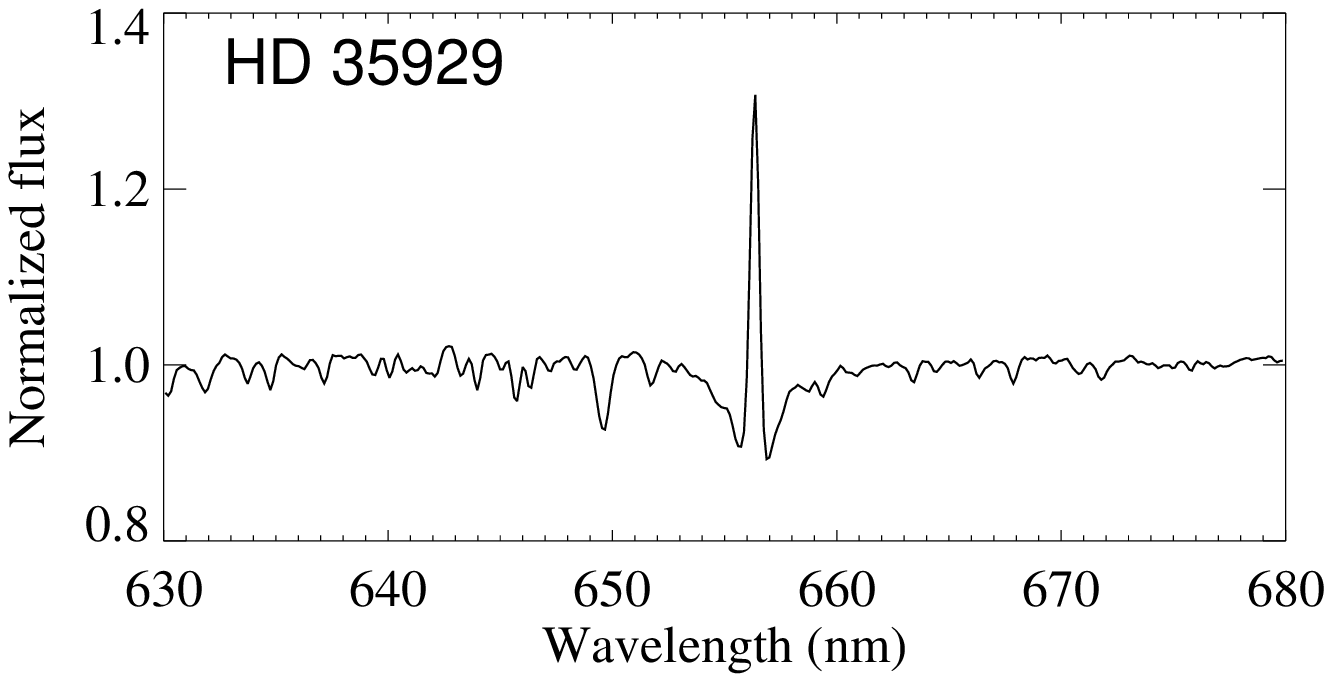}}
\resizebox{0.45\textwidth}{!}{\includegraphics{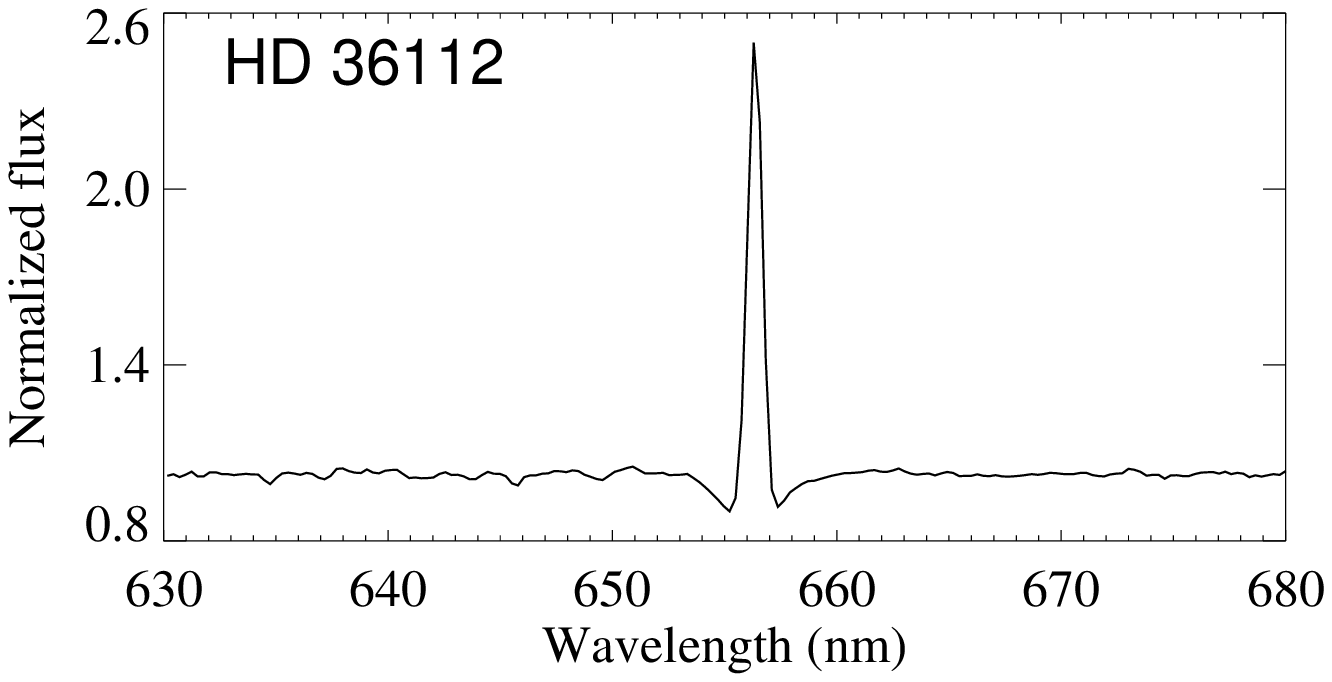}}
\caption{Observed Spectra of Herbig Ae/Be stars. In the spectra of stars with spectral type later than A7, a blend of CaI(18) $\lambda$6493.8 $\AA$ + FeI(168)$\lambda$ 6495 $\AA$ is seen in absorption. }
\label{spectra}
\end{figure}

\begin{figure}
\centering
\figurenum{1}
\resizebox{0.45\textwidth}{!}{\includegraphics{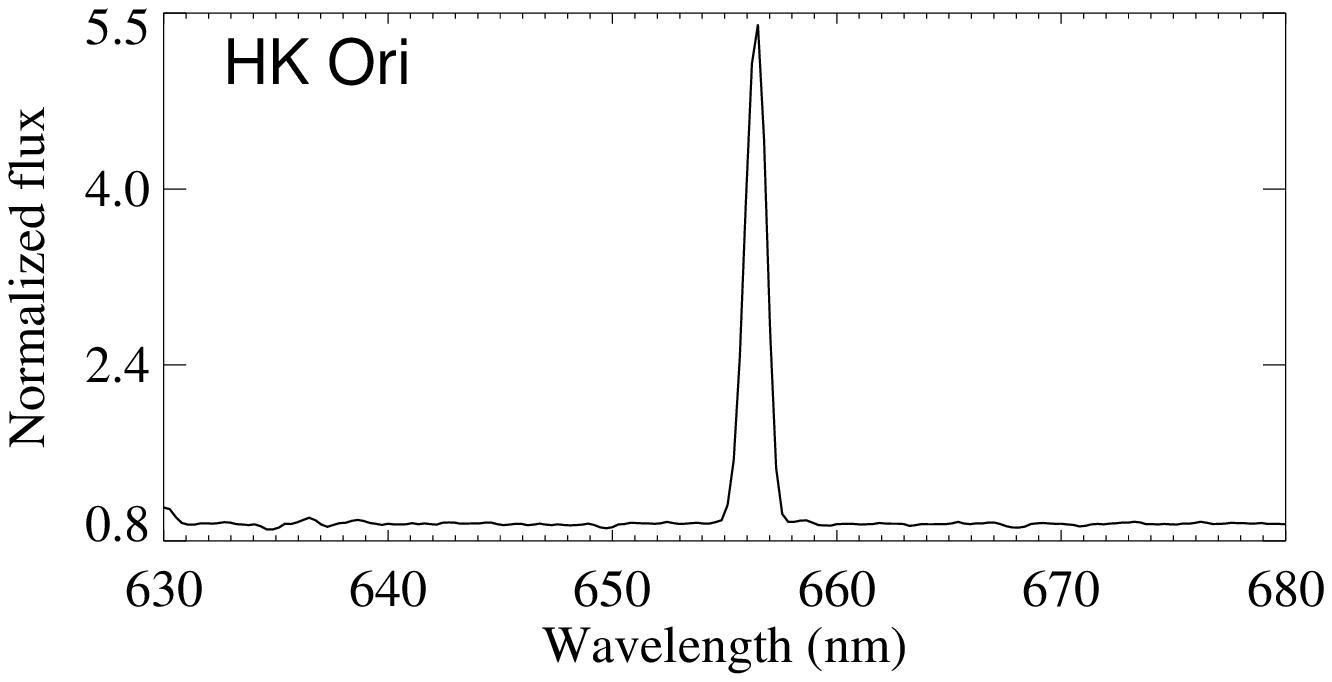}}
\resizebox{0.45\textwidth}{!}{\includegraphics{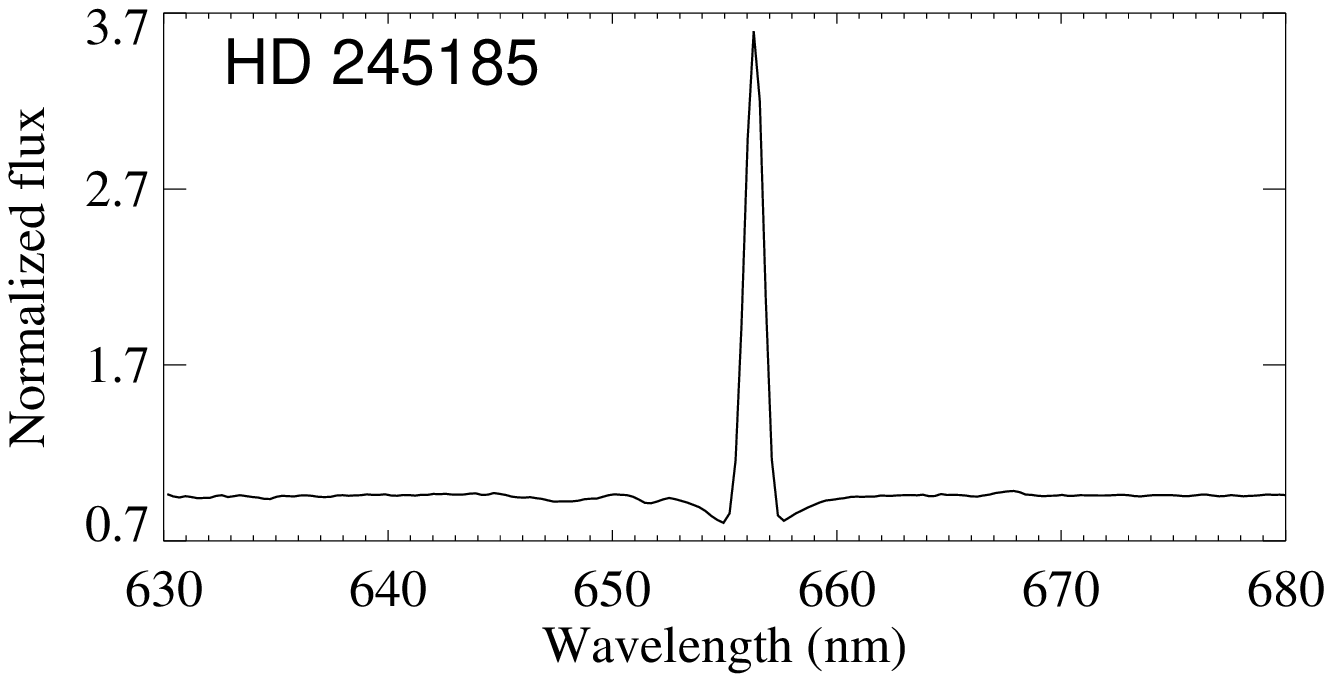}}
\resizebox{0.45\textwidth}{!}{\includegraphics{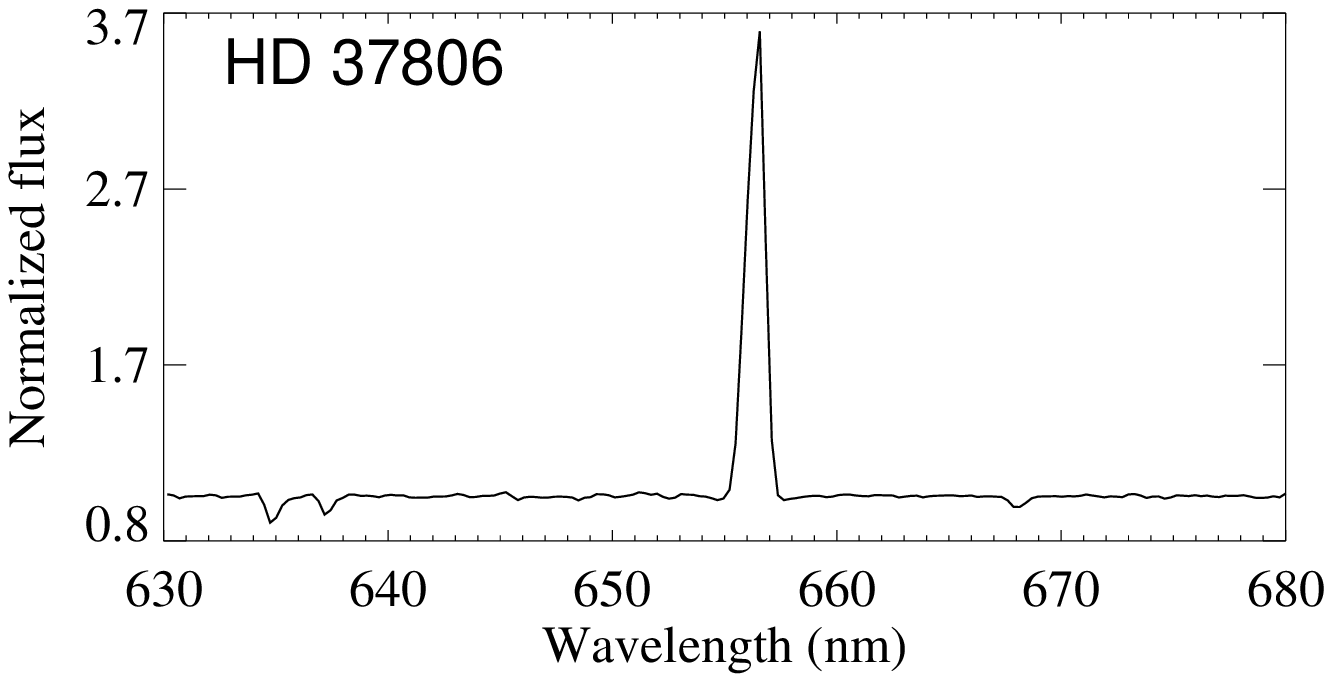}}
\resizebox{0.45\textwidth}{!}{\includegraphics{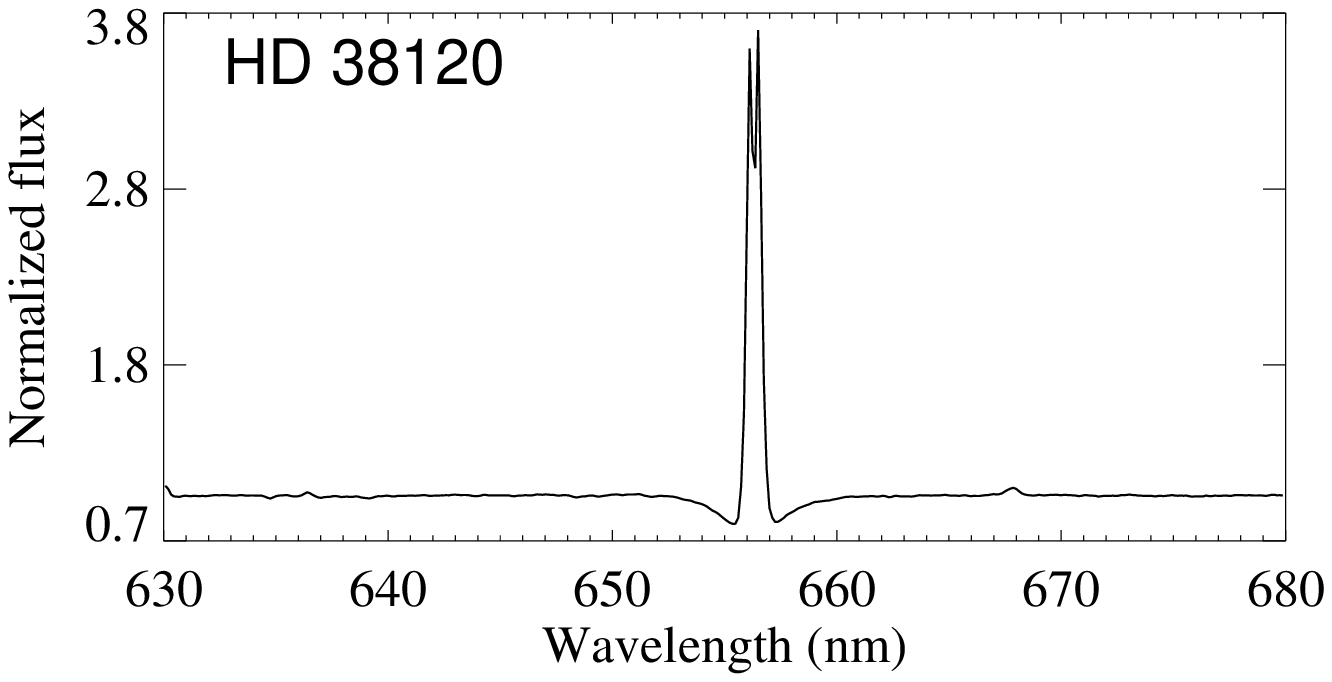}}
\resizebox{0.45\textwidth}{!}{\includegraphics{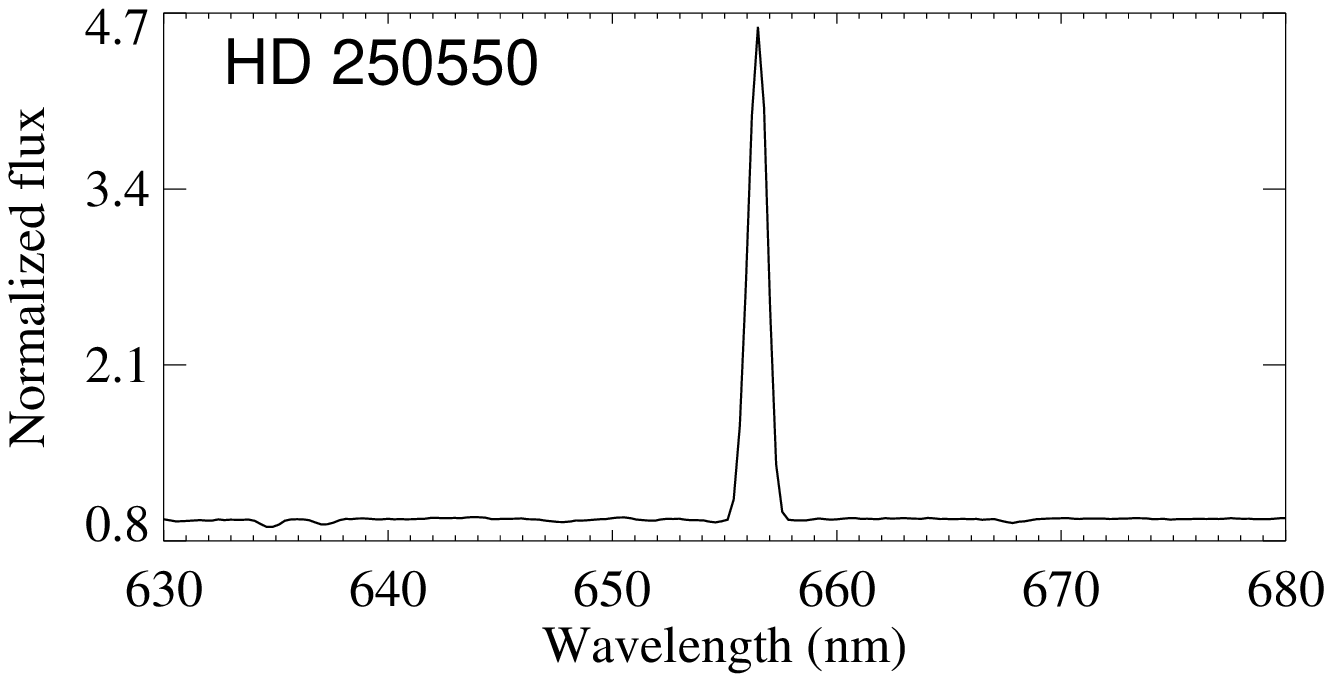}}
\resizebox{0.45\textwidth}{!}{\includegraphics{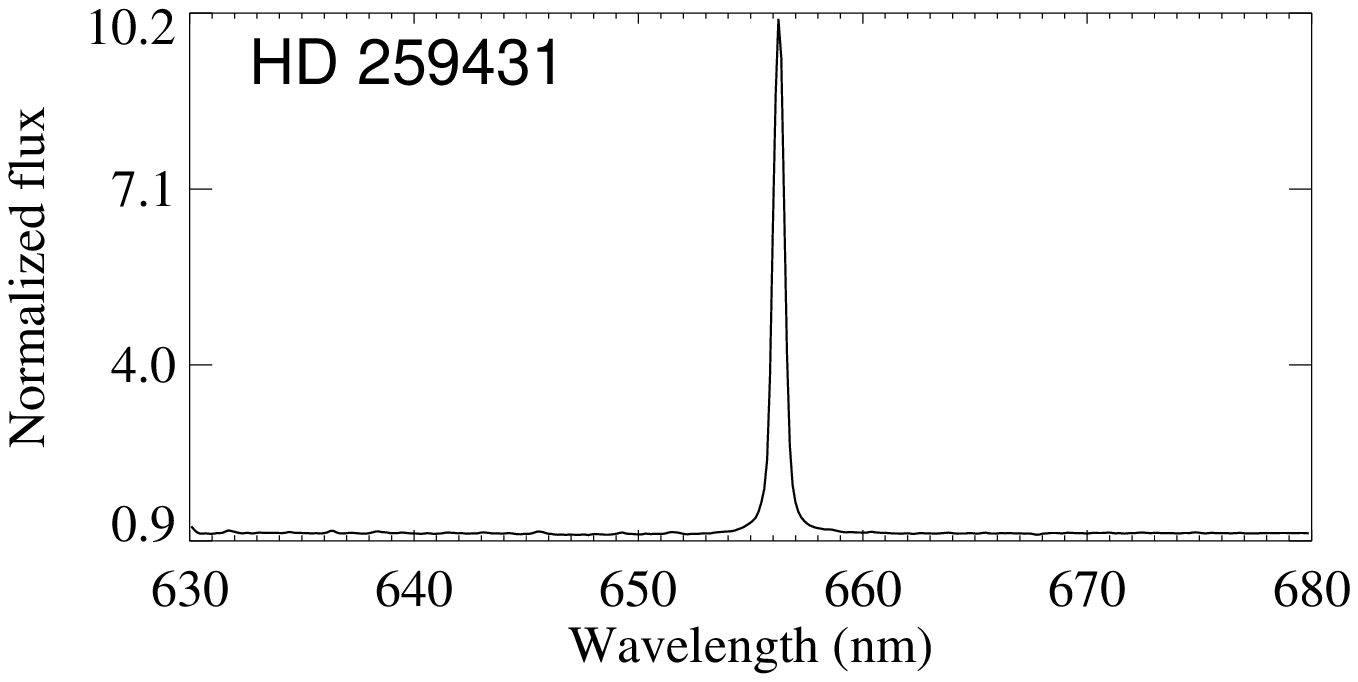}}
\resizebox{0.45\textwidth}{!}{\includegraphics{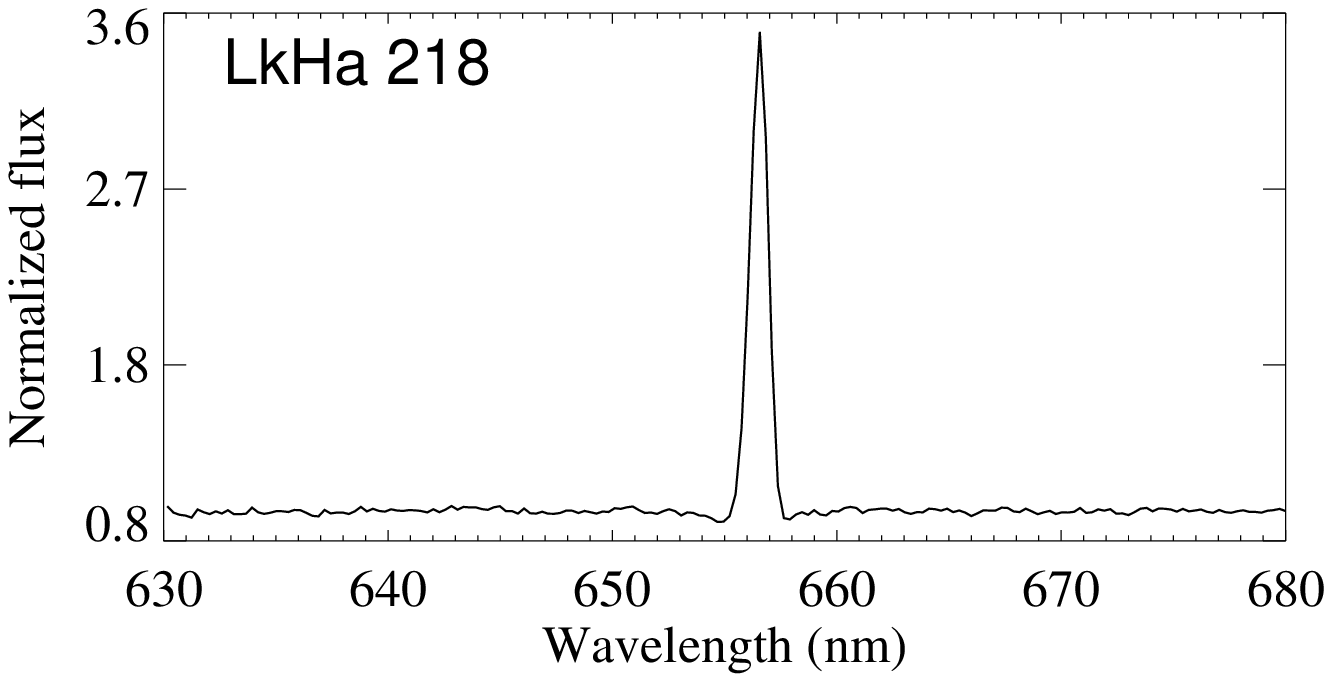}}
\resizebox{0.45\textwidth}{!}{\includegraphics{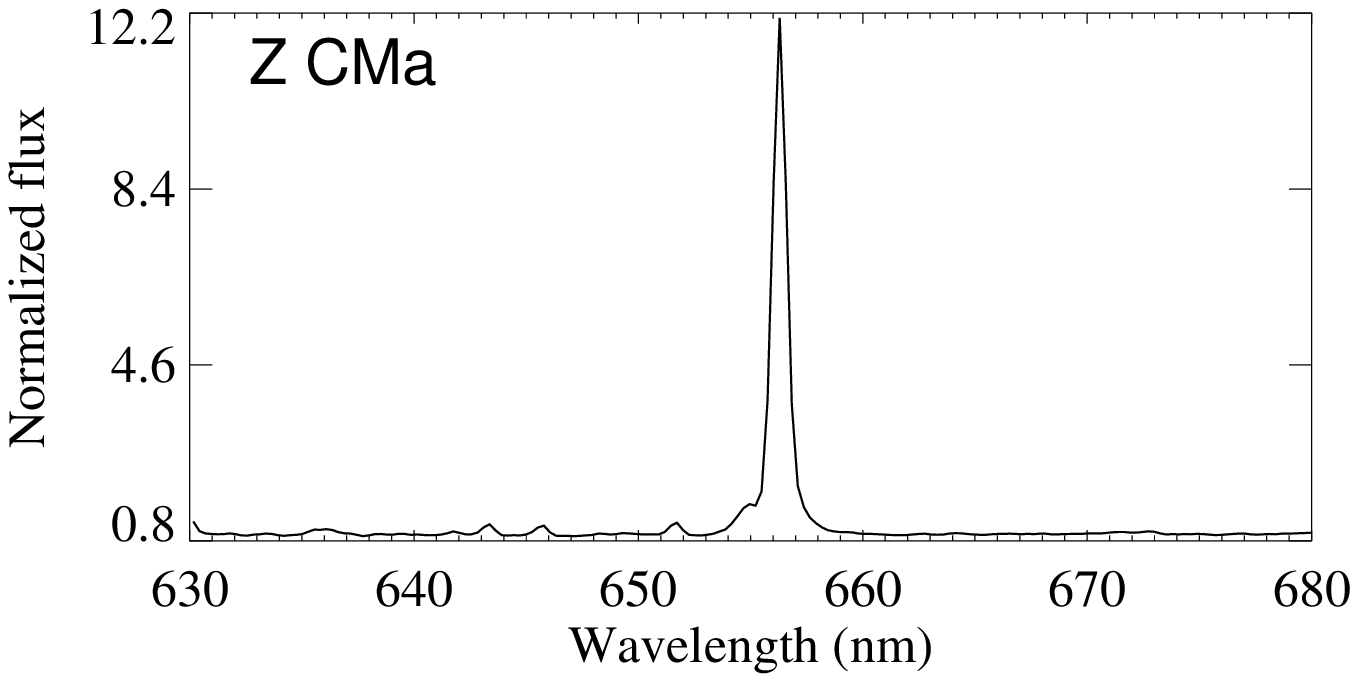}}
\resizebox{0.45\textwidth}{!}{\includegraphics{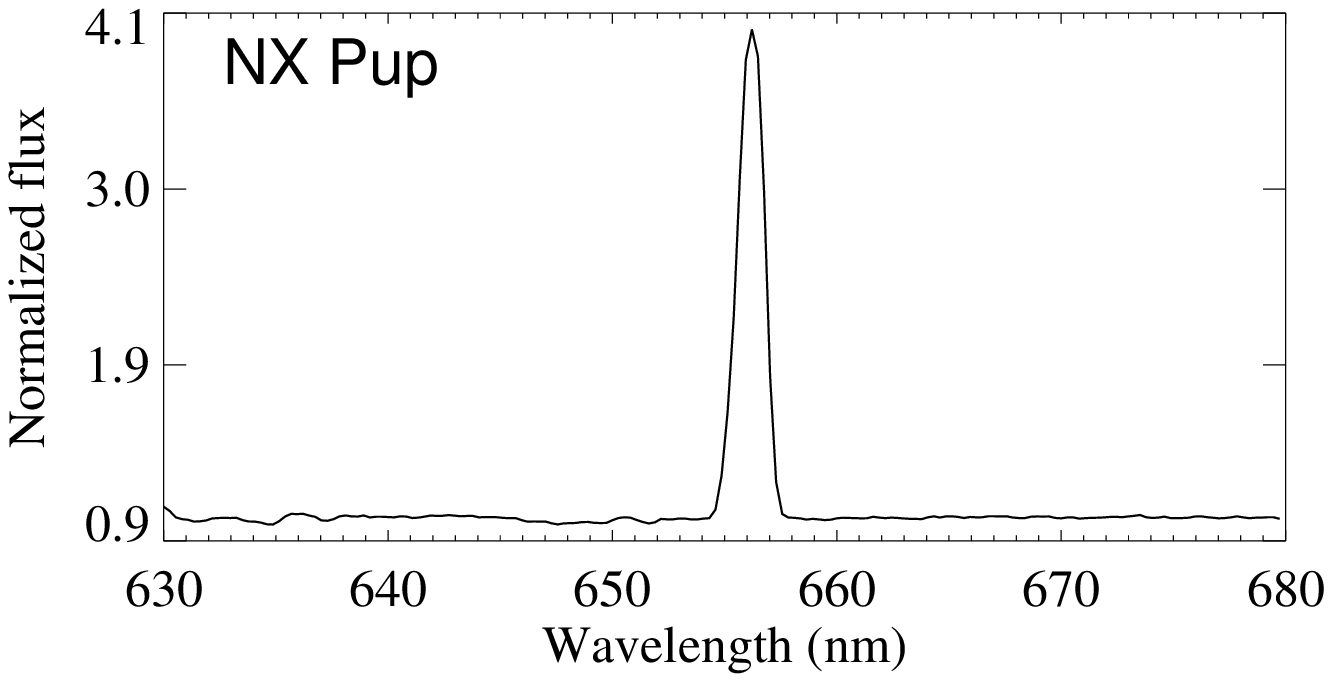}}
\resizebox{0.45\textwidth}{!}{\includegraphics{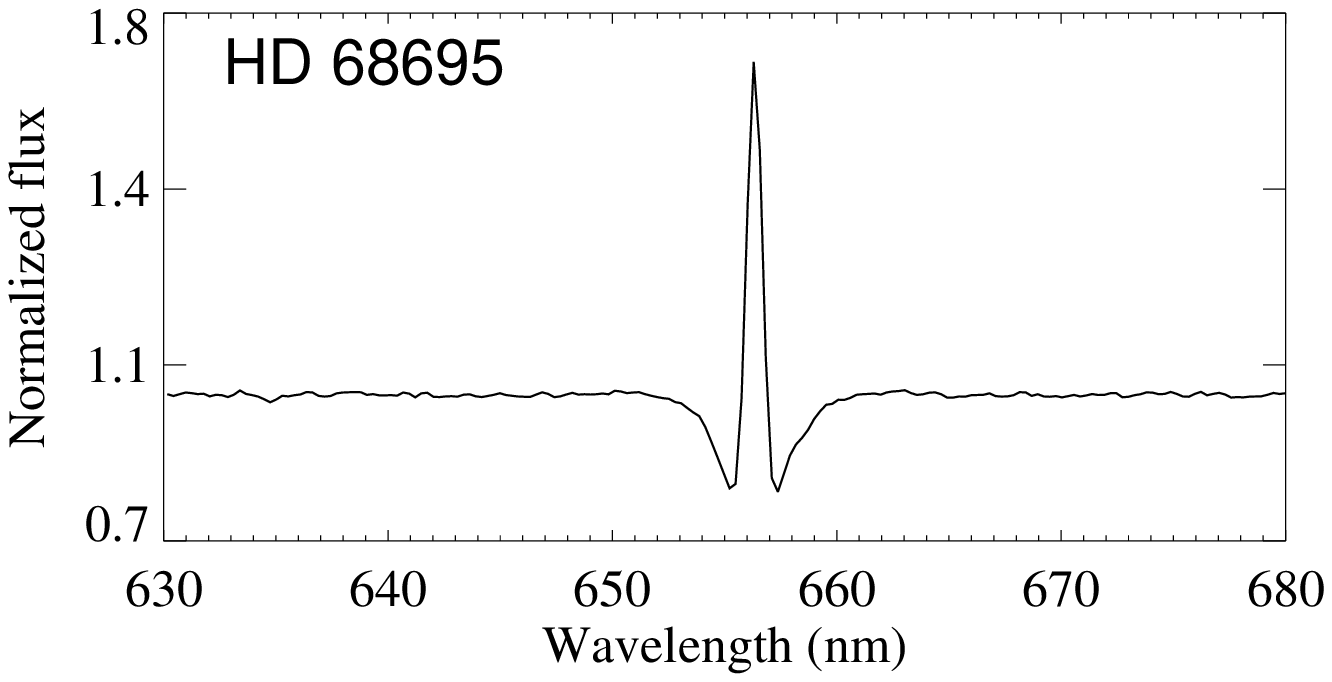}}
\caption{Continued... In the spectra of Z CMa, FeII(40)$\lambda$6516 $\AA$ is seen in emission.}
\end{figure}

\begin{figure}
\centering
\figurenum{1}
\resizebox{0.45\textwidth}{!}{\includegraphics{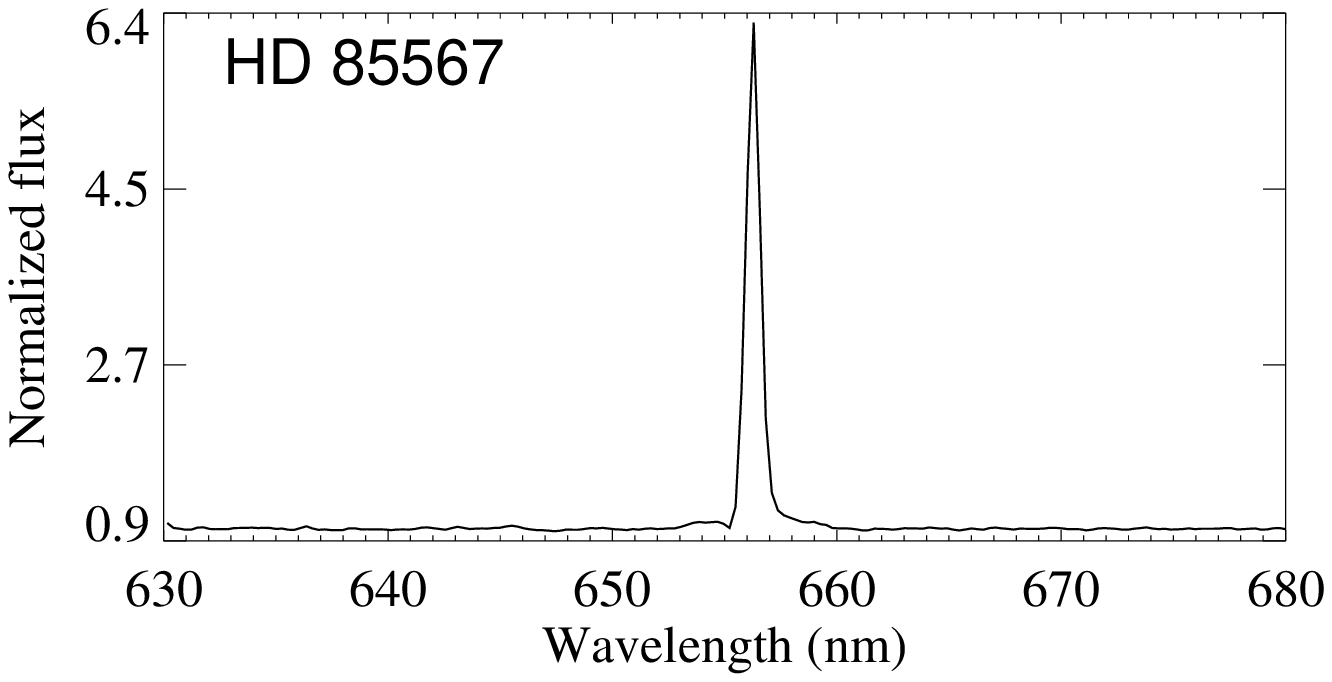}}
\resizebox{0.45\textwidth}{!}{\includegraphics{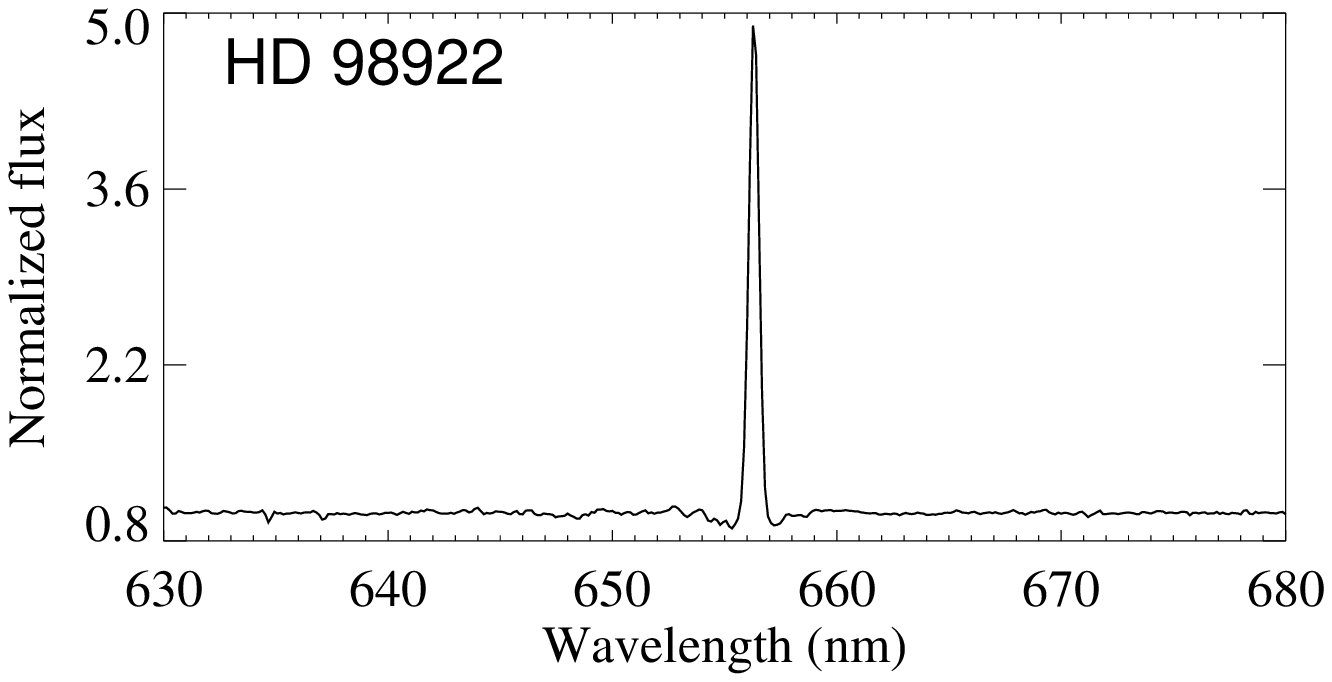}}
\resizebox{0.45\textwidth}{!}{\includegraphics{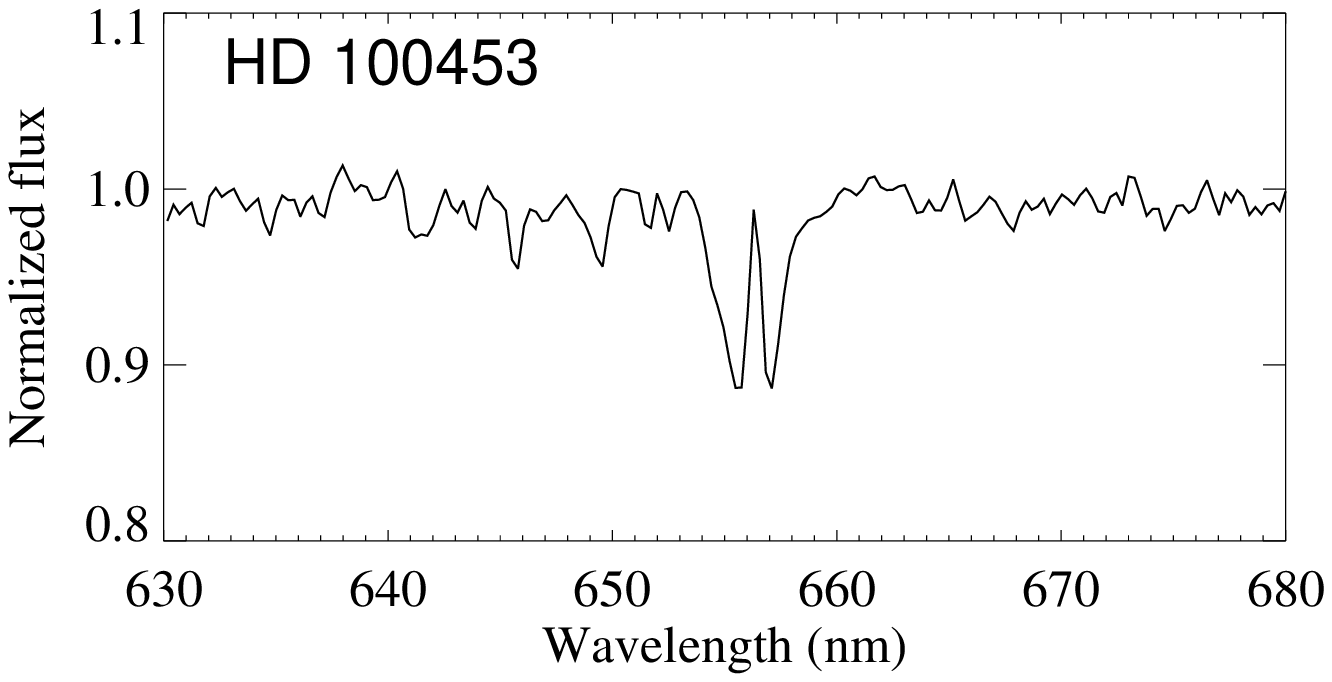}}
\resizebox{0.45\textwidth}{!}{\includegraphics{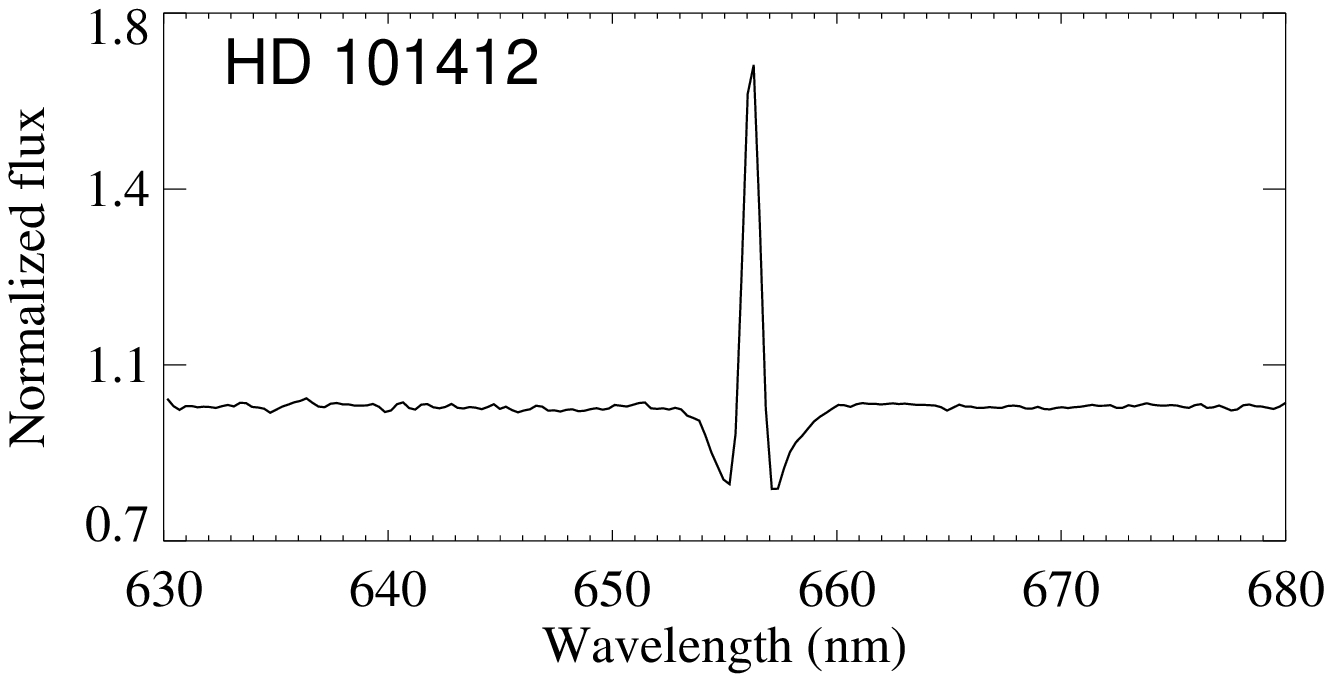}}
\resizebox{0.45\textwidth}{!}{\includegraphics{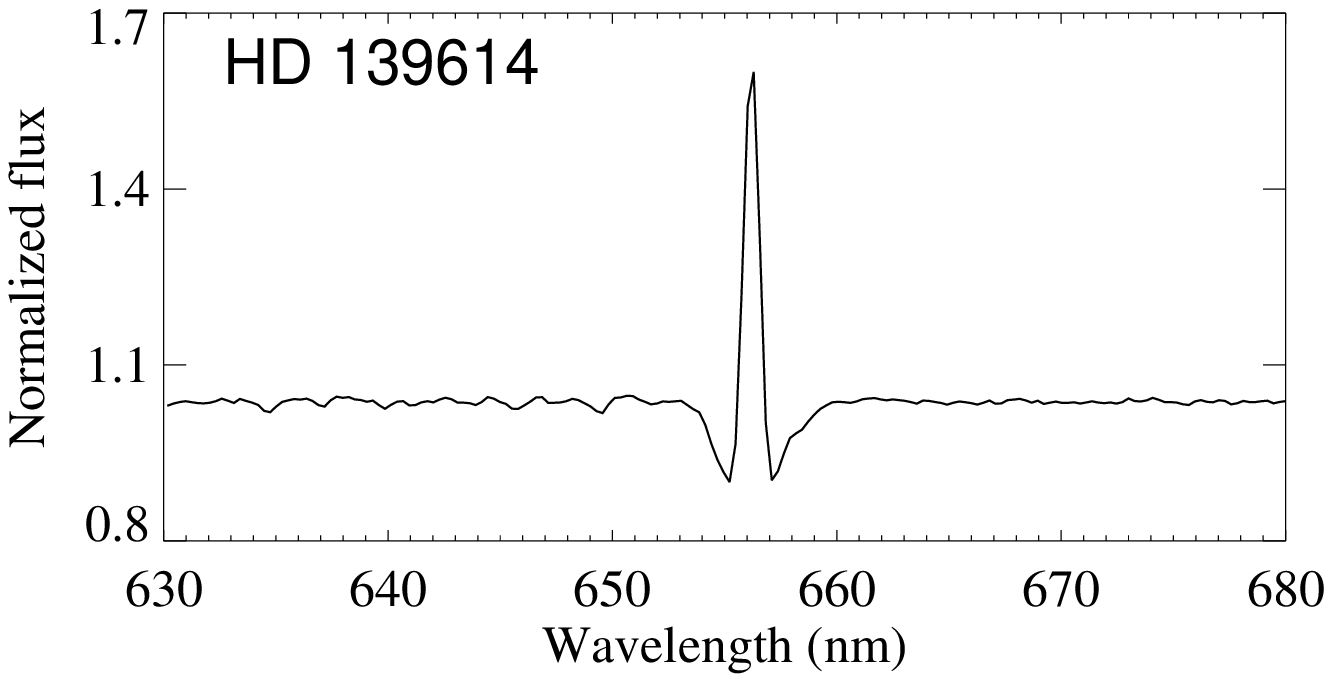}}
\resizebox{0.45\textwidth}{!}{\includegraphics{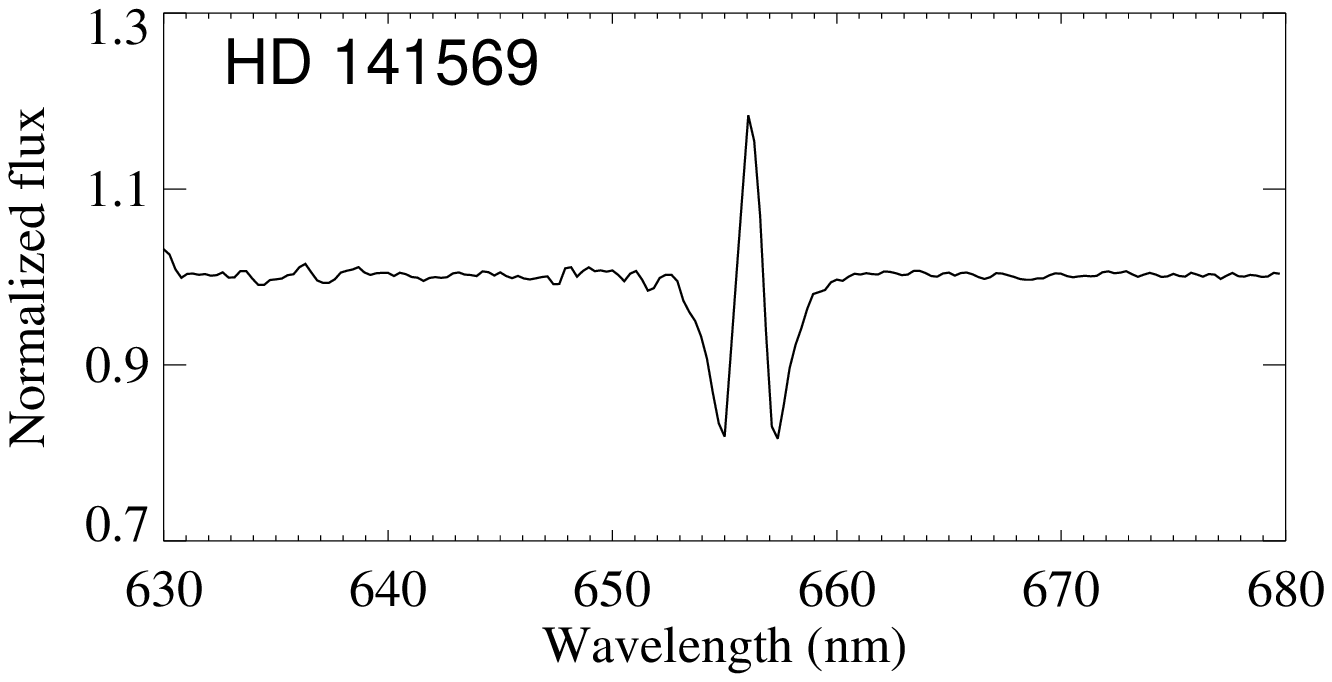}}
\resizebox{0.45\textwidth}{!}{\includegraphics{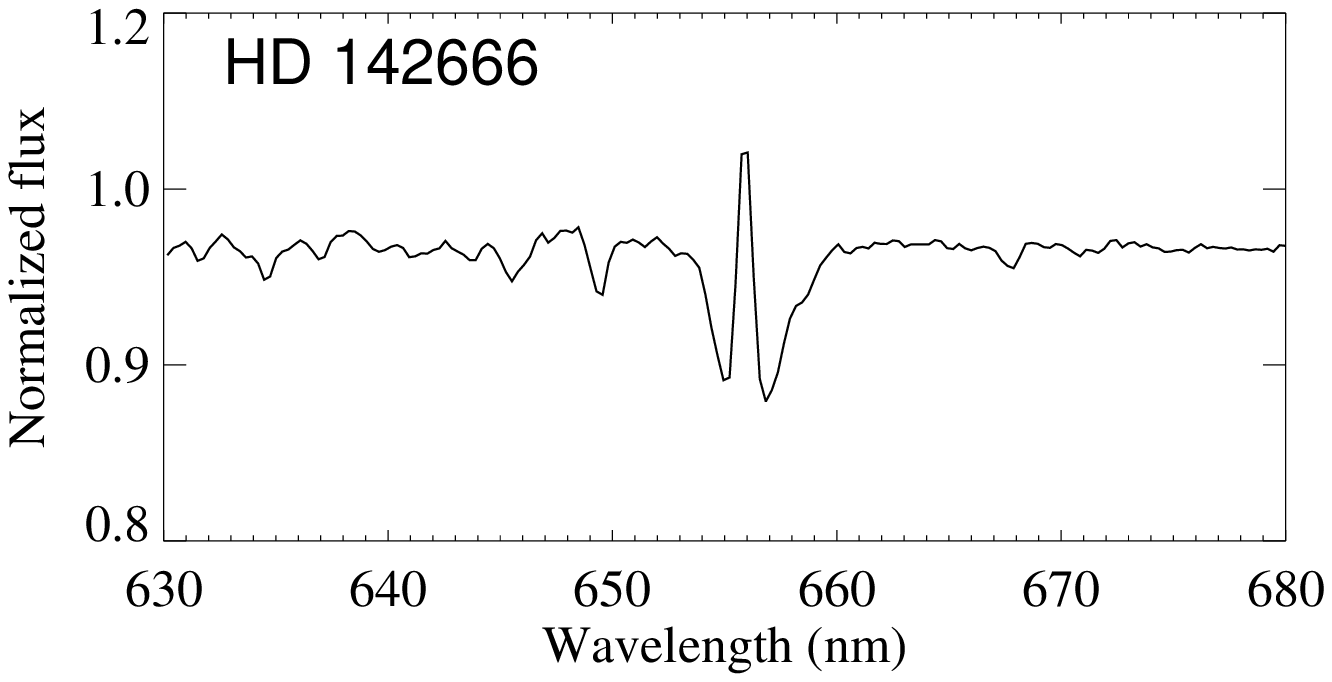}}
\resizebox{0.45\textwidth}{!}{\includegraphics{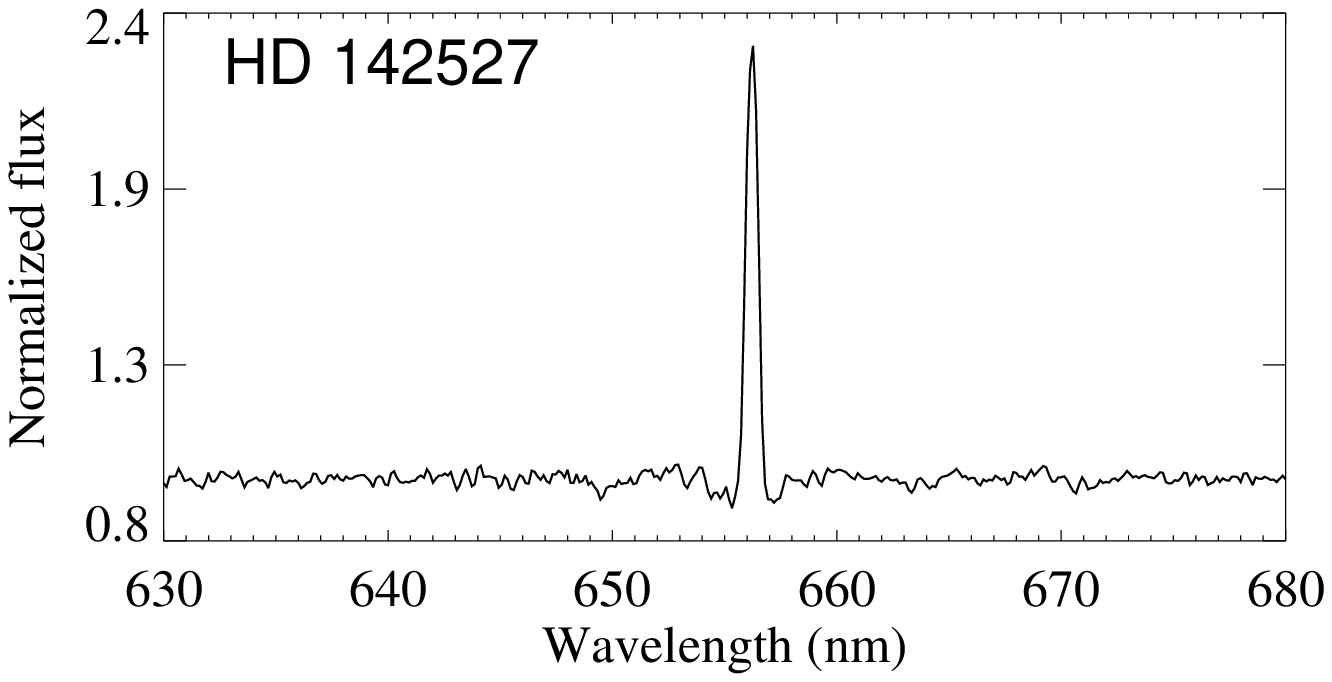}}
\resizebox{0.45\textwidth}{!}{\includegraphics{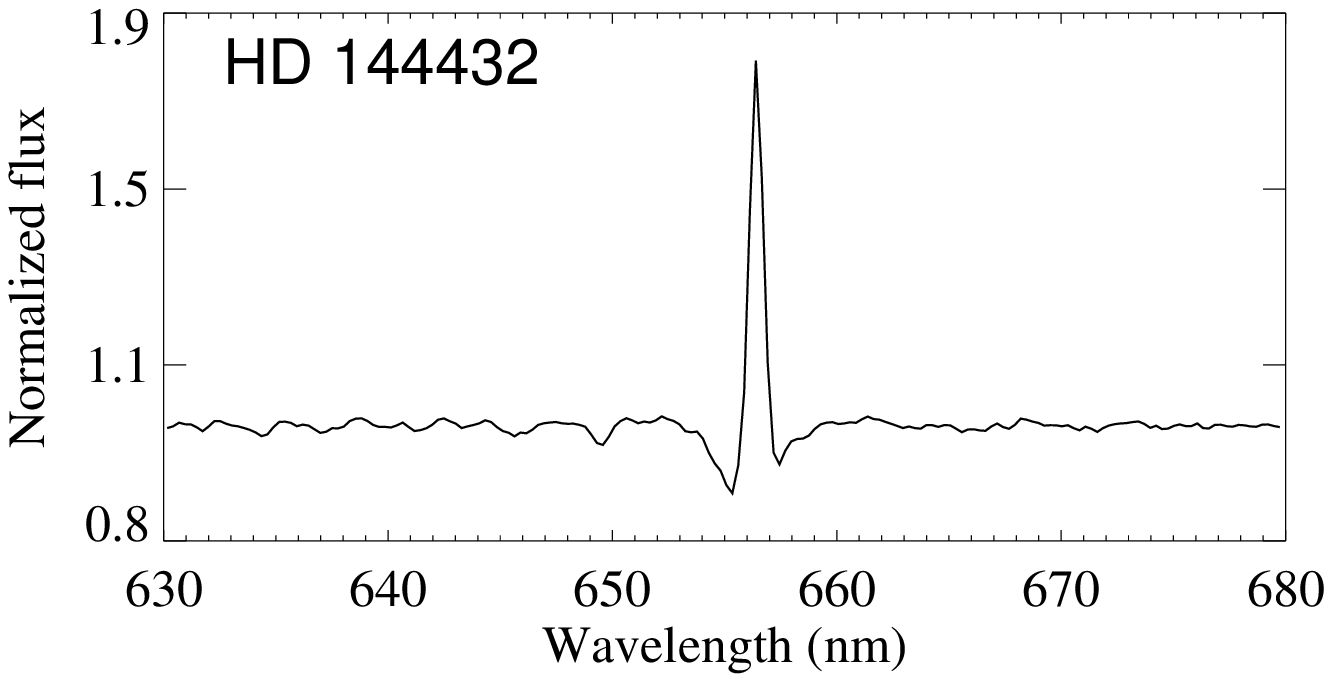}}
\resizebox{0.45\textwidth}{!}{\includegraphics{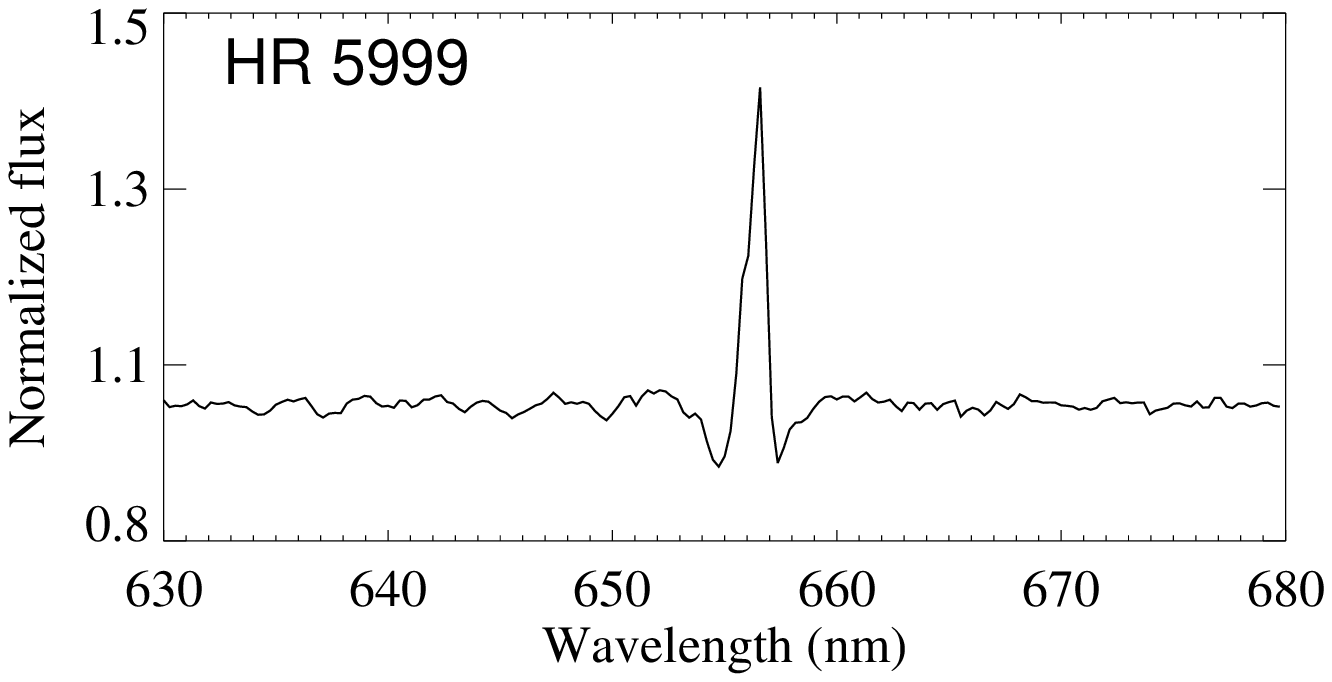}}
\caption{Continued... }
\end{figure}

\begin{figure}
\centering
\figurenum{1}
\resizebox{0.45\textwidth}{!}{\includegraphics{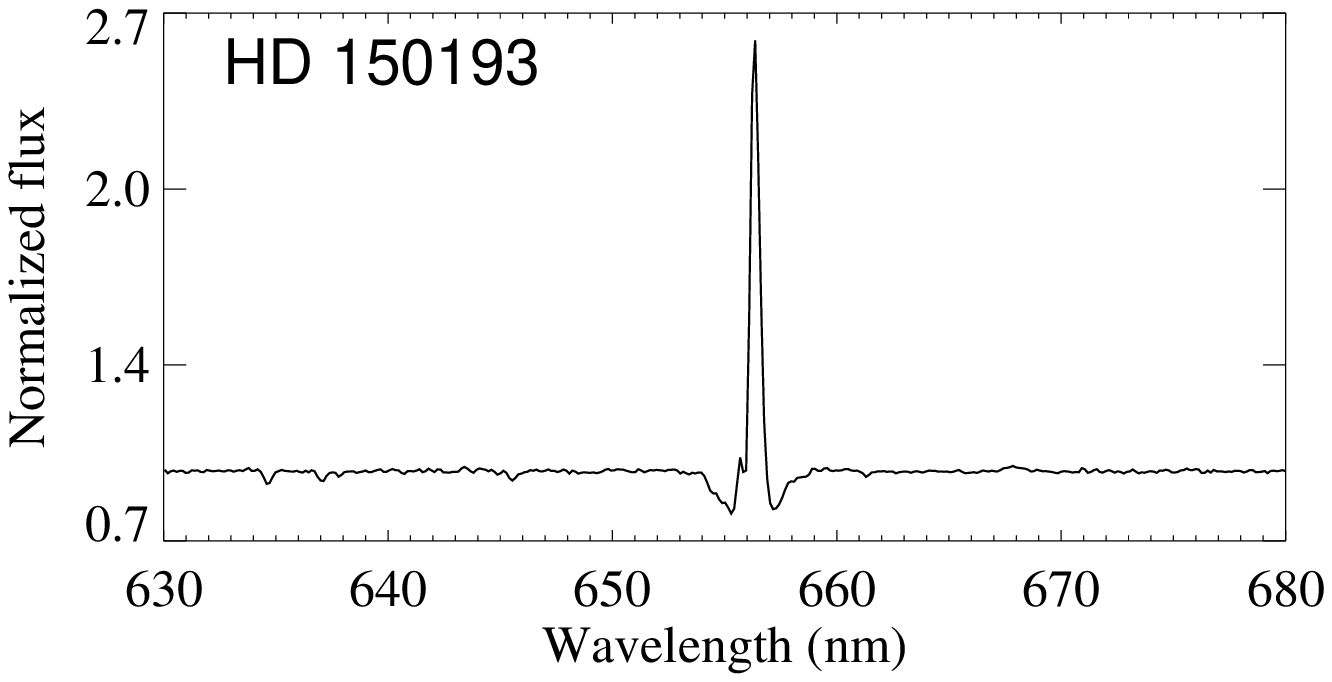}}
\resizebox{0.45\textwidth}{!}{\includegraphics{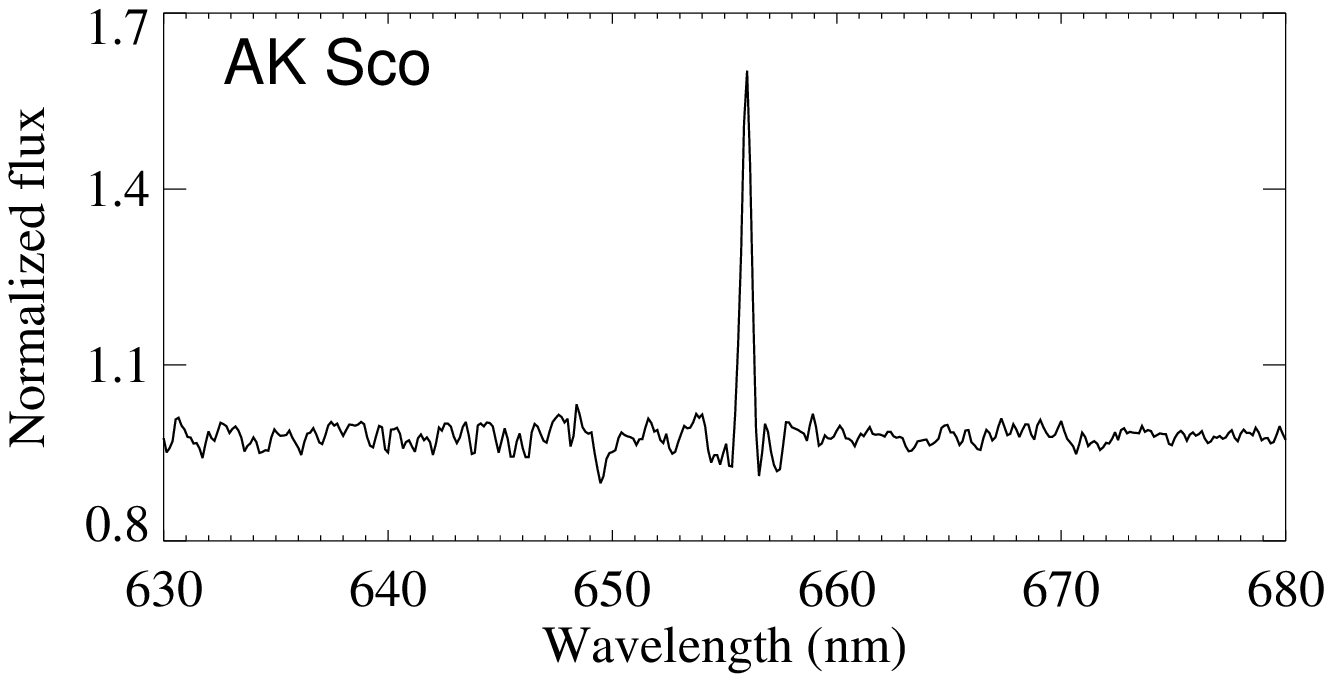}}
\resizebox{0.45\textwidth}{!}{\includegraphics{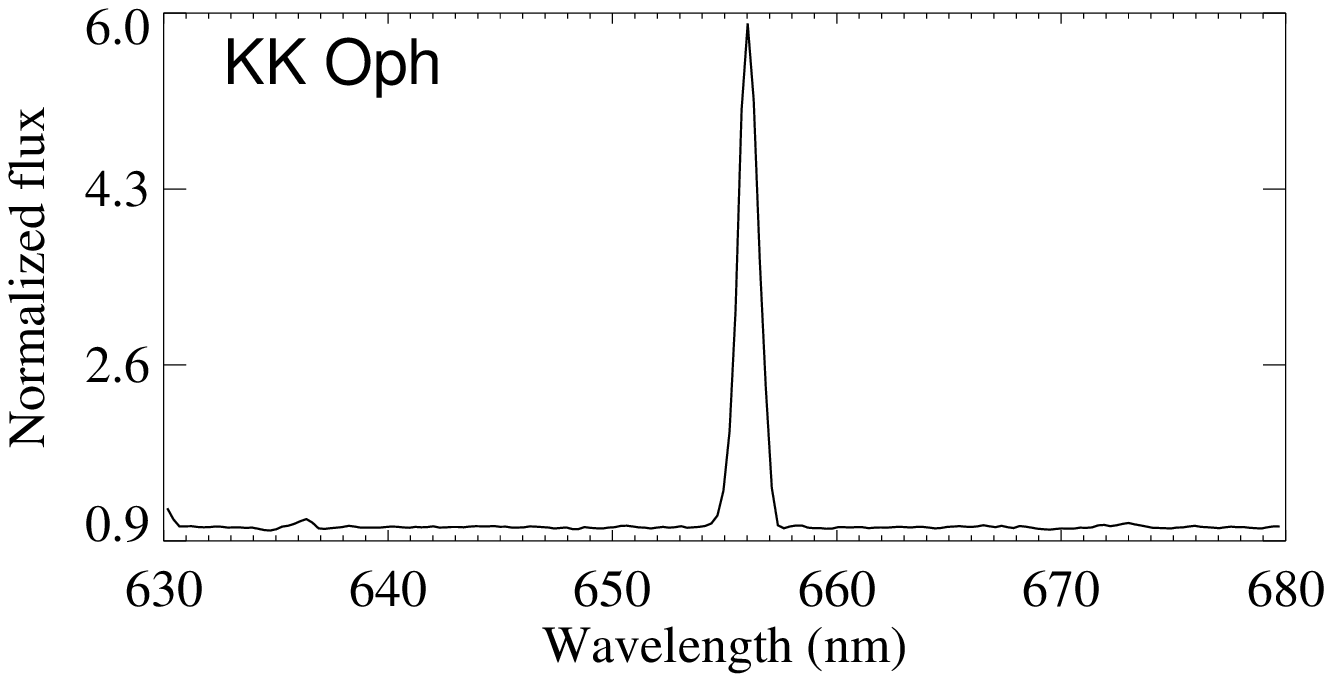}}
\resizebox{0.45\textwidth}{!}{\includegraphics{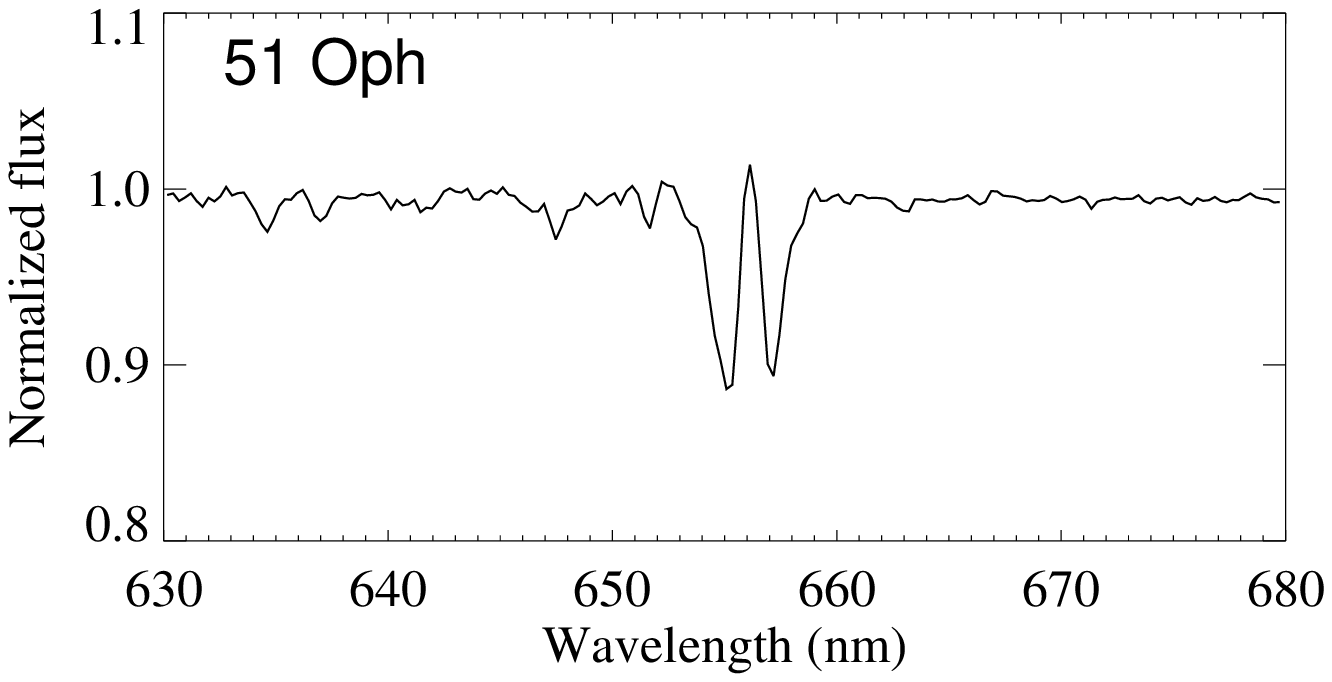}}
\resizebox{0.45\textwidth}{!}{\includegraphics{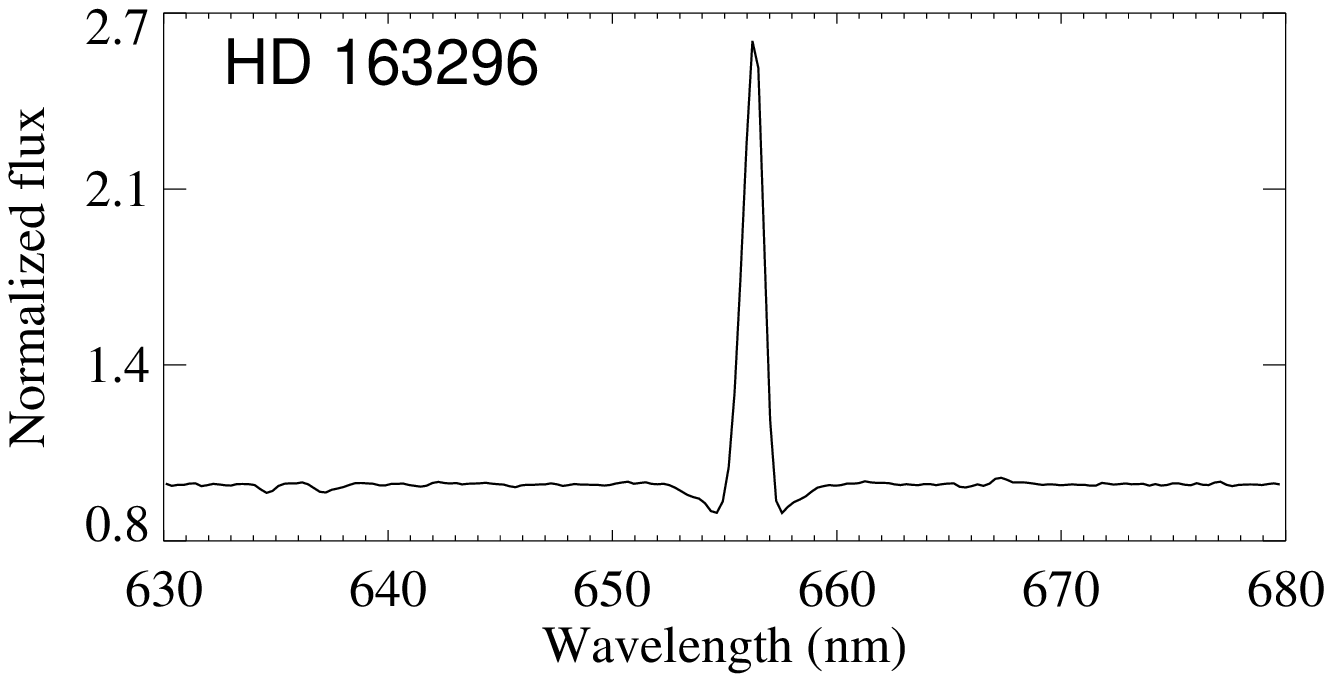}}
\resizebox{0.45\textwidth}{!}{\includegraphics{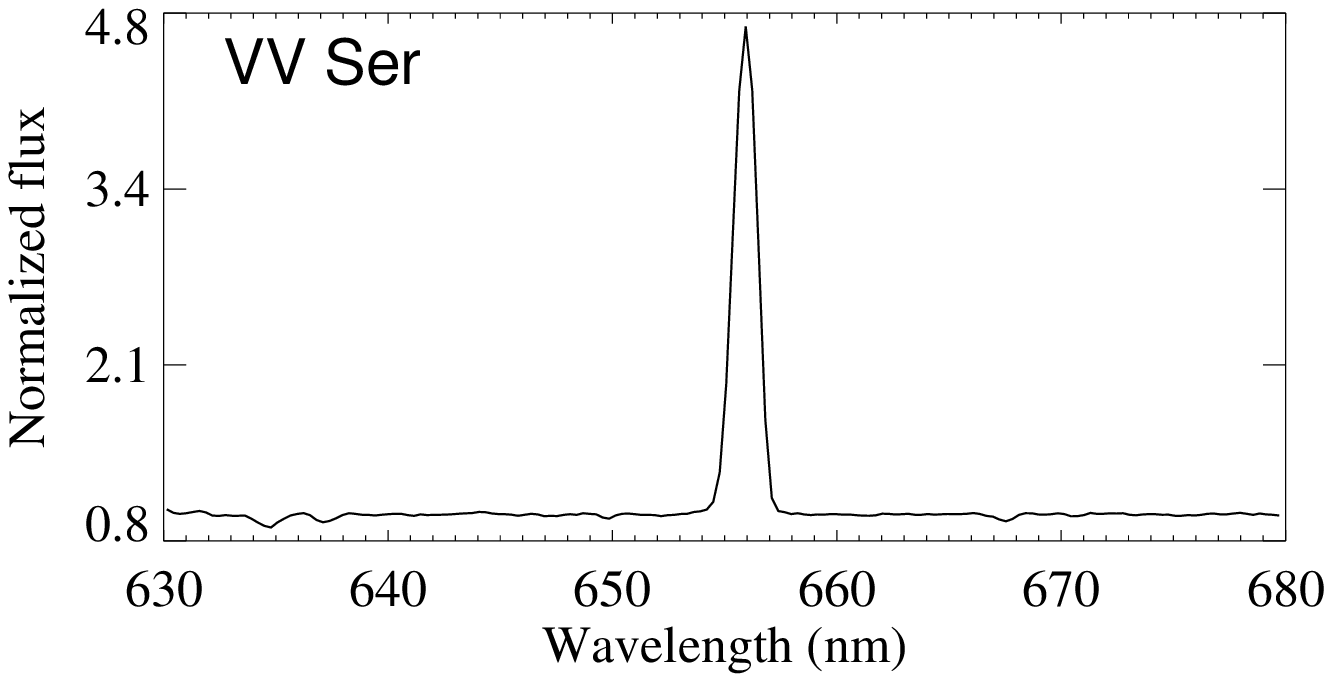}}
\resizebox{0.45\textwidth}{!}{\includegraphics{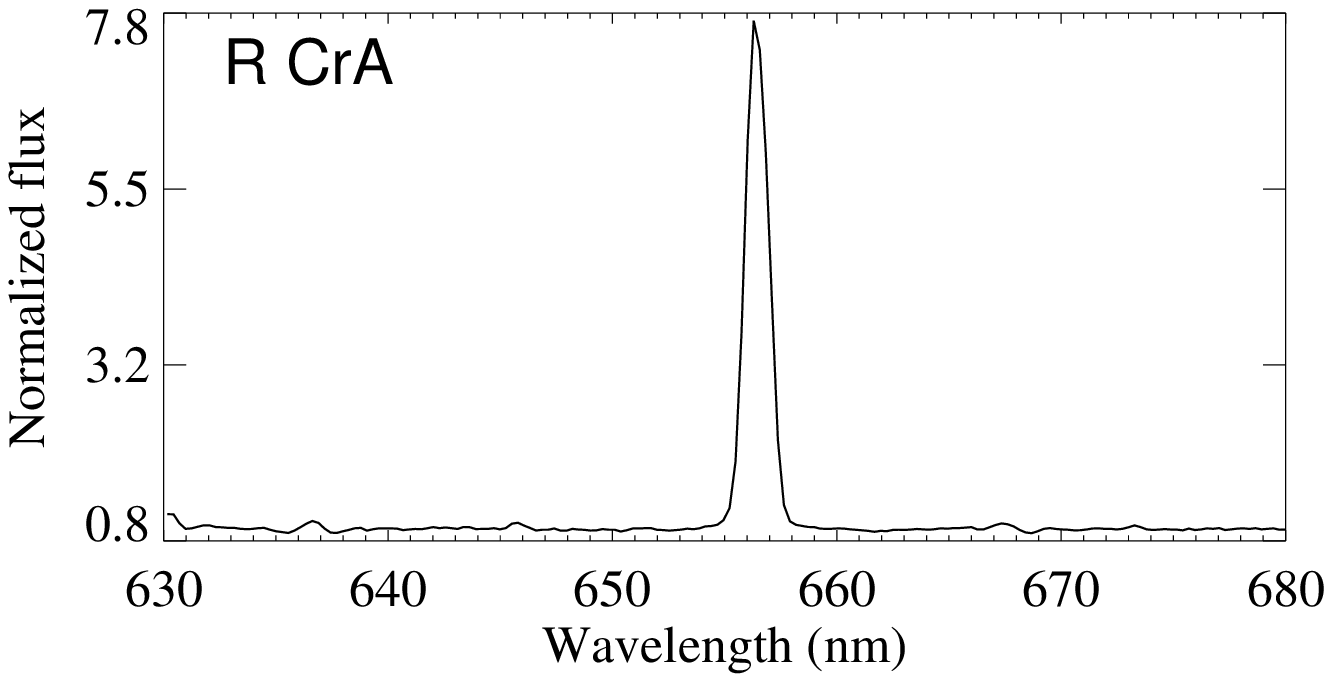}}
\resizebox{0.45\textwidth}{!}{\includegraphics{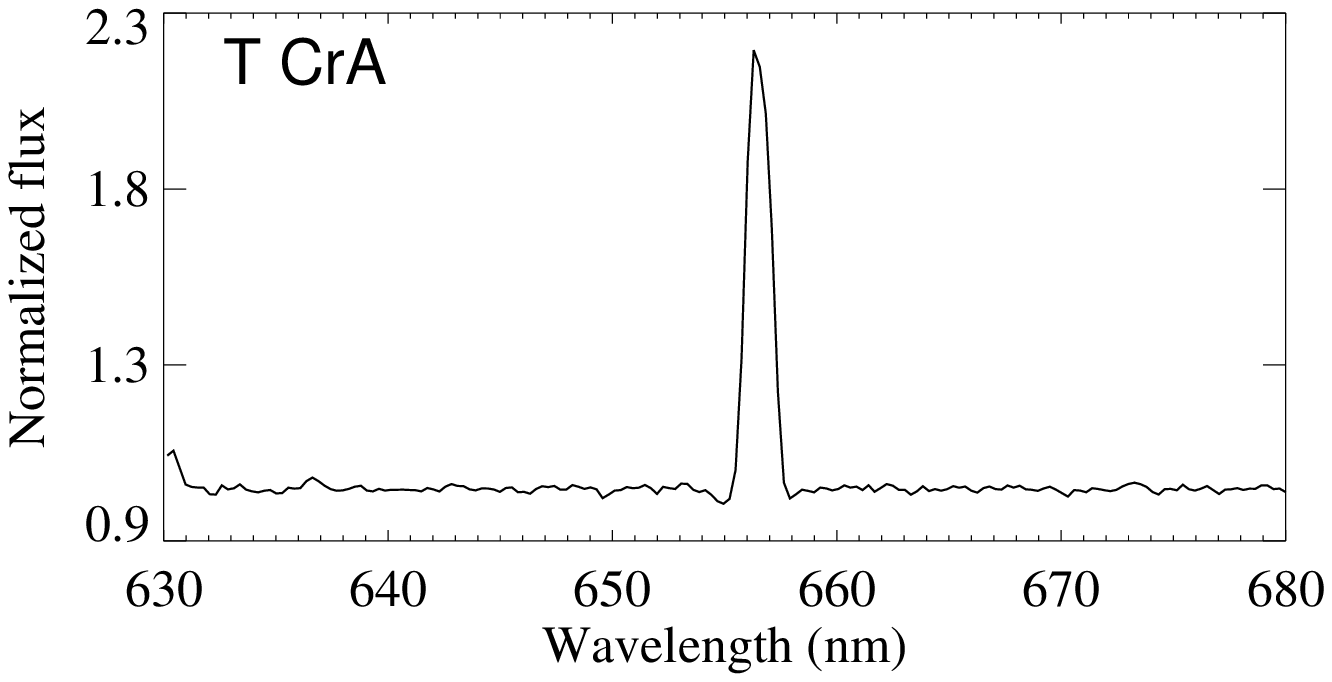}}
\resizebox{0.45\textwidth}{!}{\includegraphics{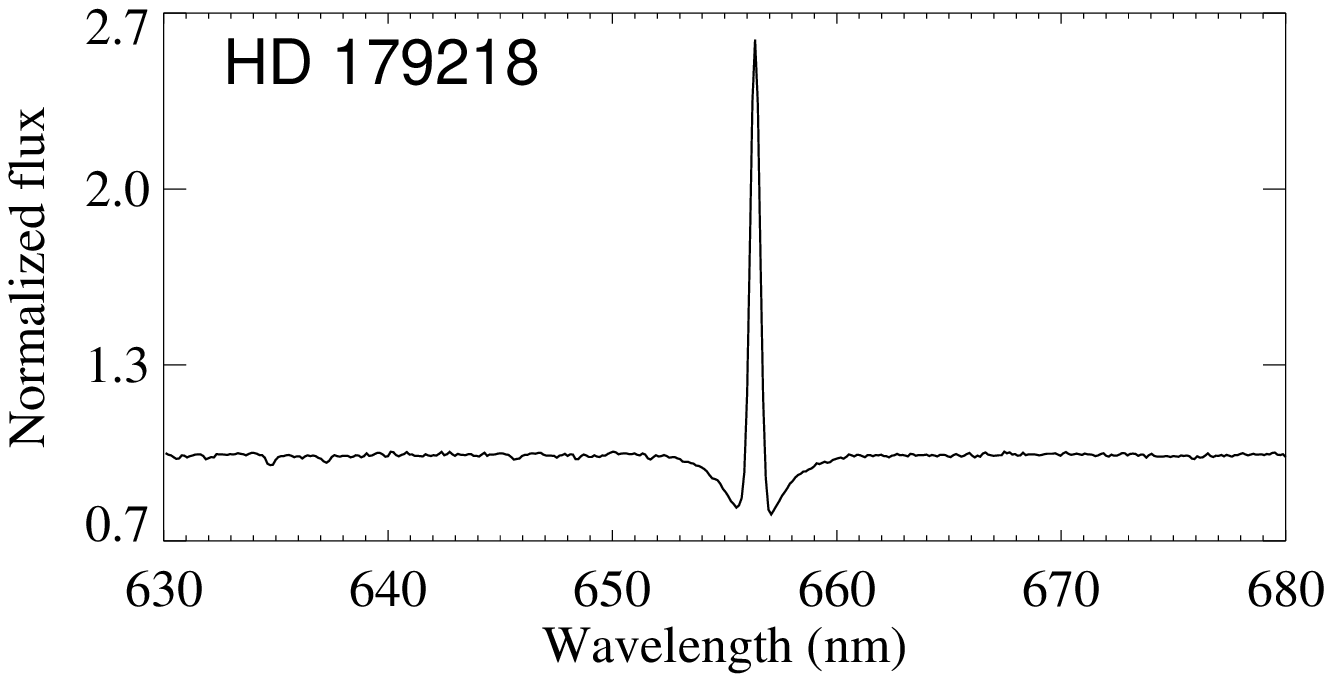}}
\resizebox{0.45\textwidth}{!}{\includegraphics{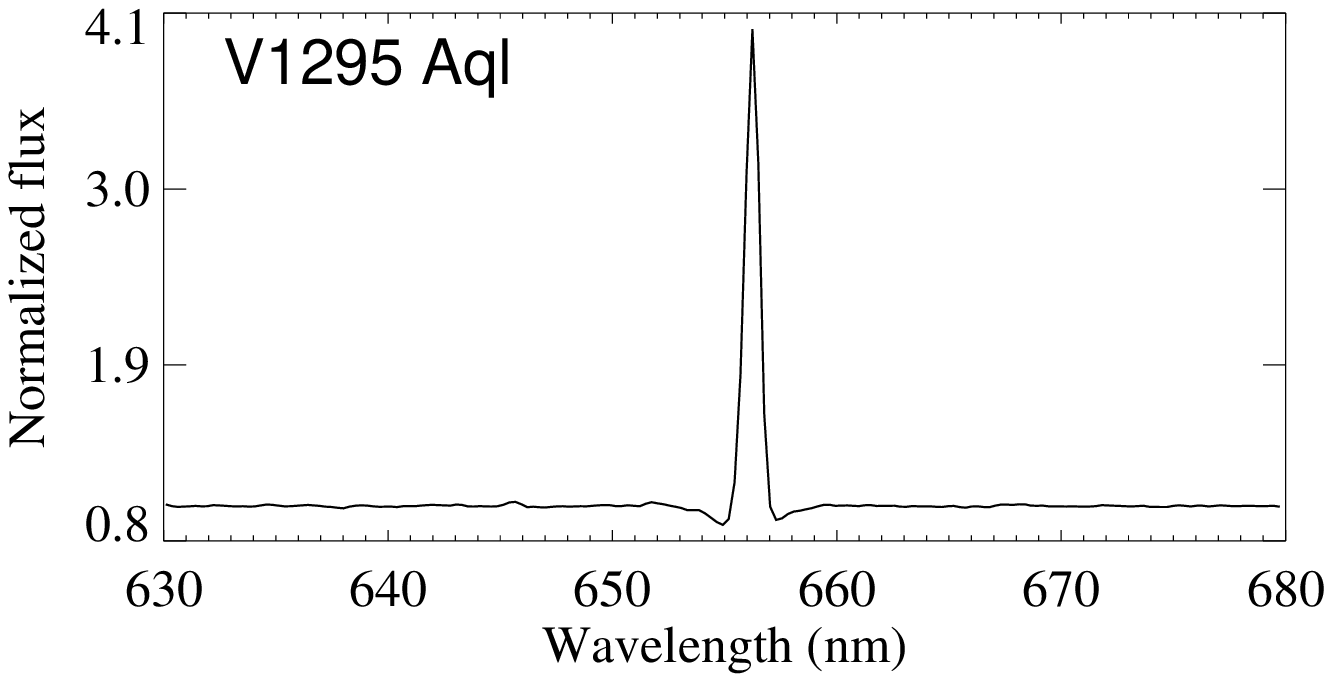}}
\caption{Continued... }
\end{figure}

\begin{figure}
\centering
\figurenum{1}
\resizebox{0.45\textwidth}{!}{\includegraphics{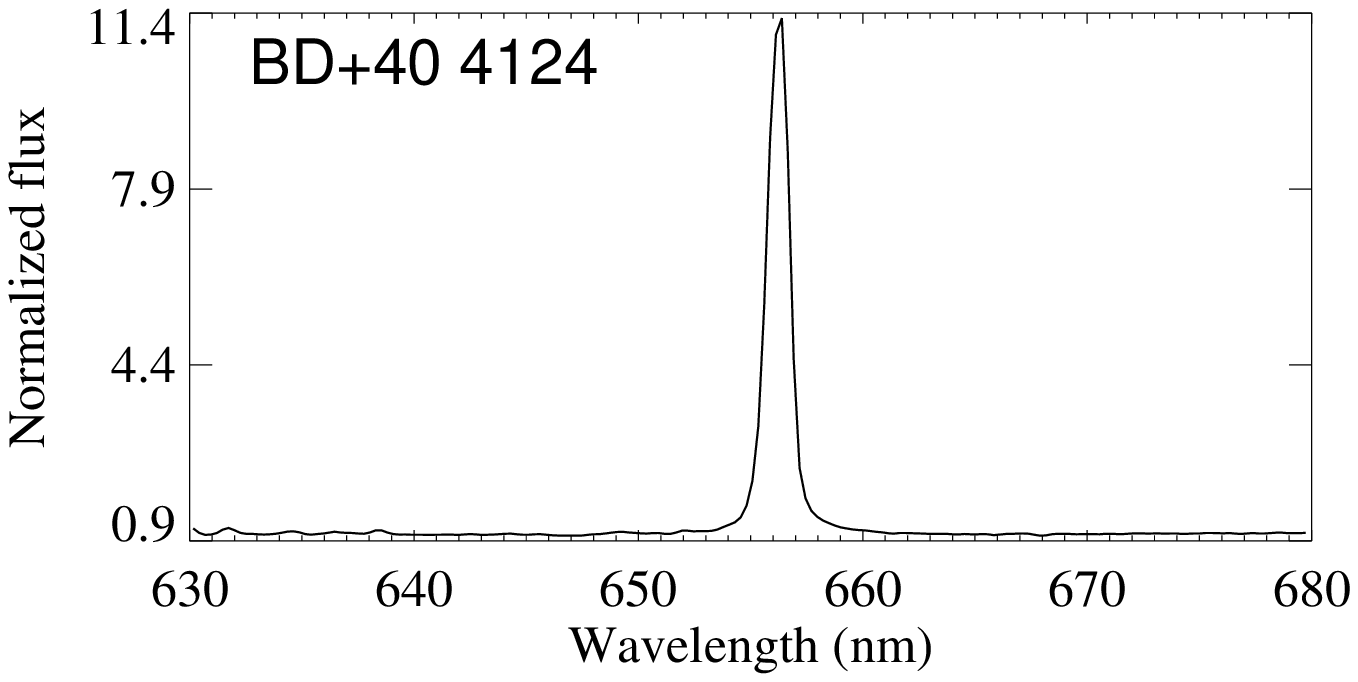}}
\resizebox{0.45\textwidth}{!}{\includegraphics{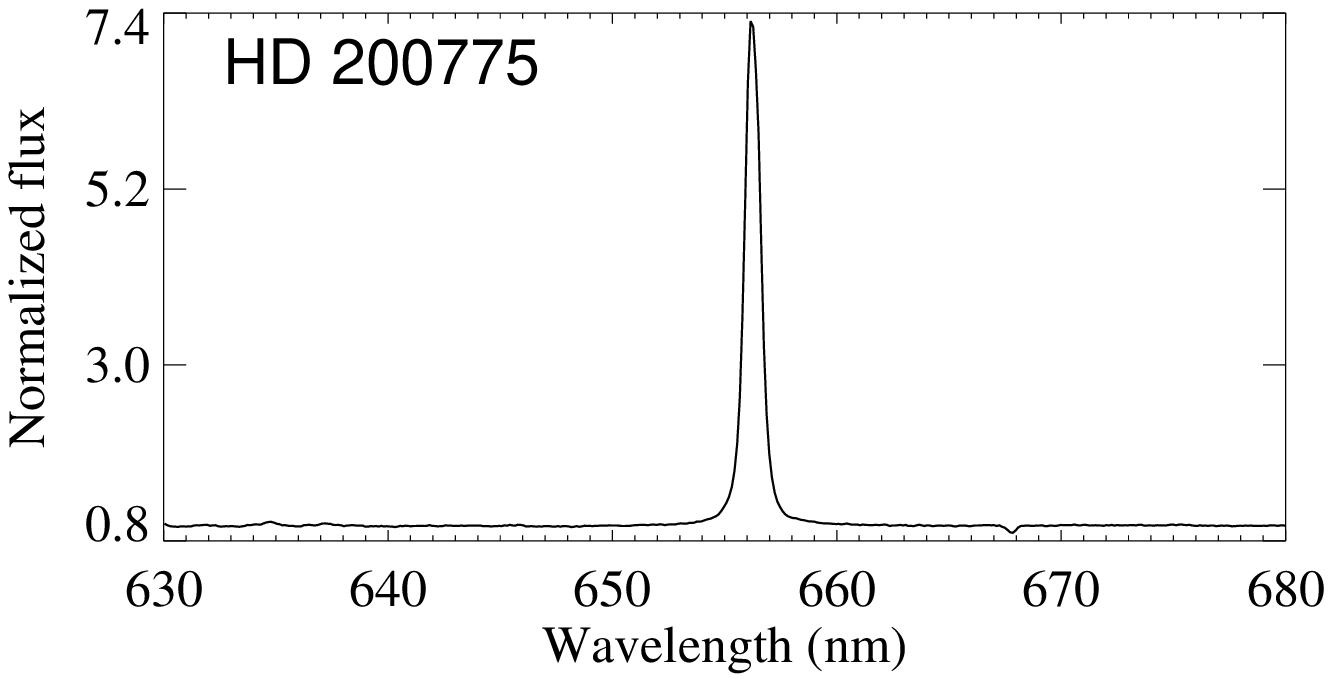}}
\resizebox{0.45\textwidth}{!}{\includegraphics{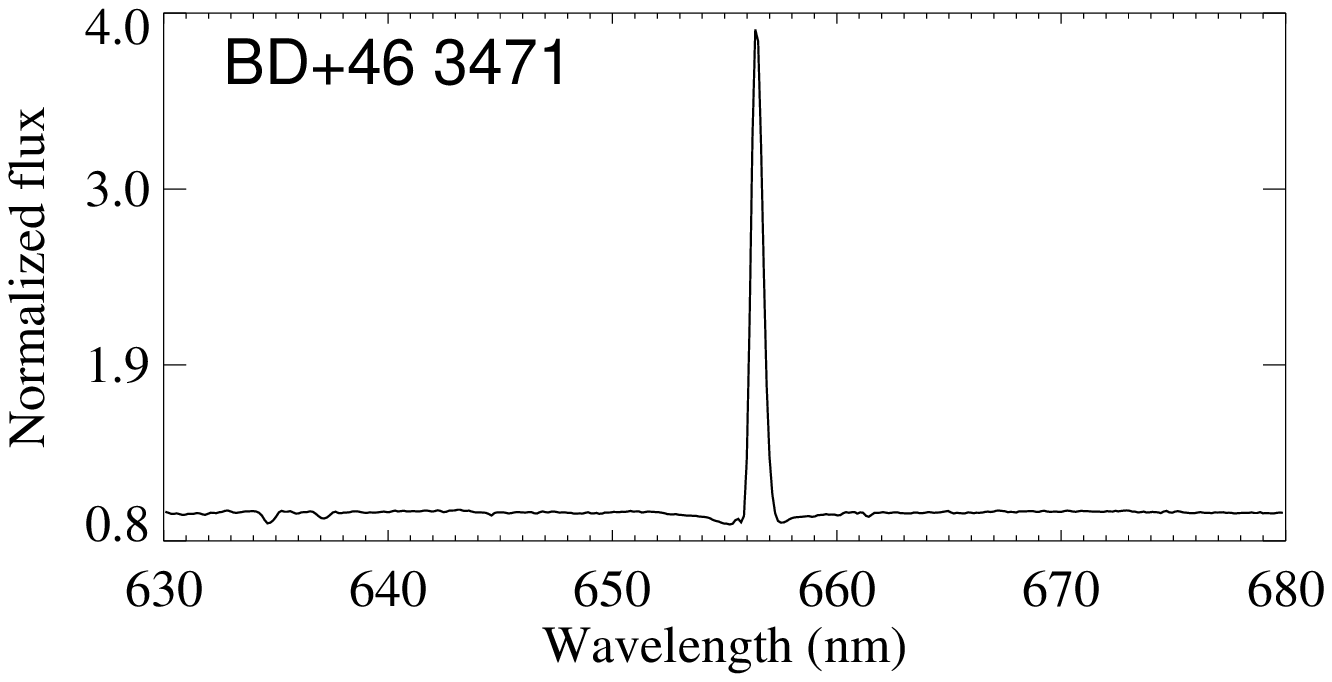}}
\resizebox{0.45\textwidth}{!}{\includegraphics{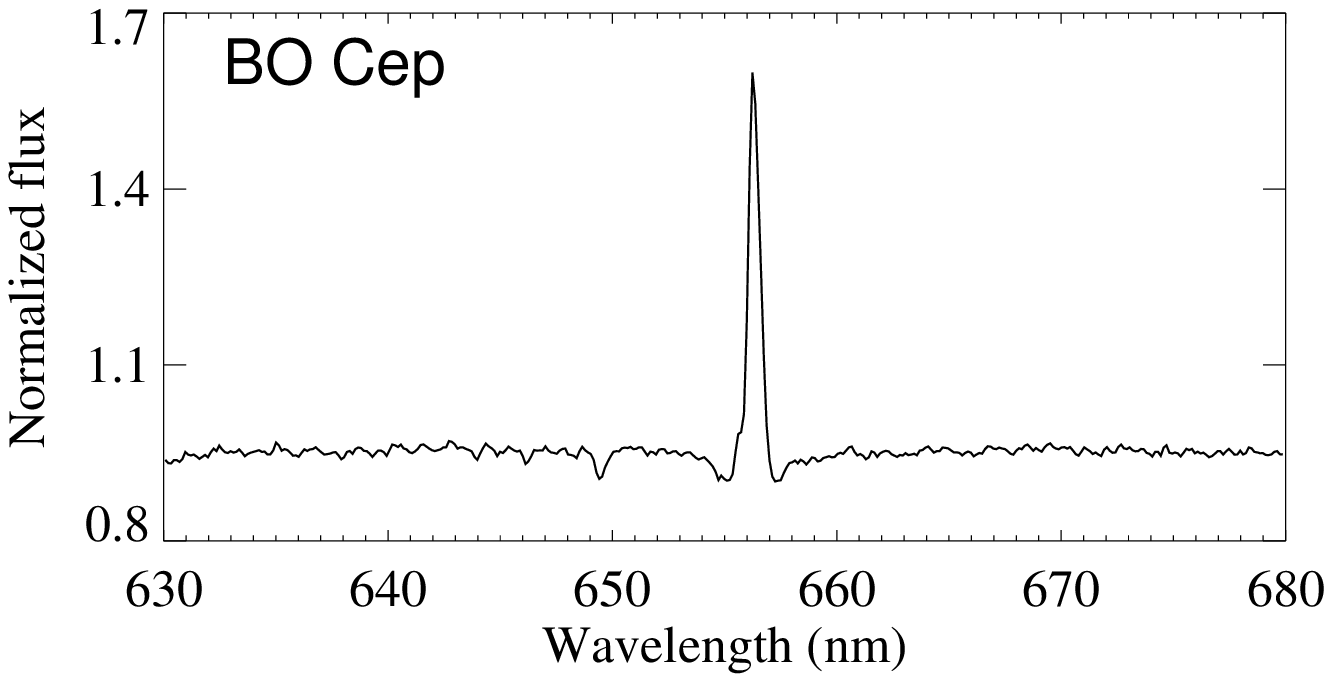}}
\resizebox{0.45\textwidth}{!}{\includegraphics{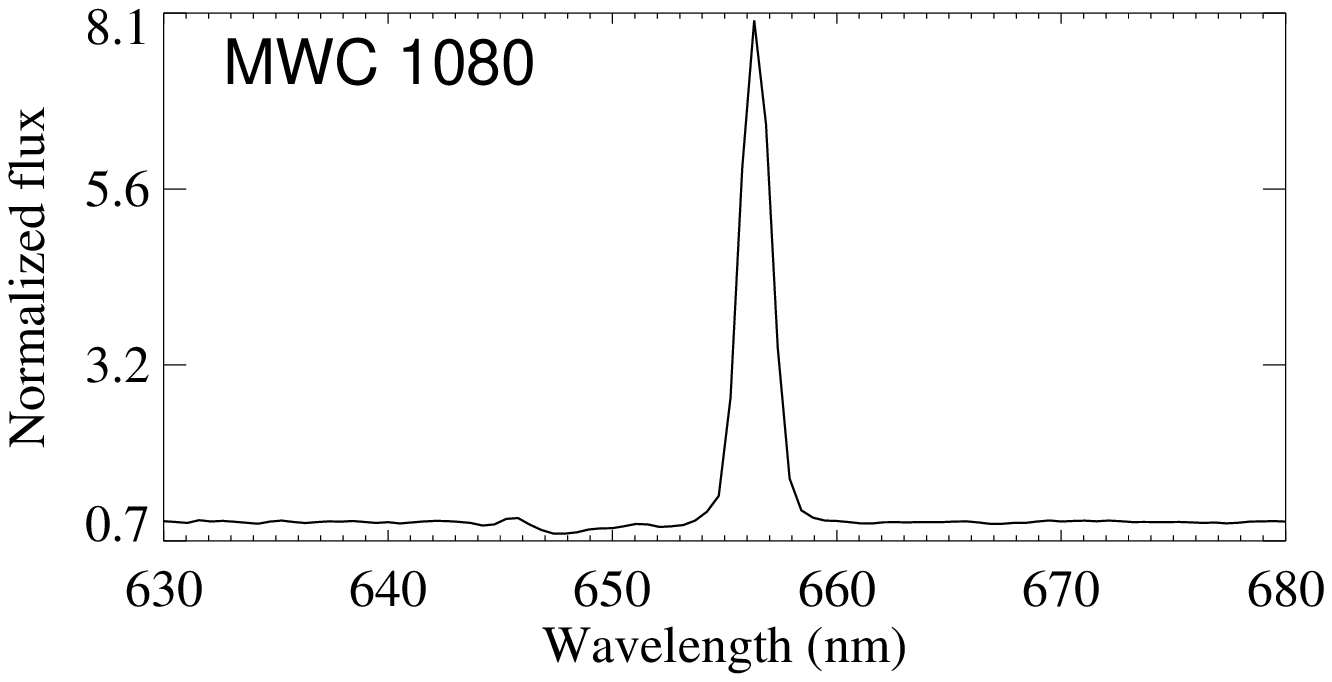}}
\caption{Continued... }
\end{figure}

\clearpage
\begin{figure}
\centering
\includegraphics[scale=0.8]{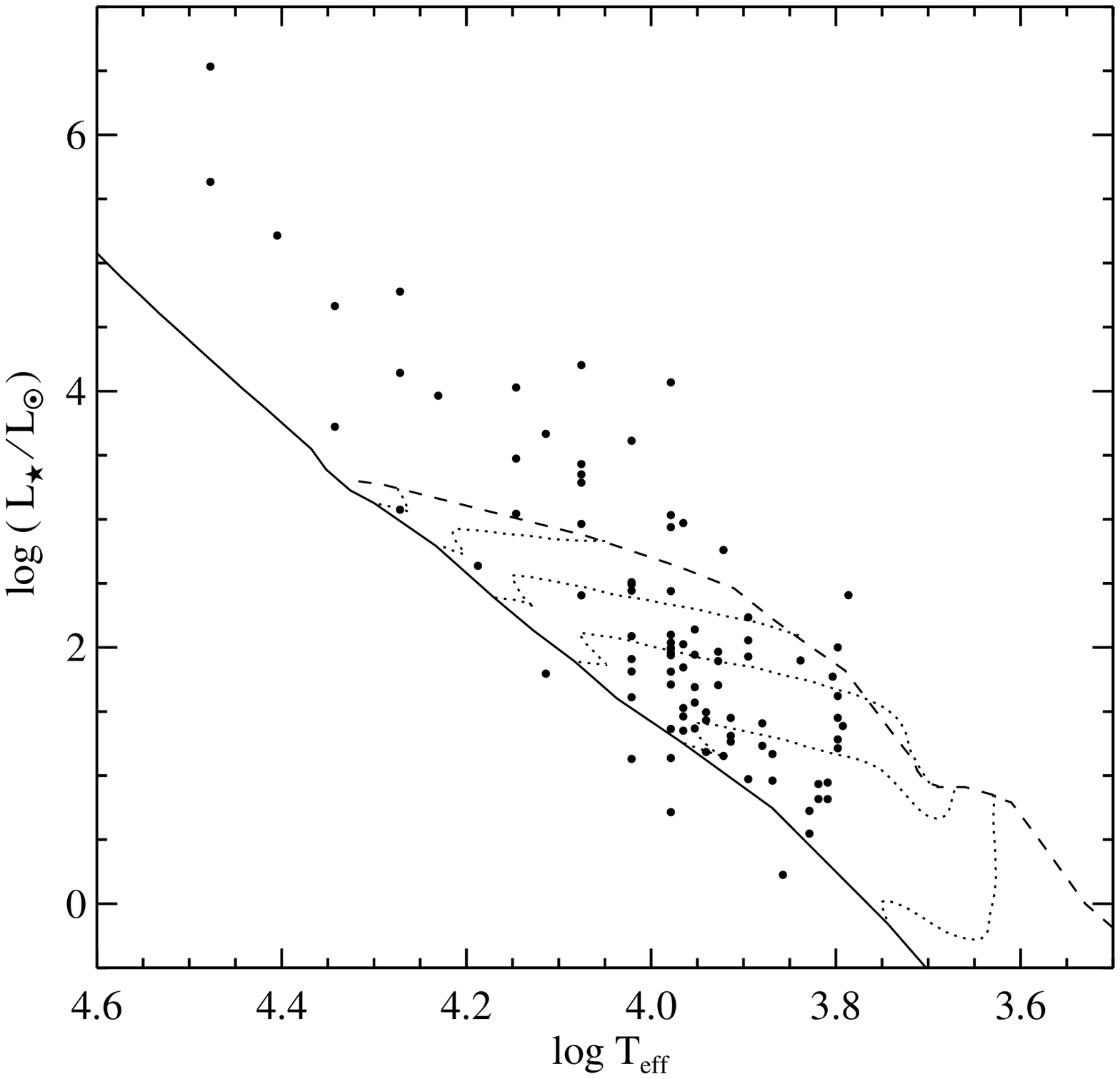}
\caption{ Locations of Herbig Ae/Be stars on the HR diagram. 
 Also plotted are the PMS evolutionary tracks (dotted lines) of
\citet{pallastahler93} . The tracks represent from bottom to top 1,  2,
3, 4, 5 \& 6 M$_{\odot}$. The stellar birth line is also plotted
(dashed line)}
\label{hr}
\end{figure}

\begin{figure}
\resizebox{0.9\textwidth}{!}{\includegraphics{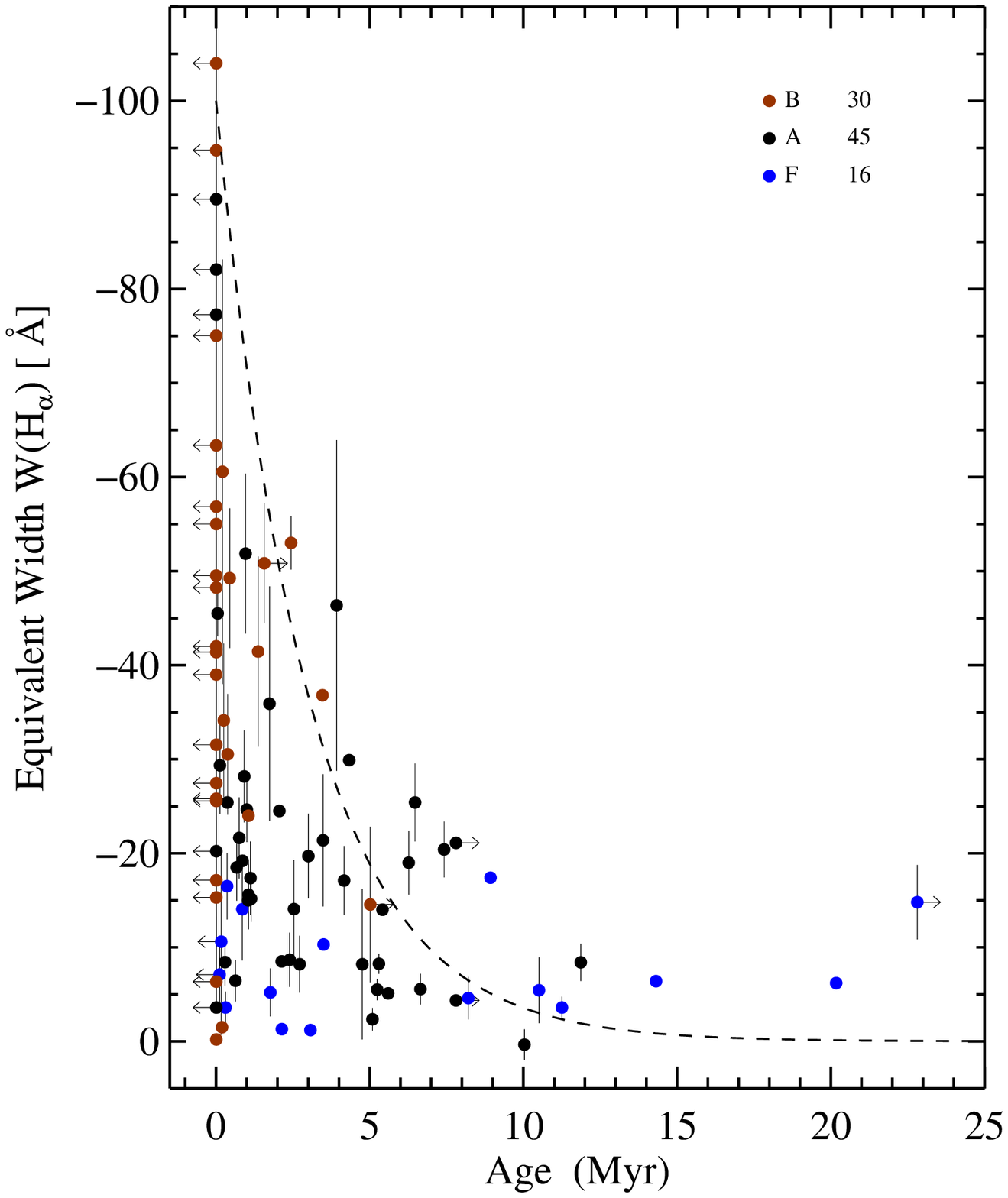}}
\caption{H$\alpha$ equivalent widths of Herbig Ae/Be stars plotted against derived 
stellar ages. Solid brown circles represent B type stars, black
circles represent A type stars and blue circles represent F type
stars.  Our sample consists of  30 B type, 45 A type and 16 F type
stars. Error bars plotted for W(H${\alpha}$) are the dispersion in the
equivalent width measurements given in Table \ref{ew}.  The dashed
line is of the functional form $W(age)=W(0)e^{-age/\tau}$ with
$W(0)=-100$ and $\tau=3Myr$.}
\label{ewage}
\end{figure}

\begin{figure}
\resizebox{\textwidth}{!}{\includegraphics{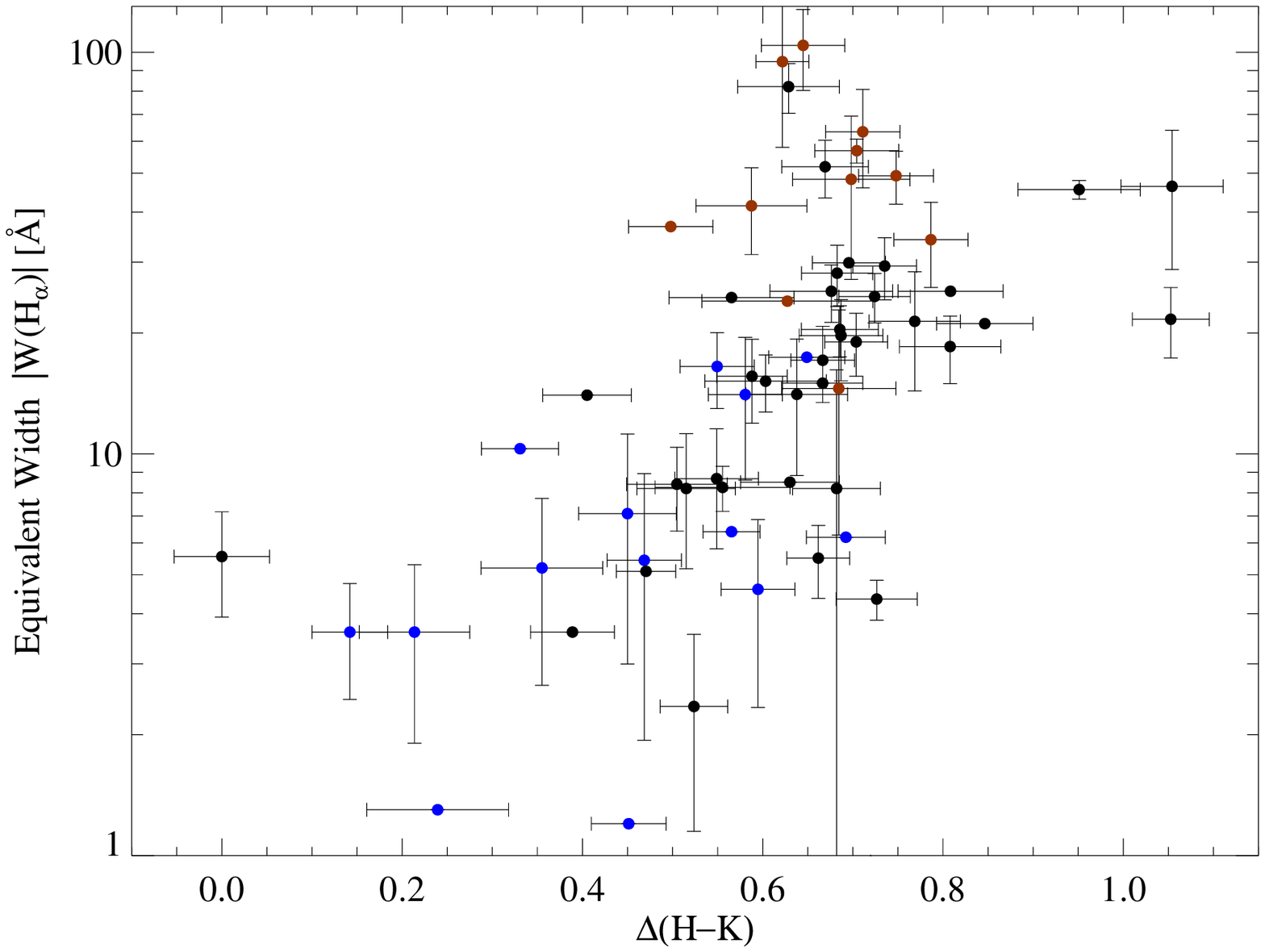}}
\caption{Absolute values of $H\alpha$ equivalent width plotted against 
color excess due to the disk $\Delta$(H-K). Error bars
plotted for W(H${\alpha}$) are the dispersion in the equivalent width
measurements given in Table 2. Errors in $\Delta$(H-K) are
computed from the observed errors in H and K$_s$ magnitudes, errors in
the estimated extinction and assuming an uncertainity of two spectral
sub-classes. Solid brown circles represent B type stars, black circles
represent A type stars and blue circles represent F type stars. }
\label{ewhk}
\end{figure}

%%%%%%%%%%%%%%%%%%%%%%%%%%%%%%%%%%%%%%%%
%%%%%%%%%Tables%%%%%%%%%%%%%%%%%%%%%%%%
%%%%%%%%%%%%%%%%%%%%%%%%%%%%%%%%%%%%%

\clearpage
\begin{deluxetable}{ccccccccc}
\tablewidth{0pt}
\tablecolumns{9}
\tablecaption{ Basic data for 91 Herbig Ae/Be stars compiled from literature. All the magnitudes and colors listed are in the Johnson system \label{data}}
\tablehead{
\colhead{Object Name} & \colhead{V} & \colhead{B-V} & \colhead{R} & \colhead{ref} &
\colhead{Sp. Type} & \colhead{ref} & \colhead{distance} & \colhead{ref}\\
\colhead{} & \colhead{mag} & \colhead{mag} & \colhead{mag} & \colhead{} &
\colhead{} & \colhead{} & \colhead{pc} & \colhead{}
}
\startdata
 LkHa 198 &  14.18 &   0.95 &  13.00 &  1 &       B9 &  4 &    600 & 11\\
    V376 Cas &  15.55 &   1.13 &  14.25 &  1 &      B5e & 19 &    630 & 19\\
      VX Cas &  11.28 &   0.32 &  10.92 &  1 &       A0 &  4 &    760 & 11\\
   BD+61 154 &  10.58 &   0.56 &   9.82 &  1 &       B8 & 31 &    650 &  7\\
     HBC 334 &  14.52 &   0.57 &  13.74 &  3 &       B3 &  4 &   1600 & 11\\
      IP Per &  10.47 &   0.33 &  10.14 & 14 &       A6 &  4 &    350 & 31\\
      XY Per &   9.21 &   0.49 &   8.72 &  1 &       A5 &  4 &    160 &  8\\
    V892 Tau &  15.17 &   1.48 &  13.06 &  1 &       B8 &  4 &    160 & 11\\
      AB Aur &   7.05 &   0.12 &   6.85 &  1 &       A1 &  4 &    144 &  5\\
    HD 31648 &   7.70 &   0.20 &   7.53 & 37 &     A3Ve & 19 &    131 &  5\\
      UX Ori &  10.40 &   0.33 &  10.02 &  1 &       A3 &  4 &    340 &  6\\
    HD 34282 &   9.89 &   0.16 &   9.74 & 37 &       A0 & 25 &    164 &  5\\
    HD 35187 &   8.17 &   0.22 &  99.99 &  2 &   A2e+A7 &  9 &    150 &  9\\
      CO Ori &  10.67 &   1.10 &   9.68 &  1 &     F7Ve & 32 &    450 & 15\\
    HD 35929 &   8.13 &   0.42 &   7.76 & 37 &    F2III & 16 &    345 & 16\\
    HD 36112 &   8.34 &   0.26 &   8.07 & 39 &    A5IVe &  6 &    205 &  5\\
      HK Ori &  11.71 &   0.56 &  11.00 &  1 &   A4+G1V & 36 &    460 & 35\\
   HD 244604 &   9.38 &   0.19 &   9.07 & 37 &   A0Vesh & 42 &    336 & 27\\
      RY Ori &  11.89 &   0.89 &  11.00 & 37 &       F7 &  4 &    460 &  4\\
     HBC 442 &  10.13 &   0.60 &   9.63 &  3 &       F8 &  4 &    460 &  4\\
   HD 245185 &   9.94 &   0.10 &   9.84 & 37 &       A1 &  4 &    400 & 11\\
       T Ori &  10.64 &   0.56 &  10.09 &  1 &       A0 &  4 &    460 &  4\\
    HD 36910 &  10.26 &   0.79 &   9.51 &  1 &       F3 &  4 &    100 &  5\\
    V380 Ori &  10.49 &   0.54 &   9.68 &  1 &      A1e &  6 &    510 &  6\\
    V586 Ori &   9.75 &   0.13 &   9.55 &  1 &      A0V & 27 &    510 & 27\\
      BF Ori &  10.69 &   0.35 &  10.31 & 37 &    A2IVe & 32 &    430 & 17\\
   HD 245906 &  10.58 &   0.39 &  99.99 &  2 &       A6 &  4 &   2000 & 18\\
      RR Tau &  12.08 &   0.75 &  11.32 &  1 &       A0 &  4 &    800 & 11\\
    V350 Ori &  11.47 &   0.48 &  11.02 & 37 &       A1 &  4 &    460 &  4\\
    HD 37806 &   7.95 &   0.04 &   7.84 & 37 &    A2Vpe &  6 &    470 &  6\\
    HD 38120 &   9.01 &   0.06 &   8.88 & 25 &       B9 & 25 &    460 & 10\\
   HD 250550 &   9.54 &   0.07 &   9.34 &  1 &       B9 &  4 &    700 & 11\\
    LkHa 208 &  11.65 &   0.44 &  11.16 &  1 &       A7 &  4 &   1000 & 11\\
    Lkha 338 &  15.12 &   0.94 &  99.99 &  4 &       B9 &  4 &    830 & 20\\
    Lkha 339 &  13.65 &   0.84 &  12.77 &  1 &       A1 &  4 &    830 & 20\\
      VY Mon &  13.47 &   1.55 &  11.74 &  1 &       B8 & 11 &    800 & 11\\
    LkHa 215 &  10.54 &   0.54 &   9.87 &  1 &       B6 &  4 &    800 & 22\\
   HD 259431 &   8.73 &   0.27 &   8.21 &  1 &       B6 &  4 &    800 & 11\\
       R Mon &  12.14 &   0.72 &  11.13 &  1 &   B8IIIe & 32 &    800 & 11\\
     HBC 217 &  11.95 &   0.56 &  11.50 &  3 &       F7 &  4 &    910 &  4\\
     LkHa 25 &  12.77 &   0.20 &  12.44 &  1 &       B7 &  4 &    800 & 11\\
     HBC 222 &  11.97 &   0.60 &  11.53 &  3 &       F7 &  4 &    910 &  4\\
    LkHa 218 &  11.87 &   0.43 &  11.46 &  1 &       A0 &  4 &   1050 & 23\\
       Z CMa &   9.47 &   1.27 &   8.32 &  1 &   B0IIIe & 33 &   1050 & 23\\
     HBC 551 &  11.81 &   0.26 &  11.42 &  1 &       B8 &  4 &   1050 & 23\\
      NX Pup &   9.98 &   0.47 &   9.37 & 37 &       A7 & 24 &    450 & 24\\
    HD 68695 &   9.82 &   0.10 &   9.71 & 25 &      A0V & 25 &    600 & 25\\
    HD 85567 &   8.54 &   0.14 &   8.22 & 37 &       B2 & 26 &   1500 & 26\\
    HD 95881 &   8.23 &   0.14 &  99.99 &  2 &       A0 & 25 &    118 &  6\\
    HD 97048 &   8.46 &   0.36 &   8.08 &  1 &   B9.5Ve &  6 &    175 &  5\\
    HD 98922 &   6.76 &   0.05 &   6.65 & 37 &       B9 & 25 &   1000 & 25\\
   HD 100453 &   7.78 &   0.29 &   7.50 & 25 &     A9Ve &  6 &    112 &  5\\
   HD 100546 &   6.73 &   0.03 &   6.64 & 37 &    B9Vne &  6 &    103 &  5\\
   HD 101412 &   9.25 &   0.18 &   9.06 & 37 &    B9.5V &  6 &    160 & 27\\
   HD 104237 &   6.55 &   0.24 &   6.30 & 37 &    A4IVe &  6 &    116 &  5\\
   HD 135344 &   8.63 &   0.51 &   8.16 & 43 &     F4Ve &  6 &    140 &  6\\
   HD 139614 &   8.40 &   0.24 &   8.18 & 25 &     A7Ve &  6 &    140 &  6\\
   HD 141569 &   7.10 &   0.10 &   6.98 & 37 &     A0Ve & 32 &     99 &  5\\
   HD 142666 &   8.67 &   0.50 &   8.20 & 25 &     A8Ve & 32 &    145 &  6\\
   HD 142527 &   8.34 &   0.71 &  99.99 &  2 &   F7IIIe &  6 &    198 &  5\\
   HD 144432 &   8.17 &   0.36 &   7.81 & 37 &    A9IVe & 32 &    145 &  6\\
     HR 5999 &   6.85 &   0.33 &   6.54 & 37 &       A7 & 25 &    208 &  5\\
   HD 150193 &   8.64 &   0.49 &   8.14 &  1 &    A2IVe & 32 &    150 &  5\\
      AK Sco &   9.00 &   0.63 &   8.48 &  1 &    F5IVe & 34 &    145 &  5\\
      KK Oph &  11.45 &   0.71 &  10.51 &  1 &     A8Ve & 32 &    160 & 28\\
      51 Oph &   4.78 &   0.03 &   4.72 & 37 & B9.5IIIe & 32 &    131 &  5\\
   HD 163296 &   6.88 &   0.09 &   6.77 &  1 &    A1Vep & 32 &    122 &  5\\
   HD 169142 &   8.15 &   0.28 &   7.86 & 25 &     A5Ve &  6 &    145 &  6\\
      VV Ser &  11.92 &   0.93 &  10.82 &  1 &       B6 &  4 &    330 & 27\\
      AS 310 &  12.45 &   1.06 &  11.30 &  1 &       B1 &  4 &   2500 & 11\\
       R CrA &  12.20 &   1.09 &  10.62 &  1 &       A0 & 25 &    130 & 27\\
       T CrA &  12.84 &   1.18 &  11.84 &  1 &       F0 & 17 &    130 & 27\\
   HD 179218 &   7.39 &   0.08 &   7.29 & 25 &    A0IVe & 32 &    244 &  5\\
   HD 344361 &  10.74 &   0.44 &  10.33 &  1 &       A3 &  4 &    440 & 27\\
      PX Vul &  11.54 &   0.83 &  10.96 & 41 &       F3 &  4 &    420 & 22\\
   V1295 Aql &   7.73 &   0.13 &   7.66 & 41 &    A2IVe & 32 &    290 & 29\\
  BD+40 4124 &  10.69 &   0.78 &   9.64 &  1 &       B3 &  4 &    980 &  4\\
   V1686 Cyg &  14.07 &   1.44 &  12.59 &  1 &       F9 &  4 &    980 & 27\\
    LkHa 134 &  11.35 &   0.66 &  10.57 &  1 &       B2 & 24 &    700 & 24\\
    LkHA 168 &  13.48 &   1.26 &  12.28 &  1 &    F6III & 44 &    700 & 44\\
   HD 200775 &   7.37 &   0.41 &   6.86 &  1 &       B3 &  4 &    429 &  5\\
    Lkha 324 &  12.65 &   1.12 &  11.58 &  1 &       B8 &  4 &    780 &  4\\
    V645 Cyg &  13.47 &   1.10 &  12.29 &  1 &       A0 & 24 &   3500 & 27\\
    V361 Cep &  10.18 &   0.39 &   9.64 &  1 &       B4 &  4 &   1250 & 30\\
    LkHa 234 &  12.21 &   0.90 &  11.22 &  1 &       B7 &  4 &   1250 & 30\\
  BD+46 3471 &  10.13 &   0.40 &   9.67 &  1 &       A0 &  4 &   1200 & 27\\
      BH Cep &  11.16 &   0.65 &  10.63 &  1 &       F5 &  4 &    450 &  4\\
      BO Cep &  11.60 &   0.56 &  11.06 &  1 &       F4 &  4 &    400 &  4\\
      SV Cep &  10.98 &   0.39 &  10.56 &  1 &       A0 &  4 &    440 & 27\\
    LkHa 233 &  13.56 &   0.84 &  12.67 &  1 &       A4 &  4 &    880 & 11\\
    MWC 1080 &  11.52 &   1.34 &   9.99 &  1 &     B0eq & 31 &   2200 & 27\\

\enddata
\tablerefs{\small{
(1) \citet{hs99}; (2) \citet{tych00}; (3) \citet{herbigbell88}; (4) \citet {hernand04}; (5) \citet{esa97}; (6) \citet{boekel05};  (7) \citet{van98}; (8) \citet{berrilli92}; (9) \citet{dunkin98}; (10) \citet{hernand05}; (11) \citet{testi98}; (12) \citet{yonekura97}; (13) \citet{preib03}; (14) \citet{mirosh01}; (15) \citet{calvet04}; (16) \citet{mirosh04}; (17) \citet{hamaguchi05}; (18) \citet{kawa98}; (19) \citet{acke04}; (20) \citet{hr76} ; (21) \citet{kut80}; (22) \citet{herbst82}; (23) \citet{shev99}; (24) \citet{corray98} (25) \citet{vieira03}; (26) \citet{mirosh01b}; (27) \citet{acke05}; (28) \citet{leinert04}; (29) \citet{eisner04}; (30) \citet{shev89}; (31) \citet{hill92}; (32) \citet{mora01}; (33) \citet{van04}; (34) \citet{alencar03}; (35) \citet{leinert97};(36) \citet{baines04}; (37) \citet{dewinter01}; (38) \citet{lahulla85}; (39) \citet{beskro99}; (40) \citet{fuji02}; (41) \citet{oped01}; (42) \citet{malf98}; (43) \citet{coul95}; (44) \citet{terra94}. } }

\end{deluxetable}

\clearpage

\begin{deluxetable}{ccccccc}
\tabletypesize{\small}
\tablewidth{0pt}
\tablecolumns{7}
\tablecaption{ Log of spectroscopic observations. \label{log}}
\tablehead{
\colhead{Object} & \colhead{RA (J2000)} & \colhead{Dec (J2000)} & 
\colhead{Date} & \colhead{Spectrograph/} & 
\colhead{UT start} & \colhead{Exp. time}\\ 
\colhead{} & \colhead{hh mm ss.s} & \colhead{dd mm ss} & 
\colhead{(yy-mm-dd)} & \colhead{Telescope} & 
\colhead{(hh:mm)} & \colhead{(s)} 
 }
\startdata
 BD+61 154 &  00 43 18.2 &  +61 54 40 &   2003-10-22 &   HFOSC/HCT &   18:20 &    300\\
      IP Per &  03 40 46.9 &  +32 31 53 &   2003-10-22 &   HFOSC/HCT &   19:12 &    300\\
      XY Per &  03 49 36.3 &  +38 58 55 &   2003-01-27 &     OMR/VBT &   13:42 &    600\\
      AB Aur &  04 55 45.8 &  +30 33 04 &   2003-10-24 &   HFOSC/HCT &   18:38 &     30\\
    HD 31648 &  04 58 46.2 &  +29 50 36 &   2002-11-29 &     OMR/VBT &   18:30 &    300\\
      UX Ori &  05 04 29.9 &  -03 47 14 &   2001-03-14 &     OMR/VBT &   14:01 &    900\\
    HD 34282 &  05 16 00.4 &  -09 48 35 &   2002-11-29 &     OMR/VBT &   20:48 &    900\\
    HD 35187 &  05 24 01.1 &  +24 57 37 &   2004-01-14 &   HFOSC/HCT &   15:11 &    600\\
    HD 35929 &  05 27 42.7 &  -08 19 38 &   2003-10-22 &   HFOSC/HCT &   20:40 &    210\\
    HD 36112 &  05 30 27.5 &  +25 19 57 &   2003-01-28 &     OMR/VBT &   14:16 &    300\\
      HK Ori &  05 31 28.0 &  +12 09 10 &   2005-02-07 &     OMR/VBT &   16:23 &   1800\\
   HD 245185 &  05 35 09.6 &  +10 01 51 &   2005-02-05 &     OMR/VBT &   17:30 &   1800\\
    HD 37806 &  05 41 02.2 &  -02 43 00 &   2003-01-28 &     OMR/VBT &   15:41 &    600\\
    HD 38120 &  05 43 11.8 &  -04 59 49 &   2003-10-22 &   HFOSC/HCT &   21:25 &    390\\
   HD 250550 &  06 01 59.9 &  +16 30 56 &   2005-02-07 &     OMR/VBT &   19:36 &   1200\\
   HD 259431 &  06 33 05.1 &  +10 19 19 &   2003-10-24 &   HFOSC/HCT &   22:58 &    150\\
    LkHa 218 &  07 02 42.5 &  -11 26 11 &   2005-02-07 &     OMR/VBT &   20:18 &   1800\\
       Z CMa &  07 03 43.1 &  -11 33 06 &   2003-01-27 &     OMR/VBT &   18:15 &    420\\
      NX Pup &  07 19 28.2 &  -44 35 11 &   2000-01-20 &     OMR/VBT &   18:00 &    900\\
    HD 68695 &  08 11 44.5 &  -44 05 08 &   2005-03-15 &     OMR/VBT &   13:54 &   2400\\
    HD 85567 &  09 50 28.5 &  -60 58 02 &   2003-01-25 &     OMR/VBT &   19:50 &    900\\
    HD 98922 &  11 22 31.6 &  -53 22 11 &   2002-04-09 &     OMR/VBT &   16:56 &    600\\
   HD 100453 &  11 33 05.5 &  -54 19 28 &   2003-01-28 &     OMR/VBT &   20:34 &    900\\
   HD 101412 &  11 39 44.4 &  -60 10 27 &   2005-03-15 &     OMR/VBT &   17:38 &   2400\\
   HD 139614 &  15 40 46.3 &  -42 29 53 &   2005-03-15 &     OMR/VBT &   18:54 &   2400\\
   HD 141569 &  15 49 57.7 &  -03 55 16 &   2002-05-19 &     OMR/VBT &   18:19 &    300\\
   HD 142666 &  15 56 40.0 &  -22 01 40 &   2005-03-15 &     OMR/VBT &   20:22 &    900\\
   HD 142527 &  15 56 41.8 &  -42 19 23 &   2002-04-09 &     OMR/VBT &   19:45 &    900\\
   HD 144432 &  16 06 57.9 &  -27 43 09 &   2002-05-19 &     OMR/VBT &   18:36 &    600\\
     HR 5999 &  16 08 34.2 &  -39 06 18 &   2002-05-19 &     OMR/VBT &   18:59 &    600\\
   HD 150193 &  16 40 17.9 &  -23 53 45 &   2002-04-11 &     OMR/VBT &   21:01 &    600\\
      AK Sco &  16 54 44.8 &  -36 53 18 &   2002-04-09 &     OMR/VBT &   20:21 &   1800\\
      KK Oph &  17 10 08.0 &  -27 15 18 &   2003-05-30 &     OMR/VBT &   18:46 &   1800\\
      51 Oph &  17 31 24.9 &  -23 57 45 &   2002-05-19 &     OMR/VBT &   19:22 &    180\\
   HD 163296 &  17 56 21.2 &  -21 57 21 &   2002-05-19 &     OMR/VBT &   20:41 &    600\\
      VV Ser &  18 28 47.8 &  +00 08 40 &   2003-05-30 &     OMR/VBT &   20:06 &   1800\\
       R CrA &  19 01 53.6 &  -36 57 07 &   2006-05-06 &     OMR/VBT &   21:05 &   1800\\
       T CrA &  19 01 58.7 &  -36 57 49 &   2006-05-06 &     OMR/VBT &   20:23 &   1800\\
   HD 179218 &  19 11 11.2 &  +15 47 15 &   2003-10-24 &   HFOSC/HCT &   15:13 &    300\\
   V1295 Aql &  20 03 02.5 &  +05 44 16 &   2002-05-19 &     OMR/VBT &   22:36 &    600\\
  BD+40 4124 &  20 20 28.2 &  +41 21 51 &   2002-05-19 &     OMR/VBT &   23:01 &    600\\
   HD 200775 &  21 01 36.9 &  +68 09 47 &   2003-10-22 &   HFOSC/HCT &   15:18 &     30\\
  BD+46 3471 &  21 52 34.0 &  +47 13 43 &   2003-10-22 &   HFOSC/HCT &   15:57 &    210\\
      BO Cep &  22 16 54.0 &  +70 03 45 &   2003-10-22 &   HFOSC/HCT &   16:32 &    450\\
    MWC 1080 &  23 17 25.5 &  +60 50 43 &   2004-10-24 &     OMR/VBT &   19:20 &    600\\

\enddata

\end{deluxetable}

\clearpage
\begin{deluxetable}{cccccccc}
\tablewidth{0pt}
\tablecolumns{8}
\tablecaption{Masses and Ages of  Herbig Ae/Be stars. \label{age}}
\tablehead{
\colhead{Object Name} & \colhead{log T$_{eff}$} & \colhead{E(B-V)} & \colhead{R$_V$} &
\colhead{A$_V$} & \colhead{log L$_{\star}$} & \colhead{Mass} & \colhead{Age} \\
\colhead{} & \colhead{K} & \colhead{mag} & \colhead{mag} & \colhead{mag} &
\colhead{(L$_{\odot}$)} & \colhead{(M$_{\odot}$)} & \colhead{Myr} }

\startdata
  LkHa 198 &   4.02 &   1.01 &   6.44 &   6.27 &   2.49 &      4.25 &      0.21 \\
    V376 Cas &   4.19 &   1.27 &   5.68 &   6.95 &   2.64 &      4.62 &      0.38 \\
      VX Cas &   3.98 &   0.32 &   5.62 &   1.73 &   1.96 &      3.00 &      1.12 \\
   BD+61 154 &   4.08 &   0.65 &   6.69 &   4.19 &   3.29 & $>$  5.11 & $<$  0.01 \\
     HBC 334 &   4.27 &   0.76 &   6.16 &   4.51 &   3.07 &      5.99 &      0.01 \\
      IP Per &   3.92 &   0.17 &   4.67 &   0.76 &   1.15 &      2.00 &      6.27 \\
      XY Per &   3.91 &   0.35 &   4.68 &   1.57 &   1.31 &      2.00 &      5.24 \\
    V892 Tau &   4.08 &   1.57 &   7.90 &  11.99 &   3.35 & $>$  5.11 & $<$  0.01 \\
      AB Aur &   3.97 &   0.09 &  10.29 &   0.90 &   1.84 &      2.77 &      1.74 \\
    HD 31648 &   3.94 &   0.11 &   1.56 &   0.16 &   1.18 &      1.99 &      7.42 \\
      UX Ori &   3.94 &   0.24 &   6.11 &   1.41 &   1.43 &      2.09 &      4.76 \\
    HD 34282 &   3.98 &   0.16 &   3.66 &   0.56 &   0.71 & $<$  2.08 & $>$  7.81 \\
    HD 35187 &   3.95 &   0.16 &   5.00 &   0.77 &   1.37 &      2.00 &      5.60 \\
      CO Ori &   3.80 &   0.60 &   4.35 &   2.51 &   2.00 & $>$  3.58 & $<$  0.12 \\
    HD 35929 &   3.84 &   0.07 &   5.00 &   0.34 &   1.90 &      3.41 &      0.31 \\
    HD 36112 &   3.91 &   0.12 &   4.48 &   0.52 &   1.45 &      2.17 &      4.17 \\
      HK Ori &   3.93 &   0.44 &   7.61 &   3.23 &   1.89 &      3.01 &      0.96 \\
   HD 244604 &   3.98 &   0.19 &   9.21 &   1.69 &   1.99 &      3.05 &      1.05 \\
      RY Ori &   3.80 &   0.39 &   6.13 &   2.30 &   1.45 &      2.49 &      1.77 \\
     HBC 442 &   3.79 &   0.08 &   5.00 &   0.38 &   1.39 &      2.41 &      2.14 \\
   HD 245185 &   3.97 &   0.07 &   5.00 &   0.34 &   1.35 &      2.07 &      6.48 \\
       T Ori &   3.98 &   0.56 &   4.71 &   2.54 &   2.10 &      3.34 &      0.67 \\
    HD 36910 &   3.83 &   0.42 &   4.39 &   1.77 &   0.55 &      1.38 &     20.17 \\
    V380 Ori &   3.97 &   0.51 &   8.93 &   4.41 &   2.97 & $>$  4.93 & $<$  0.01 \\
    V586 Ori &   3.98 &   0.13 &   8.12 &   1.02 &   1.94 &      3.00 &      1.14 \\
      BF Ori &   3.95 &   0.29 &   5.40 &   1.51 &   1.57 &      2.50 &      2.72 \\
   HD 245906 &   3.92 &   0.23 &   5.00 &   1.11 &   2.76 & $>$  4.84 & $<$  0.01 \\
      RR Tau &   3.98 &   0.75 &   5.03 &   3.62 &   2.44 &      4.26 &      0.13 \\
    V350 Ori &   3.97 &   0.45 &   4.26 &   1.84 &   1.46 &      2.22 &      4.33 \\
    HD 37806 &   3.95 &   0.00 &   5.00 &   0.00 &   2.14 &      3.58 &      0.37 \\
    HD 38120 &   4.02 &   0.12 &   5.78 &   0.67 &   2.09 &      3.12 &      1.06 \\
   HD 250550 &   4.02 &   0.13 &   9.31 &   1.17 &   2.44 &      4.13 &      0.25 \\
    LkHa 208 &   3.89 &   0.25 &   6.69 &   1.61 &   1.93 &      3.24 &      0.63 \\
    Lkha 338 &   4.02 &   1.00 &   5.00 &   4.81 &   1.81 &      2.60 &      2.44 \\
    Lkha 339 &   3.97 &   0.81 &   5.32 &   4.15 &   2.02 &      3.18 &      0.86 \\
      VY Mon &   4.08 &   1.64 &   5.65 &   8.92 &   4.20 & $>$  5.11 & $<$  0.01 \\
    LkHa 215 &   4.15 &   0.67 &   5.83 &   3.76 &   3.47 & $>$  5.43 & $<$  0.01 \\
   HD 259431 &   4.15 &   0.40 &   8.62 &   3.34 &   4.03 & $>$  5.43 & $<$  0.01 \\
       R Mon &   4.08 &   0.81 &   7.24 &   5.66 &   3.43 & $>$  5.11 & $<$  0.01 \\
     HBC 217 &   3.80 &   0.06 &   5.00 &   0.29 &   1.21 &      2.13 &      3.50 \\
     LkHa 25 &   4.11 &   0.31 &   6.64 &   1.98 &   1.80 & $<$  3.35 & $>$  1.57 \\
     HBC 222 &   3.80 &   0.10 &   5.00 &   0.48 &   1.28 &      2.23 &      3.07 \\
    LkHa 218 &   3.98 &   0.43 &   4.40 &   1.82 &   2.04 &      3.16 &      0.92 \\
       Z CMa &   4.48 &   1.57 &   3.69 &   5.55 &   5.63 & $>$  6.00 & $<$  0.01 \\
     HBC 551 &   4.08 &   0.35 &   6.47 &   2.18 &   2.41 &      3.92 &      0.45 \\
      NX Pup &   3.89 &   0.28 &   9.01 &   2.44 &   2.24 &      4.07 &      0.05 \\
    HD 68695 &   3.98 &   0.10 &   4.35 &   0.42 &   1.81 &      2.64 &      2.14 \\
    HD 85567 &   4.34 &   0.36 &   6.43 &   2.23 &   4.66 & $>$  6.00 & $<$  0.01 \\
    HD 95881 &   3.98 &   0.14 &   5.00 &   0.67 &   1.14 & $<$  2.08 & $>$  7.81 \\
    HD 97048 &   4.02 &   0.42 &   4.39 &   1.77 &   1.91 &      3.00 &      1.37 \\
    HD 98922 &   4.02 &   0.11 &   5.13 &   0.54 &   3.61 & $>$  4.95 & $<$  0.01 \\
   HD 100453 &   3.87 &   0.02 &   5.00 &   0.10 &   0.96 &      1.66 &     10.03 \\
   HD 100546 &   4.02 &   0.09 &   5.13 &   0.44 &   1.61 &      2.50 &      3.46 \\
   HD 101412 &   4.02 &   0.24 &   3.50 &   0.81 &   1.13 & $<$  2.39 & $>$  5.02 \\
   HD 104237 &   3.93 &   0.12 &   5.13 &   0.59 &   1.71 &      2.58 &      2.06 \\
   HD 135344 &   3.82 &   0.12 &   3.17 &   0.36 &   0.93 &      1.69 &      8.93 \\
   HD 139614 &   3.89 &   0.05 &   5.00 &   0.24 &   0.97 &      1.75 &     11.87 \\
   HD 141569 &   3.98 &   0.10 &   5.13 &   0.49 &   1.36 &      2.18 &      6.65 \\
   HD 142666 &   3.88 &   0.27 &   4.26 &   1.10 &   1.23 &      2.00 &      5.09 \\
   HD 142527 &   3.80 &   0.21 &   5.00 &   1.01 &   1.62 &      2.90 &      0.86 \\
   HD 144432 &   3.87 &   0.09 &   5.13 &   0.44 &   1.17 &      2.00 &      5.30 \\
     HR 5999 &   3.89 &   0.14 &   4.01 &   0.54 &   2.06 &      3.55 &      0.30 \\
   HD 150193 &   3.95 &   0.43 &   4.95 &   2.05 &   1.69 &      2.50 &      2.40 \\
      AK Sco &   3.81 &   0.21 &   1.76 &   0.35 &   0.82 &      1.50 &     10.51 \\
      KK Oph &   3.88 &   0.48 &   8.85 &   4.11 &   1.41 &      2.17 &      3.93 \\
      51 Oph &   4.02 &   0.09 &   2.52 &   0.22 &   2.51 &      4.29 &      0.20 \\
   HD 163296 &   3.97 &   0.06 &   5.13 &   0.30 &   1.53 &      2.49 &      3.00 \\
   HD 169142 &   3.91 &   0.14 &   4.57 &   0.61 &   1.26 &      2.00 &      5.42 \\
      VV Ser &   4.15 &   1.06 &   5.87 &   5.99 &   3.04 & $>$  5.43 & $<$  0.01 \\
      AS 310 &   4.40 &   1.32 &   4.78 &   6.06 &   5.21 & $>$  6.00 & $<$  0.01 \\
       R CrA &   3.98 &   1.09 &   8.48 &   8.94 &   2.94 & $>$  4.94 & $<$  0.01 \\
       T CrA &   3.86 &   0.87 &   3.60 &   3.00 &   0.23 & $<$  1.45 & $>$ 22.81 \\
   HD 179218 &   3.98 &   0.08 &   5.13 &   0.39 &   1.99 &      3.04 &      1.06 \\
   HD 344361 &   3.94 &   0.35 &   4.01 &   1.35 &   1.49 &      2.35 &      3.48 \\
      PX Vul &   3.83 &   0.46 &   0.86 &   0.38 &   0.72 &      1.50 &     14.31 \\
   V1295 Aql &   3.95 &   0.07 &   5.00 &   0.34 &   1.94 &      3.03 &      1.00 \\
  BD+40 4124 &   4.27 &   0.97 &   6.42 &   6.00 &   4.78 & $>$  5.99 & $<$  0.01 \\
   V1686 Cyg &   3.79 &   0.89 &   6.10 &   5.22 &   2.41 & $>$  3.46 & $<$  0.17 \\
    LkHa 134 &   4.34 &   0.88 &   5.13 &   4.34 &   3.72 & $>$  6.00 & $<$  0.01 \\
    LkHA 168 &   3.80 &   0.80 &   4.94 &   3.79 &   1.77 &      3.30 &      0.36 \\
   HD 200775 &   4.27 &   0.60 &   5.00 &   2.88 &   4.14 & $>$  5.99 & $<$  0.01 \\
    Lkha 324 &   4.08 &   1.21 &   4.36 &   5.06 &   2.96 & $>$  5.11 & $<$  0.01 \\
    V645 Cyg &   3.98 &   1.10 &   5.56 &   5.88 &   4.07 & $>$  4.94 & $<$  0.01 \\
    V361 Cep &   4.23 &   0.55 &   5.98 &   3.17 &   3.96 & $>$  5.81 & $<$  0.01 \\
    LkHa 234 &   4.11 &   1.01 &   5.29 &   5.13 &   3.67 & $>$  5.29 & $<$  0.01 \\
  BD+46 3471 &   3.98 &   0.40 &   5.91 &   2.28 &   3.03 & $>$  4.94 & $<$  0.01 \\
      BH Cep &   3.81 &   0.23 &   1.72 &   0.38 &   0.95 &      1.73 &      8.21 \\
      BO Cep &   3.82 &   0.17 &   4.67 &   0.76 &   0.82 &      1.50 &     11.25 \\
      SV Cep &   3.98 &   0.39 &   5.33 &   2.00 &   1.71 &      2.50 &      2.54 \\
    LkHa 233 &   3.93 &   0.72 &   5.56 &   3.85 &   1.97 &      3.19 &      0.76 \\
    MWC 1080 &   4.48 &   1.64 &   5.23 &   8.24 &   6.53 & $>$  6.00 & $<$  0.01 \\  
   
\enddata

\end{deluxetable}

\clearpage

\begin{deluxetable}{ccccccc}
\tablewidth{0pt}
\tablecolumns{7}
\tablecaption{H${\alpha}$ equivalent widths of Herbig Ae/Be stars. Negative values of equivalent widths signifies emission. \label{ew}}
\tablehead{
\colhead{Object Name} & \multicolumn{6}{c}{ H$\alpha$ equivalent width ($\AA$)} \\ \cline{2-7} 
\colhead{} & \colhead{This work} & \colhead{HB88} & \colhead{Her04} & \colhead{CR98} & 
\colhead{AVD05}  & \colhead{BPL96} }

\startdata
 LkHa 198 &    ... &   -85.00 &   -56.20 &   -40.50 &    ... &    ...\\
    V376 Cas &    ... &   -37.00 &   -22.00 &   -33.40 &   -29.70 &    ...\\
      VX Cas &    ... &   -20.00 &   -19.20 &    ... &   -12.90 &    ...\\
   BD+61 154 &   -68.00 &   -78.00 &    ... &   -44.10 &    ... &    ...\\
     HBC 334 &    ... &    ... &    -0.20 &    ... &    ... &    ...\\
      IP Per &   -16.60 &    ... &   -21.40 &    ... &    ... &    ...\\
      XY Per &    -6.30 &    ... &    -4.70 &    ... &    ... &    ...\\
    V892 Tau &    ... &   -13.00 &   -17.80 &   -20.60 &    ... &    ...\\
      AB Aur &   -44.00 &   -27.00 &   -28.20 &   -26.30 &   -54.00 &    ...\\
    HD 31648 &   -22.50 &    ... &    ... &    ... &   -18.30 &    ...\\
      UX Ori &    -5.10 &   -20.00 &    -2.30 &    ... &    ... &    -5.40\\
    HD 34282 &    -4.00 &    ... &    ... &    ... &    -4.70 &    ...\\
    HD 35187 &    -5.10 &    ... &    ... &    ... &    ... &    ...\\
      CO Ori &    ... &   -10.00 &    ... &    ... &    ... &    -4.20\\
    HD 35929 &    -2.40 &    ... &    ... &    ... &    -4.80 &    ...\\
    HD 36112 &   -14.50 &    ... &    ... &    ... &   -19.70 &    ...\\
      HK Ori &   -51.00 &   -56.00 &   -49.00 &   -63.10 &    ... &   -40.20\\
   HD 244604 &    ... &    ... &    ... &    ... &   -15.00 &    ...\\
      RY Ori &    ... &    -3.40 &    -7.00 &    ... &    ... &    ...\\
     HBC 442 &    ... &    ... &    -1.30 &    ... &    ... &    ...\\
   HD 245185 &   -25.50 &    ... &   -21.20 &    ... &   -29.50 &    ...\\
       T Ori &    ... &   -16.00 &   -21.00 &    ... &    ... &    ...\\
    HD 36910 &    ... &    ... &    -6.20 &    ... &    ... &    ...\\
    V380 Ori &    ... &   -81.00 &    ... &   -71.00 &    ... &   -94.20\\
    V586 Ori &    ... &   -16.00 &    ... &    ... &   -17.10 &   -12.40\\
      BF Ori &    ... &   -10.00 &    -6.70 &    -3.70 &    -9.30 &   -11.30\\
   HD 245906 &    ... &    ... &    -3.60 &    ... &    ... &    ...\\
      RR Tau &    ... &   -33.00 &   -25.70 &    ... &    ... &    ...\\
    V350 Ori &    ... &    ... &   -29.90 &    ... &    ... &    ...\\
    HD 37806 &   -25.40 &    ... &    ... &    ... &    ... &    ...\\
    HD 38120 &   -24.00 &    ... &    ... &    ... &    ... &    ...\\
   HD 250550 &   -37.60 &   -40.00 &   -24.80 &    ... &    ... &    ...\\
    LkHa 208 &    ... &    -8.00 &    -4.90 &    ... &    ... &    ...\\
    Lkha 338 &    ... &   -55.00 &   -51.00 &    ... &    ... &    ...\\
    Lkha 339 &    ... &   -19.00 &   -19.40 &    ... &    ... &    ...\\
      VY Mon &    ... &   -50.00 &   -28.00 &    ... &    ... &    ...\\
    LkHa 215 &    ... &   -25.00 &   -25.70 &   -26.70 &    ... &    ...\\
   HD 259431 &   -62.00 &   -55.00 &   -57.50 &   -52.90 &    ... &    ...\\
       R Mon &    ... &   -85.00 &    ... &   -93.10 &    ... &   -47.00\\
     HBC 217 &    ... &    ... &   -10.30 &    ... &    ... &    ...\\
     LkHa 25 &    ... &   -47.00 &   -47.30 &   -58.20 &    ... &    ...\\
     HBC 222 &    ... &    ... &    -1.20 &    ... &    ... &    ...\\
    LkHa 218 &   -26.50 &   -22.00 &   -32.30 &   -31.90 &    ... &    ...\\
       Z CMa &  -106.00 &   -10.00 &    ... &   -24.80 &   -45.20 &   -20.90\\
     HBC 551 &    ... &   -44.00 &   -54.50 &    ... &    ... &    ...\\
      NX Pup &   -44.00 &    ... &    ... &   -44.20 &    ... &   -48.30\\
    HD 68695 &    -8.50 &    ... &    ... &    ... &    ... &    ...\\
    HD 85567 &   -42.00 &    ... &    ... &    ... &    ... &    ...\\
    HD 95881 &    ... &    ... &    ... &    ... &   -21.10 &    ...\\
    HD 97048 &    ... &   -30.00 &    ... &   -51.10 &   -36.00 &   -48.70\\
    HD 98922 &   -27.00 &    ... &    ... &    ... &   -27.90 &    ...\\
   HD 100453 &    -0.80 &    ... &    ... &    ... &     1.50 &    ...\\
   HD 100546 &    ... &    ... &    ... &    ... &   -36.80 &    ...\\
   HD 101412 &    -8.70 &    ... &    ... &    ... &   -20.40 &    ...\\
   HD 104237 &    ... &    ... &    ... &    ... &   -24.50 &    ...\\
   HD 135344 &    ... &    ... &    ... &    ... &   -17.40 &    ...\\
   HD 139614 &    -7.00 &    ... &    ... &    ... &    -9.80 &    ...\\
   HD 141569 &    -4.40 &    ... &    ... &    ... &    -6.70 &    ...\\
   HD 142666 &    -1.50 &    ... &    ... &    ... &    -3.20 &    ...\\
   HD 142527 &   -10.20 &    ... &    ... &    ... &   -17.90 &    ...\\
   HD 144432 &    -7.50 &    ... &    ... &    ... &    -9.00 &    ...\\
     HR 5999 &    -5.70 &    -7.00 &    ... &    ... &   -11.00 &   -10.00\\
   HD 150193 &   -11.20 &    ... &    ... &    -7.80 &    -5.00 &   -10.70\\
      AK Sco &    -4.10 &    ... &    ... &    ... &    -9.40 &    -2.80\\
      KK Oph &   -59.10 &   -22.00 &   -59.40 &   -44.90 &    ... &    ...\\
      51 Oph &    -1.50 &    ... &    ... &    ... &    ... &    ...\\
   HD 163296 &   -22.10 &    ... &    ... &   -22.50 &   -14.50 &    ...\\
   HD 169142 &    ... &    ... &    ... &    ... &   -14.00 &    ...\\
      VV Ser &   -46.80 &   -22.00 &   -61.10 &   -31.80 &   -81.30 &   -46.50\\
      AS 310 &    ... &    -5.00 &    -7.70 &    ... &    ... &    ...\\
       R CrA &   -85.50 &    ... &    ... &    ... &   -44.30 &  -102.00\\
       T CrA &   -15.10 &    ... &    ... &    ... &   -10.70 &   -18.60\\
   HD 179218 &   -13.00 &    ... &    ... &    ... &   -18.20 &    ...\\
   HD 344361 &    ... &   -30.00 &   -14.40 &    ... &   -24.00 &   -17.10\\
      PX Vul &    ... &    ... &    -6.40 &    ... &    ... &    ...\\
   V1295 Aql &   -22.20 &    ... &    ... &    ... &   -27.10 &    ...\\
  BD+40 4124 &  -120.00 &   -94.00 &  -108.00 &   -31.80 &  -119.90 &    ...\\
   V1686 Cyg &    ... &    ... &    -3.60 &    ... &   -17.60 &    ...\\
    LkHa 134 &    ... &    ... &    ... &   -31.54 &    ... &    ...\\
    LkHA 168 &    ... &   -14.00 &   -19.00 &    ... &    ... &    ...\\
   HD 200775 &   -61.00 &   -35.00 &   -59.30 &   -17.70 &   -74.60 &    ...\\
    Lkha 324 &    ... &    ... &   -15.30 &    ... &    ... &    ...\\
    V645 Cyg &    ... &    ... &    ... &   -57.40 &  -121.70 &    ...\\
    V361 Cep &    ... &   -36.00 &   -28.00 &   -12.70 &    ... &    ...\\
    LkHa 234 &    ... &   -44.00 &   -68.90 &   -52.10 &    ... &    ...\\
  BD+46 3471 &   -21.10 &   -20.00 &   -18.60 &   -19.70 &   -21.70 &    ...\\
      BH Cep &    ... &    -3.00 &    -6.20 &    ... &    ... &    ...\\
      BO Cep &    -4.00 &    -4.50 &    -2.30 &    ... &    ... &    ...\\
      SV Cep &    ... &   -20.00 &   -12.10 &    ... &   -10.10 &    ...\\
    LkHa 233 &    ... &   -18.00 &   -20.50 &   -26.40 &    ... &    ...\\
    MWC 1080 &  -124.00 &   -75.00 &    ... &   -94.20 &  -122.80 &    ...\\
   
\enddata
\tablerefs{\small{
HB88: \citet{herbigbell88};  Her04: \citet{hernand04};    CR98: \citet{corray98}; 
  AVD05: \citet{acke05};   BPL96: \citet{bpl96} }}
\end{deluxetable}

\clearpage

\begin{deluxetable}{ccccccc}
\tablewidth{0pt}
\tablecolumns{7}
\tablecaption{2MASS magnitudes of Herbig Ae/Be stars. \label{2mass}}
\tablehead{
\colhead{Object Name}   & \colhead{2MASS}      &  \colhead{$J$}  & \colhead{$H$} & 
\colhead{ $K_s$}  &\colhead{ $\Delta(H-K)$}  & \colhead{e$_{\Delta(H-K)}$} \\
\colhead{}   &\colhead{ Designation} &\colhead{ mag}  &\colhead{mag} & \colhead{mag}&
 \colhead{mag}  & \colhead{mag} } 
        
\startdata
 BD+61 154 &   00431825+6154402 &   8.14 &   6.92 &   5.90 &   0.71 &   0.04\\
      IP Per &   03404696+3231537 &   9.14 &   8.41 &   7.59 &   0.70 &   0.03\\
      XY Per &   03493638+3858556 &   7.65 &   6.92 &   6.09 &   0.66 &   0.03\\
    HD 31648 &   04584626+2950370 &   6.87 &   6.26 &   5.53 &   0.69 &   0.04\\
      UX Ori &   05042998-0347142 &   8.71 &   8.04 &   7.21 &   0.68 &   0.05\\
    HD 34282 &   05160047-0948353 &   9.26 &   8.47 &   7.68 &   0.73 &   0.05\\
    HD 35187 &   05240118+2457370 &   6.95 &   6.48 &   5.91 &   0.47 &   0.03\\
      CO Ori &   05273833+1125389 &   7.98 &   7.21 &   6.51 &   0.45 &   0.05\\
    HD 35929 &   05274279-0819386 &   7.21 &   6.97 &   6.67 &   0.21 &   0.06\\
    HD 36112 &   05302753+2519571 &   7.22 &   6.56 &   5.80 &   0.67 &   0.04\\
      HK Ori &   05312805+1209102 &   9.41 &   8.31 &   7.34 &   0.67 &   0.05\\
   HD 244604 &   05315724+1117414 &   8.61 &   7.96 &   7.12 &   0.67 &   0.04\\
      RY Ori &   05320993-0249467 &   9.44 &   8.88 &   8.28 &   0.36 &   0.07\\
     HBC 442 &   05341416-0536542 &   8.99 &   8.63 &   8.28 &   0.24 &   0.08\\
   HD 245185 &   05350960+1001515 &   9.29 &   8.76 &   8.02 &   0.68 &   0.07\\
       T Ori &   05355043-0528349 &   8.27 &   7.24 &   6.22 &   0.81 &   0.06\\
    HD 36910 &   05355845+2444542 &   7.93 &   7.06 &   6.17 &   0.69 &   0.04\\
    V380 Ori &   05362543-0642577 &   8.11 &   6.96 &   5.95 &   0.63 &   0.06\\
    V586 Ori &   05365925-0609164 &   8.95 &   8.41 &   7.69 &   0.60 &   0.07\\
      BF Ori &   05371326-0635005 &   9.11 &   8.56 &   7.90 &   0.52 &   0.05\\
   HD 245906 &   05393048+2619552 &   9.13 &   8.51 &   7.98 &   0.39 &   0.05\\
      RR Tau &   05393051+2622269 &   9.69 &   8.42 &   7.39 &   0.74 &   0.04\\
    V350 Ori &   05401176-0942110 &  10.02 &   9.24 &   8.37 &   0.70 &   0.04\\
    HD 37806 &   05410229-0243006 &   7.12 &   6.25 &   5.40 &   0.81 &   0.06\\
    HD 38120 &   05431188-0459499 &   8.43 &   7.85 &   7.16 &   0.63 &   0.09\\
   HD 250550 &   06015998+1630567 &   8.47 &   7.53 &   6.63 &   0.79 &   0.04\\
   HD 259431 &   06330519+1019199 &   7.45 &   6.67 &   5.73 &   0.70 &   0.05\\
     HBC 217 &   06404218+0933374 &  10.76 &  10.28 &   9.84 &   0.33 &   0.04\\
     HBC 222 &   06405118+0944461 &  10.74 &  10.23 &   9.67 &   0.45 &   0.04\\
    LkHa 218 &   07024252-1126118 &  10.26 &   9.42 &   8.58 &   0.68 &   0.04\\
     HBC 551 &   07040669-1126084 &  10.76 &  10.02 &   9.12 &   0.75 &   0.04\\
      NX Pup &   07192826-4435114 &   8.58 &   7.29 &   6.08 &   0.95 &   0.07\\
    HD 68695 &   08114457-4405087 &   9.35 &   8.78 &   8.08 &   0.63 &   0.05\\
    HD 95881 &   11015764-7130484 &   7.38 &   6.66 &   5.73 &   0.85 &   0.05\\
    HD 97048 &   11080329-7739174 &   7.27 &   6.67 &   5.94 &   0.59 &   0.06\\
   HD 100453 &   11330559-5419285 &   6.95 &   6.39 &   5.60 &   0.72 &   0.05\\
   HD 100546 &   11332542-7011412 &   6.42 &   5.96 &   5.42 &   0.50 &   0.05\\
   HD 101412 &   11394445-6010278 &   8.63 &   8.22 &   7.47 &   0.68 &   0.06\\
   HD 104237 &   12000511-7811346 &   5.81 &   5.25 &   4.58 &   0.57 &   0.07\\
   HD 135344 &   15154844-3709160 &   7.28 &   6.59 &   5.84 &   0.65 &   0.04\\
   HD 139614 &   15404638-4229536 &   7.67 &   7.33 &   6.75 &   0.50 &   0.06\\
   HD 141569 &   15495775-0355162 &   6.87 &   6.86 &   6.82 &   0.00 &   0.05\\
   HD 142666 &   15564002-2201400 &   7.35 &   6.74 &   6.08 &   0.52 &   0.04\\
   HD 142527 &   15564188-4219232 &   6.50 &   5.71 &   4.98 &   0.58 &   0.04\\
   HD 144432 &   16065795-2743094 &   7.09 &   6.54 &   5.89 &   0.56 &   0.07\\
   HD 150193 &   16401792-2353452 &   6.95 &   6.21 &   5.48 &   0.55 &   0.05\\
      AK Sco &   16544485-3653185 &   7.68 &   7.06 &   6.50 &   0.47 &   0.04\\
      KK Oph &   17100811-2715190 &   9.07 &   7.23 &   5.79 &   1.05 &   0.06\\
   HD 163296 &   17562128-2157218 &   6.20 &   5.53 &   4.78 &   0.69 &   0.05\\
   HD 169142 &   18242978-2946492 &   7.31 &   6.91 &   6.41 &   0.41 &   0.05\\
      VV Ser &   18284786+0008397 &   8.67 &   7.43 &   6.32 &   0.70 &   0.07\\
   HD 179218 &   19111124+1547155 &   6.99 &   6.64 &   6.00 &   0.59 &   0.04\\
   HD 344361 &   19255874+2112313 &   9.09 &   8.18 &   7.28 &   0.77 &   0.05\\
      PX Vul &   19264025+2353508 &   9.32 &   8.55 &   7.91 &   0.57 &   0.03\\
   V1295 Aql &   20030250+0544166 &   7.19 &   6.65 &   5.86 &   0.72 &   0.04\\
  BD+40 4124 &   20202825+4121514 &   7.90 &   6.79 &   5.77 &   0.62 &   0.03\\
    LkHA 168 &   20520604+4417160 &  10.27 &   9.23 &   8.34 &   0.55 &   0.04\\
      BH Cep &   22014287+6944364 &   9.69 &   8.99 &   8.31 &   0.59 &   0.04\\
      BO Cep &   22165406+7003450 &  10.32 &   9.85 &   9.58 &   0.14 &   0.04\\
      SV Cep &   22213319+7340270 &   9.35 &   8.56 &   7.74 &   0.64 &   0.06\\
    LkHa 233 &   22344101+4040045 &  11.29 &  10.31 &   8.92 &   1.05 &   0.04\\
    MWC 1080 &   23172558+6050436 &   7.46 &   5.98 &   4.83 &   0.64 &   0.05\\     
\enddata

\end{deluxetable}

\end{document}